\documentstyle[12pt]{article}
\title{Linear Boltzmann equation as the long time dynamics
of an electron weakly coupled to a
phonon field}
\author{L\'aszl\'o Erd\H os 
\thanks{Partially supported by NSF grant DMS-9970323}
\\ School of Mathematics, GeorgiaTech \\ Atlanta, GA 30332, USA \\
 {\verb -lerdos@math.gatech.edu-}}

\date{Jul 1, 2001}

\newtheorem{theorem}{Theorem}[section]
\newtheorem{proposition}[theorem]{Proposition}

\newtheorem{lemma}[theorem]{Lemma}

\newtheorem{definition}[theorem]{Definition}

\newcommand{\rd}{{\rm d}}
\newcommand{\be}{\begin{equation}}
\newcommand{\ee}{\end{equation}}
\newcommand{\bey}{\begin{eqnarray}}
\newcommand{\eey}{\end{eqnarray}}

\newcommand{\sfrac}[2]{{\textstyle \frac{#1}{#2}}}

\newcommand{\tJ}{{\widetilde J}}
\newcommand{\tA}{{\widetilde A}}
\newcommand{\tR}{{\widetilde R}}
\newcommand{\tS}{{\widetilde S}}
\newcommand{\tb}{{\tilde b}}
\newcommand{\tc}{{\tilde c}}

\newcommand{\ts}{{\tilde s}}
\newcommand{\tn}{{\tilde n}}

\newcommand{\tI}{{\tilde I}}
\newcommand{\tN}{{\widetilde N}}
\newcommand{\tM}{{\widetilde M}}
\newcommand{\tm}{{\widetilde m}}
\renewcommand{\tt}{t}  

\newcommand{\ta}{{\tilde\alpha}}
\newcommand{\tbeta}{{\tilde\beta}}
\newcommand{\tsi}{{\tilde\sigma}}
\newcommand{\ttau}{{\tilde\tau}}

\newcommand{\tchi}{{\tilde\chi}}
\newcommand{\ttheta}{{\tilde\theta}}

\newcommand{\tbk}{{\tilde{k}}}
\newcommand{\tbp}{{\tilde{ p}}}

\newcommand{\tmu}{{\tilde\mu}}

\renewcommand{\a}{\alpha}

\newcommand{\g}{\gamma}
\newcommand{\e}{\varepsilon}
\newcommand{\s}{\sigma}
\newcommand{\la}{\lambda}
\newcommand{\om}{{\omega}}

\newcommand{\bE}{{\bf E}}

\newcommand{\cM}{{\cal M}}
\newcommand{\cX}{{\cal X}}

\newcommand{\bR}{{\bf R}}

\newcommand{\bN}{{\bf N}}
\newcommand{\bZ}{{\bf Z}}

\newcommand{\bw}{ w}
\newcommand{\bp}{ p}
\newcommand{\bq}{ q}
\newcommand{\br}{ r}
\newcommand{\bv}{ v}
\newcommand{\bu}{ u}
\newcommand{\bk}{ k}
\newcommand{\bm}{ m}
\newcommand{\bx}{ x}
\newcommand{\by}{ y}
\newcommand{\bz}{ z}

\newcommand{\bP}{P}
\newcommand{\bU}{ U}
\newcommand{\bX}{ X}
\newcommand{\bV}{ V}

\newcommand{\Tr}{\mbox{Tr}}

\newcommand{\wt}{\widetilde}
\newcommand{\wh}{\widehat}
\newcommand{\ov}{\overline}

\newcommand{\bxi}{\xi}

\newcommand{\cG}{{\cal G}}
\newcommand{\cS}{{\cal S}}

\newcommand{\cF}{{\cal F}}
\newcommand{\cA}{{\cal A}}
\newcommand{\cB}{{\cal B}}
\newcommand{\cE}{{\cal E}}
\newcommand{\cP}{{\cal P}}
\newcommand{\cD}{{\cal D}}

\newcommand{\cH}{{\cal H}}

\newcommand{\cN}{{\cal N}}
\newcommand{\cO}{{\cal O}}
\newcommand{\ua}{\underline{a}}
\newcommand{\ub}{\underline{b}}
\newcommand{\ualpha}{\underline{\alpha}}
\newcommand{\usi}{\underline{\sigma}}
\newcommand{\ubp}{\underline{\bp}}
\newcommand{\ubk}{\underline{\bk}}

\newcommand{\utbp}{\underline{\tbp}}
\newcommand{\utbk}{\underline{\tbk}}
\newcommand{\um}{\underline{m}}
\newcommand{\utm}{\underline{\tm}}
\newcommand{\utau}{\underline{\tau}}
\newcommand{\uttau}{\underline{\ttau}}
\newcommand{\Om}{\Omega}

\input epsf

\begin{document}
\maketitle

\centerline{\it Dedicated to Domokos Sz\'asz on the occassion of his $60^{th}$
birthday}

\begin{abstract}
We consider the long time evolution of a quantum particle weakly
interacting with a  phonon field. We show that
in the weak coupling limit the Wigner distribution of
the electron density matrix converges to the solution
of the linear Boltzmann equation globally in time.
The collision
kernel is identified as the sum of an emission
and an absorption term that depend on the equilibrium distribution of
the free phonon modes.
\end{abstract}



\medskip\noindent
{\it Running title:} Boltzmann equation from phonons.

\newpage

\tableofcontents

\section{Introduction}

Quantum evolution of a few particles can  effectively be computed by
using the Schr\"odinger equation. Applying the same
first principles to macroscopic systems with many degrees of freedom,
 however, is practically impossible. 
Macroscopic theories have to be developed
which retain the relevant features of the original problems but are simple
enough to be  computationally feasible.

Due to thermal energy,
lattice vibrations in metals can result in random deviations
from the periodic background potential. This vibration
is usually modelled by many independent harmonic oscillators.
These are called phonons, 
and they can be considered bosonic particles.

The interaction between the phonons and the electron
is  a collision process in which a
phonon can be emitted or absorbed,
subject to momentum and energy conservation.
The number of  phonons is not conserved, hence they
are described by second quantization.

Our goal is to show that the
long time Schr\"odinger evolution of a quantum
particle (electron) in a phonon field can be described on large scales
by the Boltzmann equation in $d\ge 3$ dimensions.
We use two sets of variables; $(\bx,t)$ stand for the microscopic space
and time, and let
$$
	(\bX, T):= (\e\bx, \e t)
$$
be the macroscopic variables. Here $\e$ is the scaling
parameter separating microscopic and macroscopic scales and
we will consider the $\e\to0$  {\it scaling limit}.
The velocity is unscaled.
The quantum dynamics is given in  microscopic variables.
The resulting Boltzmann equation gives the evolution of the time
dependent phase space density, $F_T(X,V)$, in macroscopic variables,
hence it contains only the macroscopic features of the evolution.
We recall that the linear Boltzmann equation with dispersion
relation $e(\bk)$ and collision kernel
$\sigma(\bV, \bU)$ is
\bey
\lefteqn{\partial_T F_T(X,V) + \nabla e(V)\cdot\nabla_X F_T(X, V)}\qquad
        \qquad \nonumber\\
        & = &\displaystyle{\int
	 \Big[ \sigma(\bV, \bU) F_T(X, U) - \sigma(\bU, \bV)F_T(X, V)
        \Big] \rd U} \\
        & = & \displaystyle{
	\int  \sigma(\bV, \bU) F_T(X, U)\rd U - \sigma_0(V) \, F_T(X, V)}
\label{eq:Beq}
\eey
with the total cross section $\sigma_0(V) : = \int \sigma(\bU, \bV)\rd \bU$.

\bigskip

{\it Acknowledgements.}
This work initially was a joint project with H.-T. Yau and several
ideas presented here have been developed in collaboration with him. 
 I would like to thank him for the invaluable discussions and encouragement
through the entire work.
 
Part of this project was completed during several visits
at the Erwin Schr\"odinger Institute, Vienna, and at the Center
of Theoretical Studies, Hsinchu, Taiwan. The author is grateful
for the hospitality and financial support.

\subsection{Definitions}

For convenience, we fix a convention to avoid carrying factors of $2\pi$
along the Fourier transforms.
We {\it define} the measure $\rd \bx$ on $\bR^d$ to be
the Lebesgue measure {\it divided} by $(2\pi)^{d/2}$.
With this convention the $d$-dimensional Fourier
transform (usually denoted by hat) is defined as
$$
	\wh f(\bp)  = \cF f(\bp) := \int_{\bR^d}
	 f(\bx) e^{-i\bp\cdot \bx} \rd \bx
$$
and its inverse
$$
	f(\bx) = \cF^{-1} \wh f(\bx) = 
	\int_{\bR^d} \wh f(\bp) e^{i\bp\cdot\bx}\rd \bp 
	\; .
$$
In particular, $\int_{\bR^d} e^{i\bp\cdot\bx} \rd \bx = \delta(\bp)$,
 where $\delta(\bp)$ is
understood with respect to the $\rd \bp$ measure. This
convention applies only to $d$-dimensional integrals, in one
dimension $\rd \alpha$ still denotes the usual Lebesgue measure and
$\int_{-\infty}^\infty e^{i\alpha t} \rd\alpha = 2\pi\delta(t)$. 
For functions $\psi \in C^1(\bR^d)$ we also define
the measure $\delta(\psi(\bp))\rd\bp$  supported on the
level set $\{ \psi =0 \}$ by
$$
	\int F(\bp) \delta(\psi(\bp))\rd\bp
	: = \int_{\psi^{-1}(0)} {F(\bp)\over |\nabla\psi(\bp)|}
	\rd\sigma(\bp)\;, \qquad F\in C(\bR^d)\;, 
$$
where $\rd\sigma$ is the $(d-1)$-dimensional 
surface measure inherited from $\rd\bp$.

The configuration space is a big box
$\Lambda= \Lambda_L:=[-\sqrt{2\pi}L/2, \sqrt{2\pi}L/2]^d\subset\bR^d$ 
in $d\ge 3$ dimensions with $\int_\Lambda \rd \bx =L^d$.
Its dual is $\Lambda^*= \Lambda_L^{dual}: = (\sqrt{2\pi} L^{-1}\bZ)^d
\subset \bR^d$.
We use the shorter integral notation for the normalized sum on
the dual lattice
$$
	\int_{\Lambda^*}\rd\bk := 
	 {1\over L^d}\sum_{\bk\in \Lambda^*} \; .
$$
With this notation $\int_\Lambda e^{i\bk \cdot\bx} \rd \bx = \delta(\bk)$ and
$\int_{\Lambda^*}e^{i\bk\cdot\bx} \rd \bk =\delta(\bx)$ where the delta
functions are with respect to the measures $\rd\bk$ and $\rd \bx$.
In particular, $\delta (k)$ is the lattice delta function, i.e., 
$\delta(k)=L^d$ if $k=0$ and is zero otherwise.

\medskip

{\it Convention.} Letters $\bx,\by,\bz$ will usually denote
configuration variables in $\Lambda\subset \bR^d$; 
$\bk,\bp,\bq$, $\br, \bu, \bv, \bxi$ stand for momentum
variables in $\Lambda^*\subset\bR^d$; $s,t,\alpha, \beta,\e$,
$\nu,\eta,\tau, \theta, \omega$ are real scalars and $i,j,\kappa$, $\ell,m,n$
are integers. The same convention applies to capital letters.

\medskip

For simplicity,  we will omit the domains of integration in the notations.
In general, the notation $\int$ is used for $d$-dimensional
integration with respect to the modified Lebesgue measure on $\bR^d$
if the $L\to\infty$ limit has already been taken. Otherwise,
the domain of integration is $\Lambda$ or $\Lambda^*$, depending
on the integration variable.
For one dimensional integrations the limits will always be indicated.

We will take the thermodynamic limit $L\to\infty$ before any other limit
and our result will be uniform in $L$. The compact configuration space
is a mere convenience in order to define certain expressions rigorously;
for most purposes
 the reader can always think of $\bR^d$ instead of $\Lambda$ and $\Lambda^*$.

\bigskip

The electron is considered as  a spinless nonrelativistic particle
with state space  $\cH_e: = L^2_{per}(\Lambda_L)$.
 We denote the electron
dispersion relation by $e(\bk)$, $\bk\in \bR^d$,
i.e., the electron Hamiltonian, $H_e$, is multiplication by $e(\bk)$
in momentum space $(k\in \Lambda^*)$.

The state space of $m$ independent phonons is $\cH_p^m:=
\cS\Big[ \otimes_{j=1}^m L^2_{per}(\Lambda_L)\Big]$,
 where $\cS$ is the symmetrization
operator. Their Hamiltonian, in momentum representation, is
a multiplication operator by
$$
	H_{ph}^{m}(\bk_1, \bk_2, \ldots, \bk_m) =
	 \sum_{j=1}^m \omega(\bk_j)\; , \qquad \bk_j\in\Lambda^*\; ,
$$
where $\omega(\bk)$ is the phonon dispersion law defined for $\bk\in\bR^d$.

The Fock space of the phonons is $\cH_{ph} : 
= \oplus_{m=0}^\infty \cH_{ph}^m$,
and the phonon Hamiltonian is $H_{ph}= \oplus_m H_{ph}^m$.
The total Hilbert space, including the electron, is
$\cH_{tot}: =\cH_e \otimes \cH_{ph}$. The $m$-phonon sector
consists of wavefunctions $\Psi^m = \Psi^m (\bx; \bk_1, \bk_2, \ldots,  \bk_m)
\in \cH_e\otimes \cH_{ph}^m$, where $\bx$ is the electron coordinate. The
elements of $\cH_{tot}$ are denoted by $\Psi = ( \Psi^m )_{m=0}^\infty$.
The total noninteracting
Hamiltonian of the electron-phonon system is $H_e\otimes 1 + 1\otimes H_{ph}$,
which we shall write as $H_e + H_{ph}$ for brevity.

To define the interaction, we need
to define the phonon creation and annilihation operators
$c^\dagger_\bk$ and $c_\bk$ as
\bey
	(c_\bk\Psi)^m(\bx; \bk_1, \bk_2, \ldots, \bk_m) &: =&
	\sqrt{m+1} \, \Psi^{m+1}(\bx; \bk, \bk_1, \bk_2, \ldots, \bk_m)\\
	(c_\bk^\dagger\Psi)^m(\bx; \bk_1, \bk_2, \ldots, \bk_m) &: =&
	{1\over \sqrt{m}} \sum_{j=1}^m \Psi^{m-1}(\bx; \bk_1, \ldots,
	\widehat \bk_j,  \ldots, \bk_m)\delta(\bk_j-\bk)
\eey
(hat denotes omission).
These operators satisfy the standard commutation relations, i.e.,
$$
	[c_\bk, c_{\bk'}^\dagger] = \delta(\bk-\bk')
$$
and any other commutator is zero. 
In terms of these operators $H_{ph} = \int \omega(\bk) N_\bk \rd \bk$,
where $N_\bk : = c_\bk^\dagger c_\bk$ is the number operator.

\medskip

The equilibrium state of the phonons is the Gibbs state
\be
	\gamma_{ph} : = Z^{-1} \exp{\Big( -\beta H_{ph}+ 
	\mu\int  N_\bk \rd \bk \Big)}
	=  Z^{-1} \exp{ \Bigg( \int \Big[ -\beta\om(\bk)
	+\mu\Big] N_\bk \rd \bk \Bigg)}
\label{phdens}\ee
with inverse temperature $\beta>0$ and chemical potential $\mu$.
Here $Z=Z_\Lambda$ is the normalization, i.e.,
\be
	Z: = \Tr_{\cH_{ph}}  \exp{  \Bigg(\int \Big[ -\beta\om(\bk)
	+\mu\Big] N_\bk \rd \bk\Bigg) } 
\label{def:Z}\ee
such that $\hbox{Tr}_{\cH_{ph}} \gamma_{ph} =1$.
We define
\be
	\cN (\bk): =\hbox{Tr}_{\cH_{ph}}\Big( \gamma_{ph} N_\bk \Big)
	=  {e^{-\beta \omega(\bk)+\mu}\over 1-e^{-\beta \omega(\bk)+\mu}}
\label{Np}\ee
to be the expected number of phonons in the mode $\bk$.

\medskip

The interaction Hamiltonian is given by
\be
	H_{e-p} : = i\lambda\int_{\Lambda^*}
	 Q(\bk)\Big[ e^{-i\bk\cdot\bx}c_\bk^\dagger -
	e^{i\bk\cdot \bx} c_\bk\Big] \rd \bk,
\label{interaction}\ee
as a multiplication operator on $\cH_e$,
where $Q(\bk)$ models the details of the electron-phonon interaction
and $\la$ is the coupling constant. We will choose
$\lambda = \sqrt{\e}$ and this is the weakest coupling
that yields a nontrivial (non-free) macroscopic evolution on the time
scale $t=T/\e$.

\bigskip

The full dynamics is given by the Schr\"odinger equation
\be
	i\partial_t \Psi_t  = H \Psi_t,
\label{phonschr}\ee
where 
$$
	H= H_e+ H_{ph}+ H_{e-p}
$$
is the full Hamiltonian.
In particular, for
$Q(\bk)= |\bk|^{-1}$ we obtain the well-known Fr\"ohlich Hamiltonian
describing the polaron.

The Schr\"odinger equation  can be
written as an evolution equation for density matrices $\Gamma_t$
on $\cH_{tot}$:
\be
	i\partial_t \Gamma_t = [H, \Gamma_t].
\label{densev}\ee
For example, (\ref{densev}) and (\ref{phonschr}) are equivalent
if $\Gamma_t $ is
the projection operator onto the state $\Psi_t$.
 The density matrix formalism is more
general, and is appropriate to describe thermal states.

We let $\gamma_e= \gamma_{e,0}$, $\Tr\; \gamma_e=1$,
 be the initial electron density
matrix; if we start from a pure electron state $\psi_0$, then
 $\gamma_e=\gamma_{e, 0}$
is the projection onto $\psi_0$. The total initial
density matrix is 
\be
	\Gamma_0: = \gamma_e \otimes \gamma_{ph} \; .
\label{initstate}\ee
The solution of (\ref{densev}) is $\Gamma_t = e^{-itH}\Gamma_0e^{itH}$.

We are interested only in the evolution of the electron, 
i.e., we want to compute $\hbox{Tr}_\cH \Big(\Gamma_t \cO\Big)$, where $\cO$
is an observable acting on $\cH_e$. We can first take 
partial trace $\hbox{Tr}_{\cH_{ph}} = : \Tr_{ph}$ to integrate out
the phonon modes to obtain a density matrix on $\cH_e$. The phase
space properties of such density matrix are described by the Wigner
distribution.

The Wigner distribution $W_\gamma$ of any density matrix $\gamma$
on $\cH_e$ is defined as
\be
	W_\gamma(\bx, \bv): =\int_\Lambda e^{-i\bv\cdot\by} \gamma\Big(
	\bx + \frac{\by}{2}, \bx - \frac{\by}{2}\Big) \rd\by \; ,
\label{def:wig}\ee
or in momentum space
$$
	W_\gamma(\bx, \bv): =\int_\Lambda e^{i\bu\cdot\bx} \wh\gamma\Big(
	\bv + \frac{\bu}{2}, \bv - \frac{\bu}{2}\Big) \rd\by \; ,
$$
where hat on density matrices denotes Fourier transform in both variables.
We use the convention that the hat
 on functions  defined on the phase space
$L^2(\Lambda \times \Lambda^*)$  means Fourier transform in the
first variable only. In particular, the Fourier transform of the Wigner
distribution is
\be
	\wh W_\gamma(\bxi, \bv) := \int_\Lambda
	 e^{-i\xi\cdot \bx} W_\g(\bx, \bv)\rd\bx
	=\wh\gamma\Big(
	\bv + \frac{\bxi}{2}, \bv - \frac{\bxi}{2}\Big) \; .
\label{def:wigfour}
\ee
For pure states, $\gamma= |\psi\rangle\langle\psi|$, we have
$\gamma(x,y)=\psi(x)\overline{\psi}(y)$ and $W_\gamma$ is equal
to $W_\psi$, the Wigner transform of the wavefunction $\psi$,
in accordance with \cite{EY2}.

We also define the rescaled Wigner distribution to describe large
scale (macroscopic) properties of the density matrix
\be
	W^\e_\gamma (\bX, \bV): = 
	\e^{-d} W_\gamma\Big( {\bX\over \e}, \bV\Big)\; .
\label{def:reswig}\ee

\subsection{Assumptions}\label{sec:assump}

We introduce the notation  $\langle x \rangle : = (x^2 + 1)^{1/2}$.
We use indexed constants $C_1, C_2, \ldots$ to quantify 
estimates in our assumptions.

For the electron and phonon 
dispersion relations, $e(\bk), \;\om (\bk)\ge 0$, we assume symmetry,
$e(\bk)=e(-\bk)$, $\om(\bk)=\om(-\bk)$ and 
\be
	|\nabla^\ell e(\bk)| \leq C_1(1+\langle \bk \rangle^{2-\ell}) \qquad
	\ell= 0,1,\ldots, 2d\;,
\label{eq:econd}
\ee
\be
	|\nabla^\ell \om(\bk)| \leq C_2(1+\langle \bk \rangle^{2-\ell}) \qquad
	\ell= 0,1,\ldots, 2d\;.
\label{ombound}
\ee
We also assume that the functions $\bk \mapsto \Phi_\pm(\bp, \bk):=
 e(\bk+\bp) \pm \om(\bk)$ satisfy
\be
	\lim_{\bk\to\infty} \Phi_\pm(\bp, \bk) = \infty\; ,
\label{Philim}\ee
and
\be
	0< C_3\leq
	\mbox{Hess}_\bk \Phi_\pm(\bp, \bk)
	\leq C_4 
\label{hessom}
\ee 
(the upper bound follows from (\ref{eq:econd})
and (\ref{ombound})). By degree theory, 
these conditions imply that 
$\bk\mapsto\Phi_\pm(\bp, \bk)$ has only one critical point.

\medskip

Let
\be
	E_\pm(\bp, \theta,\delta):=  
	\Big\{ \bk \; : \; |\Phi_\pm(\bp, \bk) - \theta|\leq
	\delta\Big\} \;
\label{levelset}
\ee
be a small neighborhood of the level set
$k\mapsto \Phi_\pm(\bp, \bk)=\theta$.

We remark that conditions (\ref{eq:econd})--(\ref{hessom})
  also imply the existence of two
constants $\wt\varrho, \wt C>0$ depending on $C_1, \ldots, C_4$ such that
\be
	\sup_{\bp,\bq,\theta}
	\Big|\;  E_\pm(\bp, \theta,\delta)
	\cap B(\bq, \varrho)\; \Big| \leq \wt C 
	\delta \varrho^{d-1}
\label{layer}
\ee
whenever $\delta,\varrho \leq \wt\varrho$. Here
$|\cdot |$ denotes the Lebesgue measure
of a set, and $B(\bq, \varrho)$ is the ball of radius $\varrho$ about
$\bq\in \bR^d$.
We also need a certain transversality condition on two such sets.

\medskip

{\it Transversality condition:} There exist positive constants $\wt\varrho, 
C_5$ such that for any $\delta_1, \delta_2, \varrho \leq \wt \varrho$, 
any $\bp_1, \bp_2\in \bR^d$
and any $\theta_1, \theta_2\in \bR$ 
\be
	\sup_\bq \; \Big| \; E_\pm(\bp_1, \theta_1,\delta_1) 
	 \cap E_\pm(\bp_2, \theta_2,\delta_2)
	\cap B(\bq, \varrho)
	 \; \Big|\leq {C_5 \delta_1\delta_2\varrho^{d-2} \over
	|\bp_1-\bp_2|}\;. 
\label{trans}
\ee

\medskip

In particular, elementary calculation shows that 
all these conditions are satisfied in $d\ge3$
if $e(\bk) = \frac{1}{2}\bk^2$
and $\| \nabla^2\om \|_\infty$ is sufficiently small. This is
our primary example.

\bigskip

To ensure that the statistical operator is trace class,
we always assume that
\be
	\inf_\bk \om (\bk) - \mu\beta^{-1}\ge C_6 >0 \; .
\label{ass1}
\ee
Note that the number density function $\cN$ is symmetric and it
belongs to $ C^{2d}(\bR^d)$.

\bigskip

We assume that $Q(\bk)\in C^{2d}(\bR^d)$ 
is real (since $H_{e-p}$ is self-adjoint) and
 symmetric, i.e.,
\be
	Q(\bk)= Q(-\bk) = \ov{Q(\bk)} \; ,
\label{eq:Qsym}
\ee
and it has a fast decay  up to $2d$-derivatives
\be
	\max_{\ell=0,\ldots, 2d} |\nabla^\ell_\bk Q(\bk)|
	\leq C_7\langle \bk\rangle^{-2d-12} \; .
\label{Qdec}
\ee

\bigskip

The initial electron density matrix $\gamma_e=\gamma_e^\e$ will depend on $\e$
so that it has a macroscopic profile in the scaling limit.
We assume that the limit
\be
	F_0(\bX, \bV):=\lim_{\e\to0}W^\e_{\gamma_e^\e}(\bX, \bV)  
\label{initlimit}
\ee
 exists weakly in $\cS(\bR^{2d})$. In addition, we assume that
\be
	\limsup_{\e\to0}
	\int \langle \bp \rangle^{3d+12} \wh\g_e^\e (\bp, \bp) \rd \bp
	\leq C_8< \infty \; .
\label{gammadecay}
\ee
For example, $\gamma_e^\e$ can be a pure state, $\gamma_e^\e:= |\psi^\e\rangle
\langle \psi^\e|$, with a WKB wavefunction
$$
	\psi^\e(\bx ) = \e^{d/2} A(\e\bx)e^{iS(\e\bx)/\e} \; ,
$$
where $A, S \in \cS(\bR^d)$.

\subsection{Main Theorem}

\begin{theorem} Let  $\lambda = \sqrt{\e}$ and let $\Gamma_t^\e$ solve
the Schr\"odinger equation (\ref{densev}) with initial
condition $\Gamma_0^\e = \gamma_e^\e \otimes \g_{ph}$, where
the initial electron density matrix satisfies (\ref{initlimit}) and
 (\ref{gammadecay}). We let $\gamma^\e_{t}: = \Tr_{ph} \Gamma^\e_t$
be the electron density matrix at time $t$.

We assume that the electron and phonon
dispersion relations satisfy (\ref{eq:econd})--(\ref{hessom}), 
(\ref{trans}) and (\ref{ass1}), while the interaction function
$Q(\bk)$ satisfies (\ref{eq:Qsym}) and (\ref{Qdec}).
% Finally we assume either (A1) or (A2). 
Then for any $T>0$
$$
	\lim_{\e\to0}\lim_{L\to\infty}
	 W^\e_{\gamma^\e_{T/\e}}(\bX, \bV) = F_T(\bX, \bV)
$$
weakly in $\cS(\bR^{d}\times\bR^d)$, and $F_T$ satisfies the Boltzmann
equation (\ref{eq:Beq}) with initial condition $F_0$ and collision
kernel
\bey
	\sigma(\bV, \bU):  &=& 2\pi |Q(\bU-\bV)|^2 
	\Bigg\{  \Big(\cN(\bU-\bV)+1\Big)\delta\Big(e(\bV)-e(\bU)
	 +\omega(\bU-\bV)\Big) \nonumber\\
	&& +  \cN(\bU-\bV)
	\delta\Big(e(\bV)-e(\bU) - \omega(\bV-\bU)\Big)\Bigg\}\;.
\label{phonker}
\eey
These two  terms correspond to phonon emission and absorption, respectively.
\end{theorem}

{\it Remark 1.}
The Boltzmann equation is irreversible, while the Schr\"odinger
dynamics is reversible. This is not controversal,
since the time evolved quantum state cannot be reconstructed
from the solution of the Boltzmann equation. The macroscopic
equation gives only a robust information on the evolution:
it contains neither the phonons nor the microscopic details
of the electron state. The former was neglected when taking
partial trace, the latter is lost in the scaling (weak) limit.

\medskip

{\it Remark 2.}
  Scaling limit of the dynamics
of a quantum particle in a weakly coupled random potential was obtained
 for short time in  \cite{S}, \cite{HLW} and for long time in \cite{EY2}.
A similar result was obtained in \cite{EY1} in a
 different limiting regime, in the so-called low density limit.
 Despite these similarities,
the current proof differs substantially from  \cite{EY2}
(see also a short announcement \cite{E}).
We replace the idea of the partial time integration by
a more effective classification of the indirect  Feynman graphs.
This enables us to expand the Duhamel formula further than in
\cite{EY2} and still control the error terms (so-called {\it indirect}
and {\it recollision} terms).
In particular, for any $\kappa$ we will be 
able to gain an extra $\e^{\kappa}$ factor relative to the size of the
main term with the exception of $C(\kappa)^n$ pairings
in the $n$-th order terms of the Duhamel expansion
(see Propositions \ref{prop:omconst}, and \ref{prop:omnonconst}).

\medskip

{\it Remark 3.} The
evolution problem for the electron density matrix
$\gamma_t= \Tr_{ph}\Gamma_t$ is formally equivalent 
to the Schr\"odinger equation
$$
	i\partial_t \gamma_t = \Big[ H_e + \lambda V_\omega, 
	\g_t\Big]
$$
with a time dependent Gaussian random potential $V_\omega(x, t)$
with covariance (in Fourier space)
\be
	\bE \overline{\wh V_\omega(\bp, t)} 
	\wh V_\omega(\bq, s) = \delta(\bp-\bq)
	|Q(\bp)|^2\Big[\Big( \cN(\bp)+1\Big)e^{i(t-s)\omega(\bp)}
	+\cN(\bp)e^{-i(t-s)\omega(\bp)}\Big] .
\label{timcov}
\ee
This means that the formal perturbation expansions
of these two problems coincide term by term.
However, this connection is only formal and in our rigorous
proof we cannot and do not make use of it.
\medskip

{\it Remark 4.}
Long time evolution of a {\it microscopically localized}
electron weakly coupled to a phonon bath was
studied in \cite{CEFM} in the  dipole approximation (see also
references therein). The limiting equation is diffusive (Fokker-Planck)
already on the first nontrivial time scale (in the
van Hove limit). In this case there is no Boltzmann
equation before diffusion emerges.
The diffusion mechanism is quite different; it is a resonance effect
between certain phonon modes and the
eigenfrequencies of the confinement.
\medskip

{\it Remark 5.}  In a more realistic transport model, 
 electron-electron interactions should be  included.
This is a genuine many-body problem and is much more difficult.
In classical dynamics an analogous result
has been proven by O. Lanford \cite{L} in the low density
limit. There is no rigorous result in the quantum case.
\medskip

{\it Convention.} Throughout the paper we use the letter $C$
to denote various constants that depend only on the dimension $d$
 and the constants $C_1, C_2 \ldots $ quantifying the
assumptions in Section \ref{sec:assump}.
For brevity, we usually neglect certain variables
in the formulas if it does not cause confusion. In particular,
we will omit the superscript $\e$ from $\Gamma^\e_t$ and $\gamma^\e_t$.

\section{Structure of the proof}
\setcounter{equation}{0}

\subsection{Duhamel expansion}

We use the Duhamel formula (\ref{eq:duh})
for $H= H_0 + H_{e-p}$ with
$H_0= H_e+ H_{ph}=e (-i\nabla) + H_{ph}$. Recall that
$H_e$ and $H_{ph}$ commute.

We are interested in the partial trace of $\Gamma_t = e^{-itH}\Gamma_0
e^{itH}$  with respect to the phonon variables.
In principle, we could fully expand $e^{-itH}$ by Duhamel on
both sides. But this expansion does not converge for large $T = t\e$,
$\e := \lambda^2$, 
unless we give a fairly detailed classification of Feynman graphs. For each
graph of order $n$ the best generally valid theoretical
bound is $(C\lambda^2t)^n/n!$  (although
we essentially prove only $(C_a\lambda^2t)^n/[n!]^a$, $0\leq a < 1$,
in Lemma \ref{lemma:trivbound}), but then $n!$
is lost to combinatorics (number of graphs is $\sim C^n n!$)
 and the series does not converge
for $t \ge C^{-1}\lambda^{-2}$, i.e., for $T\ge C^{-1}$.
 Even if we gain an extra $t^{-\kappa}$
from most pairings (due to reabsorption and
 crossing), with some fixed $\kappa$,
it does not make the series converge.

So the idea is to stop the Duhamel expansion once the expanded part
is small enough. This  means that either there are many
absorptions-emissions
or at least one reabsorption.
However, immediate reabsorptions do contribute to the main
term, and they have to be resummed (see later in (\ref{eq:resum})).
For simplicity, we use the word collision for absorption or emission.

\medskip

The Duhamel formula for $H=H_0+H_{e-p}$ states that for any $N_0 \ge 1$
\bey
	e^{-itH} &=& \sum_{N=0}^{N_0-1}
	(-i)^N \int_0^{t*} [\rd s_j]_0^N
	e^{-is_0H_0} H_{e-p} e^{-is_1H_0} H_{e-p}\ldots
        H_{e-p}e^{-is_N H_0} \nonumber\\
	&&+ (-i)^{N_0} \int_0^{t*} [\rd s_j]_0^{N_0}
	e^{-is_0H} H_{e-p} e^{-is_1H_0} H_{e-p}\ldots
        H_{e-p}e^{-is_{N_0} H_0} \; .
\label{eq:duh}
\eey
 For brevity, we introduced the
following notation
\be
	\int_0^{t*} [\rd s_j]_0^N : =
	\int_0^t\ldots \int_0^t \Big(\prod_{j=0}^N \rd s_j\Big)
        \delta\Big( t- \sum_{j=0}^N s_j\Big) \; ,
\label{timenot}
\ee
where the star refers to the constraint $t= \sum_{j=0}^N s_j$.
 Notice that the upper integration limits   on
the right hand side of (\ref{timenot})
 could be omitted  because the
lower limits and the delta function together guarantee them.

For the threshold, we shall choose   $N_0=N_0(t)$ such that
 $t^2\ll N_0! \ll t^{2.2}$ (see (\ref{nzerochoice})).

\medskip

{\it Remark:} This is {\it not} the choice in our earlier paper \cite{EY2},
where  $N_0!\ll t$, and we
used partial time integration.
 Here we rather expand  further to avoid
partial time integration. This will require more effective
control on classically irrelevant interference terms.
\medskip

Since we have to distinguish a few recollisions, 
the Duhamel formula has to be written more carefully.
Write $H_{e-p}$ (\ref{interaction}) as follows
\be
	H_{e-p} =  i\lambda \int
	 Q(\bk)\Big[ e^{-i\bk\cdot\bx}c_\bk^\dagger -
	e^{i\bk\cdot \bx} c_\bk\Big] \rd \bk =
	 i\la  \int Q(\bk) e^{-i\bk\cdot\bx}b_\bk
	  \rd \bk
\label{Hep}\ee
with
$$
	b_\bk : = c_\bk^\dagger - c_{-\bk}, \qquad\mbox{hence}\qquad
	 b_\bk^* = c_\bk -c_{-\bk}^\dagger 
	= -b_{-\bk} \; , 
$$
 and we used the symmetry of $Q$.
Here star means the adjoint, and $b_\bk, b_\bk^*$ do {\it not} satify 
the canonical commutation relations,
in fact 
$$
	[b_\bk^{(*)}, b_\bm^{(*)}] =0 \qquad \mbox{for any} \;\;  \bk, \bm \;.
$$
We also introduce the notation
\be
	\prod_{j=1}^n A_j : = A_1 A_2 \ldots A_n
	\qquad \mbox{and} \qquad
	\prod_{j=n}^1 A_j : = A_n A_{n-1} \ldots A_2 A_1
\label{opprod}\ee
for the ordered product of noncommuting objects (operators) $A_j$.

With these notations we rewrite (\ref{eq:duh}) in the Fourier space
of $\cH_e$ as
\bey
	e^{-itH} &=&  \sum_{N=0}^{N_0-1}
	\lambda^N\int_0^{t*} [\rd s_j]_0^N \int \Big( 
	\prod_{j=1}^N \rd \bk_j\Big)
	e^{-is_0H_0} \Big( \prod_{j=1}^N Q(\bk_j)b_{\bk_j} e^{-is_jH_0}
	\Big)\nonumber\\
	&& +  \lambda^{N_0}\int_0^{t*} [\rd s_j]_0^{N_0}\int \Big( 
	\prod_{j=1}^{N_0} \rd \bk_j\Big)
	e^{-is_0H}\Big( \prod_{j=1}^{N_0} Q(\bk_j)b_{\bk_j} e^{-is_jH_0}
	\Big)\; .
\label{eq:duh1}
\eey

\subsection{Stopping rule for the  Duhamel expansion}

The key observation is that the Duhamel formula (\ref{eq:duh1})
is too rigid. There is no need to decide apriori how long
we expand each term. Once a new step of the  expansion is made,
and a factor $H_{e-p}$ added,  it
contains an integration over a new variable
$\bk_j$  which is actually a big summation (\ref{Hep}).
This means that each term ramifies into a sum of several
new terms. We can decide individually for each term whether we want 
to continue  the Duhamel expansion or stop.
The stopping rule must depend only
on the already expanded terms. We
can stop the expansion for those terms which already had
enough number of recollisions (it turns out that one is enough)
or which already had many collisions.

By {\it recollision} or {\it reabsorption}
we mean $\bk_j+\bk_\ell=0$ for some $j\neq\ell$ in (\ref{eq:duh1}).
It contains both physically relevant
(e.g. $c_\bk^\dagger c_\bk$ and $c_{-k}c_{-k}^\dagger$)
and irrelevant  (e.g., $c_\bk^\dagger c_{-\bk}^\dagger$ and  $c_{-k}c_{k}$)
interaction terms. The contribution of the irrelevant terms
will vanish in the $L\to\infty$ limit.
The recollision  $\bk_j+\bk_\ell=0$ is called {\it immediate} if $|j-\ell|=1$,
otherwise it is called {\it genuine}. The reason for this distinction is that
 immediate reabsorptions do occur
without rendering the  corresponding term small as $\e\to0$.

Hence we will stop the Duhamel expansion
if we  either see $N_0$ collisions
or if we see  one genuine  reabsorption. 
 If we did not stop, it
means {\it fully expanded} terms, which then has no reabsorption.

\bigskip

 We need several notations to define these terms rigorously.
Let $\chi(\bv)$, $v\in\Lambda^*$, be the characteristic 
function at 0, i.e., $\chi(\bv)=1$
if $\bv=0$ and $\chi(\bv)=0$ otherwise.

\begin{definition}\label{def:J}  Let $\cM(n,N)$ be the set of $(n+1)$-tuple
of nonnegative integer numbers $\um: = (m_0, m_1, \ldots, m_n)$
with the property that  $n+2|\um|= N$, $|\um|: = \sum_j m_j$.
Such an $(n+1)$-tuple is equivalent to an increasing  subsequence
$\mu= ( \mu(1), \mu(2), \ldots, \mu(n))$ 
of the numbers $\{ 1, 2, \ldots, N\}$ such that
all differences $\mu(j+1)-\mu(j)$ are odd for all $j=0,1,\ldots, N$.
 For convenience we set $\mu(0):=0$
and $\mu(n+1): = N+1$. The identification between $\um$ and $\mu$
is given by the relations $\mu(j+1)-\mu(j)=2m_j+1$
for all $j=0,1,\ldots, n$.  We will use both representations
simultaneously.
 Let
$$
	I:=I(\um):=\{ \mu(j) \; : \;
	j=1,2,\ldots, n\} =\Big\{ j+2\sum_{i=0}^{j-1} m_i  \; : \;
	j=1,2,\ldots, n\Big\} \; .
$$
For each $j=0,1,\ldots, n$ let
$$
	I_j: = I_j(\um): =\{ \mu(j) + 1, \mu(j)+3,\ldots, \mu(j)+2m_j-1\},
	\qquad (\mbox{with}\;\; |I_j|=m_j)\; ,
$$ 
and 
$$
	I_j^c: = I_j^c(\um) : = \{ \mu(j), \mu(j)+2,\ldots, \mu(j)+2m_j\},
	\qquad  (\mbox{with} \;\; |I_j^c|=m_j+1)\; ,
$$
be its complement between two consecutive elements of the subsequence
$\mu$.
Finally let
$$
	J:= J(\um):=
	\bigcup_{j=0}^n I_j\; , \qquad J^c: =  \bigcup_{j=0}^n I_j^c \; ,
$$
and clearly $\{ 0, 1, \ldots ,N \} = J \cup J^c$ with $J\cap J^c=\emptyset$,
$|J|=|\um|$ and $|J^c| = n + |\um| + 1$.
\end{definition}

The increasing subsequence $\mu$ associated with
$\um\in \cM(n,N)$ will determine the indices
of those phonon interactions which are not parts of immediate
reabsorptions, i.e., $\um$ encodes the immediate reabsorption
pattern. Here $N$ refers to the total number of phonon interactions
and $n$ of them are not part of immediate reabsorptions.
In the sequel, these two numbers always have the same parity
without further remark. The number $m_j$ denotes the
number of immediate recollision pairs between $\mu(j)$ and $\mu(j+1)$.

For any $\um\in \cM(n,N)$ we define
\be
 	\Xi(\ubk, \um) : =  
	 \prod_{b\in J(\um)}
	\chi(\bk_{b}+\bk_{b+1})
\label{def:Xi}
\ee
to be the restriction onto momenta that respect the
immediate reabsorption pattern given by $\um$. 
Here $\ubk:=(\bk_1, \ldots, \bk_N)$ stands for the collection
of the phonon variables.
The phonon momenta indexed by the subsequence $\mu$ are called {\it external},
the rest are {\it internal}. The number of external momenta is $n$ if $\um\in
\cM(n, N)$.

We then define the measure
\be
	\int^{(\um)} \rd\ubk
	: = \int \rd\ubk \;
	\Xi(\ubk,\um)
	\prod_{b\neq b'\in I(\um)}
	\Big(1-\chi(\bk_{b} + \bk_{b'})
	\Big) \; ,
\label{def:schintmu}
\ee
where  $\rd\ubk : = \prod_{j=1}^N \rd\bk_j$ for simplicity.
This measure excludes pairing among  external  momenta,
 while the internal momenta
are consecutively paired.
Notice that these restricted integrations make sense only for 
finite $\Lambda$. We also set
\be
	\int^{\#(n,N)}  \rd\ubk:
	= \sum_{\um\in \cM(n,N)}\int^{(\um)} \rd\ubk\;,
\label{def:schint}
\ee
i.e., in this measure we exclude all recollisions, which are not 
 immediate. In the sequel, genuine recollision will be simply called
recollision.

Now we  define the one recollision terms. These terms
will always be amputated, i.e., the last free
propagator $e^{-is_0H_0}$ will not be present.

For any $a\in \{ 2,3,\ldots, n\}$, we
let $\cM_a(n,N)\subset \cM(n,N)$ be the set of those $\um$'s
 such that $m_0=0$, $\mu(a)\ge 3$ and if $a=2$, we additionally
require that $m_1\ge 1$. The momenta 
with index $\mu(1)=1$ and $\mu(a)$ will form the (genuine) recollision.
We also let $I_a = I_a(\um): = I(\um)\setminus\{ 1, \mu(a)\}$ be the set of
the indices of the remaining external momenta.

For $\um\in \cM_a(n,N)$ we let
\be
	\int^{*(\um,a)} \rd\ubk
	:= \int \rd\ubk\;
	\chi(\bk_1 + \bk_{\mu(a)})\Xi (\ubk,\um)
	\prod_{b\neq b'\in I_a}
	 \Big(1-\chi(\bk_b + \bk_{b'})
	\Big) 
\label{def:starintmua}
\ee
be the measure where the immediate reabsorption pattern given by $\um$
and a single genuine recollision between the first and the 
$\mu(a)$-th phonon is enforced.
Let
\be
	\int^{*(n,N)}\rd\ubk:
	=  \sum_{a=2}^N \sum_{\um \in \cM_a(n,N)}
	 \int^{*(\um,a)} \rd\ubk  \;.
\label{def:starint}\ee
This measure expresses that there is no recollision
of external momenta apart from
the designated one between $1$ and $\mu(a)$. 

\bigskip

Now we define the (amputated) operator acting on $\cH_{tot}$
\be
	\cA (\tau, \ubk, N):
	= \lambda^N
	\int_0^{\tau*} [\rd s_j]_1^N
	 \Big(\prod_{j=1}^N Q(\bk_j) b_{\bk_j} e^{-is_jH_0} \Big) \;.
\label{def:cA}
\ee
We can integrate out this operator on different sets of $\ubk$
corresponding to various reabsorption patterns
(superscript denotes the number of  reabsorptions):
\be
	\cD_{n,N}^0(\tau) : = 
	 \int^{\#(n,N)}\rd\ubk
	\; \cA (\tau, \ubk, N),
	\qquad \cD_N^0(\tau): = \sum_{n=0}^N\cD_{n,N}^0(\tau) \; ,
\label{def:D0}
\ee
\be
	\cD_{n,N}^1(\tau) : =
	\int^{*(n,N)}\rd\ubk
	\; \cA (\tau, \ubk, N),
	\qquad \cD_N^1(\tau): = \sum_{n=2}^N\cD_{n,N}^1(\tau)\; .
\label{def:D1}
\ee
These are  amputated objects. 
Depending on whether a term is fully expanded or not, we define
non-amputated terms by letting $e^{-isH_0}$ or $e^{-isH}$ act on it.
Terms with a recollision are always amputated.

We define
\bey
	\cB (t, \ubk, N) :
	&=&  \lambda^N
	\int_0^{t*} [\rd s_j]_0^N
	e^{-is_0H_0} \Big(\prod_{j=1}^N Q(\bk_j) b_{\bk_j} e^{-is_jH_0} \Big)
	\nonumber\\
	&=& (-i)\int_0^t\rd s_0 \;  e^{-is_0H_0} \cA (t-s_0, \ubk, N) \; .
\label{def:cB}
\eey

The fully expanded terms are  
\be
	\cE_{n,N}^0 (t) : = (-i)\int_0^t 
	e^{-is_0H_0}\cD_{n,N}^0(t-s_0) \rd s_0
	= \int^{\#(n,N)} \rd\ubk\;
	 \cB (t, \ubk, N) \;,
\label{def:E0}
\ee
and we let
$$
	\cE_N^0(t) := \sum_{n=0}^N \cE_{n,N}^0(t) \; .
$$

The non-fully expanded terms are
$$
	\cH_{n,N}^0 (t) : = (-i)\int_0^t 
	e^{-is_0H}\cD_{n,N}^0(t-s_0) \rd s_0 \; ,
	\qquad \cH_N^0(t) := \sum_{n=0}^N\cH_{n,N}^0 (t)\; ,
$$
\be
	\cH_{n,N}^1 (t) : = (-i)\int_0^t 
	e^{-is_0H}\cD_{n,N}^1(t-s_0) \rd s_0\; , 
	\qquad \cH_N^1(t) := \sum_{n=2}^N\cH_{n,N}^1 (t) \;.
\label{def:H}\ee

We will use the following Duhamel expansion:
\begin{lemma} For  any fixed $N_0\ge 1$ 
we have
\be
	e^{-itH} = \cH_{N_0}^0 (t) + 
	\sum_{N=3}^{N_0}\cH_{ N}^1 (t)
	+ \sum_{N=0}^{N_0-1} \cE_{N}^0 (t) \;. 
\label{eq:duh2}
\ee
\end{lemma}

{\it Proof.}  When expanding
the Duhamel formula, at each step we  generate a new fully expanded term
and a term with one more interaction  $H_{e-p}$ and a full propagator.
Since $H_{e-p}$ contains a sum over all momenta (\ref{Hep}),
we obtain $|\Lambda^*|$ new terms.
Hence the expansion can be represented by
a  successively growing rooted tree-graph, where the vertices correspond
to interaction terms from the expanded $H_{e-p}$'s.
Each vertex  ramifies into a terminal branch (free propagator)
plus $|\Lambda^*|$ new branches whose endpoints are
labelled by the momenta of the newly expanded interaction term.
In this graph there is a unique path to the root
 from each vertex; the
vertices on this path are called the {\it predecessors} of this vertex.
In particular, every vertex (apart from the root)
 has an immediate
predecessor, called its {\it father}.

We say that a vertex forms a {\it bond} (immediate recollision)
with its father if they have the same momentum label
and if the father does not already form a bond with
its own father. The notion of the bond
is defined successively as the tree grows along further expansion.
 Vertices that are not part of a bond are called
{\it independent}.

We stop the expansion at any vertex if 

\medskip

(i) it has $N_0-1$ predecessor; or

(ii)  its momentum label coincides with the label of any of its 
independent predecessor different from its father
(genuine recollision).

\medskip

These terms are included in $\cH_{N_0}^0$ and $\cH_{N}^1$,
respectively,
while the fully expanded terms (corresponding to terminal branches)
are contained in the
last term of (\ref{eq:duh2}). 
$\;\;\;\Box$

\bigskip

Therefore we can write for any $K>1$ integer
$$
	\gamma_t= \Tr_{ph} e^{-itH} \Gamma_0 e^{itH}
	 =  \g^{main}_K(t)
	 + \g^{err}_K(t)\;,
$$
where 
\be
	\g^{main}_K(t) =  \sum_{N, \tN =0}^{K-1}\Tr_{ph} \cE_{N}^0 (t)\Gamma_0
	\Big[  \cE_{\tN}^0 (t)\Big]^*
\label{def:gammamain}
\ee
corresponds to the main term containing no reabsorption
and less than $K$ collisions, and
\be
	\g^{err}_K: = \sum_{\tN=0}^{N_0-1}\sum_{N=K}^{N_0-1}
	 \Tr_{ph} \cE_N^0 \Gamma_0 \Big[\cE_\tN^0 \Big]^*
	+ \sum_{\tN=K}^{N_0-1}\sum_{N=0}^{K-1}
	 \Tr_{ph} \cE_N^0 \Gamma_0 \Big[\cE_\tN^0 \Big]^*
	+ \Tr_{ph} 
	\cH_{N_0}^0
	\Gamma_0  \Big[ 
	\cH_{N_0}^0 \Big]^*
\label{eq:gammaerr}
\ee
$$
	+ \sum_{N, \tN=3}^{N_0}  \Tr_{ph}\cH_N^1 \Gamma_0  \Big[ 
	\cH_{\tN}^1 \Big]^* +
	\sum_{N=0}^{N_0-1} 
	\Tr_{ph} \Bigg( \cE_N^0 \Gamma_0
	\Big[ \cH_{N_0}^0 +  \sum_{\tN=3}^{N_0}   \cH_{\tN}^1\Big]^* 
	+  \Big[ \cH_{N_0}^0 + \sum_{\tN=3}^{N_0}   \cH_{\tN}^1\Big]\Gamma_0
	\Big[  \cE_{N}^0\Big]^*	\Bigg) \; .
$$

\subsection{Observables and Wigner transform}

Eventually we want to compute
$\mbox{Tr}_e\; \gamma_t \cO$
with some electron observable $\cO$ acting on $\cH_e$. For example, 
to determine the weak limit of $W_{\gamma_t}^\e$ in $\cS(\bR^{d}\times
\bR^d)$,
we have to test the macroscopic Wigner distribution (\ref{def:reswig}) against
a smooth function $J(X,V)\in \cS(\bR^{d}\times \bR^d)$.
 Let $J_\e(\bx, \bv):= \e^dJ(\bx \e, \bv)$,
hence $\wh J_\e(\xi, \bv):= \e^{-d} \wh J(\xi\e^{-1}, \bv)$,
recalling that hat means Fourier transform in the first variable.

A simple calculation shows that
$$
	\langle J, W_{\gamma_t}^\e \rangle := \int \ov{J(X,V)}
	 W_{\gamma_t}^\e(X,V) dX dV 
	= \mbox{Tr}_e \; \gamma_t \cO_\e \; ,
$$
%	= \int dX dV \ov{J(X, V)} \e^{-d}W_{\gamma(t)}( X/\e, V)
%	= \int \rd \xi \rd v 
%	\e^{-d}\ov{\wh J(\xi/\e, v)} \wh W_{\gamma(t)}(\xi, v)
%$$
%$$
%	= \int \rd \xi \rd v 
%	\e^{-d}\ov{\wh J(\xi/\e, v)} \gamma_t\Big( v - {\xi\over 2},
%	v+ {\xi\over 2}\Big) 
%$$
%$$ 
%	= \int \rd a\rd b \; \gamma_t(a, b)
%	 \e^{-d} \ov{\wh J\Big({b-a\over\e}, {a+b\over 2}\Big)} 	
%	= :\mbox{Tr}_e \; \gamma_t \cO_\e \; ,
%$$
where the observable  $\cO_\e$   is given by the kernel
\be
	\cO_\e(u, v): = \wh J_\e\Big( u-v, {u+v\over 2}\Big)\;.
%	 \e^{-d} \wh J\Big({u-v\over\e}, {u+v\over 2}\Big)
%	=: \wh J_\e( u-v, (u+v)/2) \; ,
\label{def:okernel}
\ee

To control $\mbox{Tr}_e \; \gamma_t \cO_\e $
we will need that $\cO_\e\cO_\e^*$ is a uniformly  bounded operator:
\be
	\sup_\e \| \cO_\e\cO^*_\e\|<\infty \; .
\label{eq:obs}
\ee
 In our case
it is enough if $J_\e$ satisfies 
\be
	\| J\|:=\sup_{\e}\int \sup_v |\wh J_\e(\xi, v)|\rd\xi < \infty
\label{jeps}
\ee
since $\| \cO_\e\cO^*_\e\|\leq \| J \|^2$ by an easy calculation.
%$$
%	\| \cO_\e\cO^*_\e\| \leq \sup_a \int \rd b \rd c
%	|\cO_\e(a, b)||\cO_\e(c, b)|
%	= \sup_a \int \rd b \rd c |\wh J_\e( a-b, (b+a)/2)| \,\, 
%	|\wh J_\e( c-b, (b+c)/2)|
%$$
%%$$
%	\leq \sup_a  \int \rd b |\wh J_\e( a-b, (b+a)/2)|
%	\int \rd c \sup_v |\wh J_\e( c-b, v)| 
%	\leq \| J\| \sup_a  \int \rd b |\wh J_\e( a-b, (b+a)/2)|
%$$
%$$
%	\leq \| J\| \sup_a  \int \rd b\sup_w |\wh J_\e( a-b, w)|
%	\leq \| J \|^2 \; .
%$$
The estimate (\ref{jeps}) is satisfied 
for $\wh J_\e$ if
\be
	\int \sup_v |\wh J(\xi, v)| \rd\xi < \infty \; ,
\label{jbound}\ee
in particular if $J\in \cS(\bR^{d}\times \bR^d)$.

In Sections \ref{sec:main}--\ref{sec:onereabs} we
will show that for any $T\ge 0$
\be
	\limsup_{K\to\infty} \limsup_{\e\to0} 
 	\limsup_{L\to\infty} \Big|\Tr_e \g^{err}_K(t)\cO_\e \Big| =0
\label{errz}\ee
with $t=T/\e$.
In Section \ref{sec:wign} we will  check that the weak limit of 
$W^\e_{\gamma^{main}_K(t)}$ satisfies the Boltzmann equation as 
$L\to\infty$, then $\e\to0$ and finally $K\to\infty$.
This will complete the proof of the Theorem.

\section{Estimating the error terms: statement of the Main Lemma}
\label{sec:main}

\bigskip

All error terms in $\Tr_e \gamma^{err}_K(t) \cO$ have the structure
$$
	\Tr_{e+ph}  \cG(t) \Gamma_0 \Big[ \cG(t)\Big]^*  \cO
$$
with $\cG=\cE$ or $=\cH$ with appropriate indices (see (\ref{eq:gammaerr})).
We use that $\Gamma_0\ge 0$ and a simple Schwarz inequality
of prototype
\bey
	\Big| \Tr \; A \Gamma_0 B^*  \cO\Big|
	&\leq& \Tr A\; \Gamma_0 A^* +  \Tr \; \cO^* B \Gamma_0 B^*  \cO
	\nonumber\\
	&=& \Tr\;  A\Gamma_0 A^* +  \Tr \;  \Gamma_0^{1/2} B^*  \cO \cO^* B 
	\Gamma_0^{1/2} \nonumber\\
	&\leq&  \Tr\;  A\Gamma_0 A^* +  \| \cO\cO^*\| \;
	\Tr \;  \Gamma_0^{1/2} B^*  B\Gamma_0^{1/2} 
	\nonumber\\
	&=&  \Tr\;  A\Gamma_0 A^* +  \| \cO\cO^*\| \; \Tr\;  B\Gamma_0 B^* \;.
	\nonumber
\eey
%Here we also used that $UVU^* \leq \| V \| UU^*$ in operator
%sense for any operators $U, V$.

Hence after eliminating the "cross-terms" in $\g^{err}(t)$ 
(see (\ref{eq:gammaerr}))
by Schwarz inequalities, we have the estimate for any $0< \nu <1$, $K\leq N_0$
\bey
	\Big|\Tr_e \g^{err}_K(t)\cO_\e \Big|
	&\leq& C_\cO \Tr_{e+ph}
	\Bigg\{  \nu \sum_{N=0}^{K-1}
	  \langle N \rangle^2 
	\cE_N^0 (t)\Gamma_0\Big[  \cE_N^0 (t)\Big]^*
	+\nu^{-1}\sum_{N=K}^{N_0-1}
	  \langle N \rangle^2 
	\cE_N^0 (t)\Gamma_0\Big[  \cE_N^0 (t)\Big]^*	
	\nonumber\\
	&& +\nu^{-1} N_0
	 \cH_{N_0}^0 (t)\Gamma_0
	\Big[  \cH_{N_0}^0 (t)\Big]^*
	+\nu^{-1} N_0\sum_{N=3}^{N_0}  \cH_{N}^1 (t)\Gamma_0
	\Big[  \cH_{N}^1 (t)\Big]^*
	\Bigg\}\; ,
\label{eq:errest}
\eey
with $C_\cO: = C(1+\sup_\e\| \cO_\e\cO_\e^*\|) <\infty$.

The terms in the first  two summations are fully expanded, their
estimates are easier. 
The  third and fourth terms still contain the full
propagator. We estimate them
in terms of amputated fully expanded terms by using
unitarity and paying a price of an extra factor $t$.
The prototype of such term is
\be
	\wt\gamma(t):=\mbox{Tr}_{ph} \int_0^t \rd s_0\int_0^t \rd\ts_0\;
	e^{-is_0H} A(t-s_0) \Gamma_0 B^*(t-\ts_0)e^{i\ts_0 H}\; ,
\label{gammatilde}
\ee
where $A$, $B$ denote one of the amputated $\cD_{n,N}^b$ for $b=0$ or
$b=1$ and  $n\leq N\leq N_0$ (see (\ref{def:D0})--(\ref{def:D1})). 
By a simple Schwarz estimate inside the time integrations in
(\ref{gammatilde}) we obtain
\be
	\mbox{Tr}_e \wt\gamma(t) 
	\leq  t \cdot 
	 \mbox{Tr}_e\Bigg[ \int_0^t \rd s_0 
	\mbox{Tr}_{ph} A(t-s_0) \Gamma_0 A^*(t-s_0)
	+  \int_0^\tt \rd\ts_0
	 \mbox{Tr}_{ph} 
	 B(\tt-\ts_0) \Gamma_0 B^*(\tt-\ts_0) \Bigg]\;.
\label{eq:symm}
\ee

%$$
%	\mbox{Tr}_e \wt\gamma(t) 
%	= \mbox{Tr}_{e+ph}  \int_0^t \rd s_0\int_0^t \rd\ts_0
%	e^{-is_0H} A(t-s_0) \Gamma_0 B^*(t-\ts_0)e^{i\ts_0 H}
%$$
%$$
%	\leq \int_0^t \rd s_0\int_0^t \rd\ts_0
%	 \mbox{Tr}_{e+ph} e^{-is_0H} A(t-s_0) \Gamma_0 A^*(t-s_0)e^{is_0H}
%$$
%$$
%	+  \int_0^t \rd s_0\int_0^t \rd\ts_0
%	 \mbox{Tr}_{e+ph} 
%	e^{-i\ts_0 H} B(t-\ts_0) \Gamma_0 B^*(t-\ts_0)e^{i\ts_0 H}
%$$
%$$
%	= t  
%	\int_0^t \rd s_0 \mbox{Tr}_{e+ph} A(t-s_0) \Gamma_0 A^*(t-s_0)
%	+ t \int_0^t \rd\ts_0
%	 \mbox{Tr}_{e+ph} 
%	   B(t-\ts_0)\Gamma_0
%	 B^*(t-\ts_0) \;. \;\;\Box
%$$

\bigskip
\noindent
Therefore we can continue (\ref{eq:errest})
using  the definition of $\cE_N^0,  \cD_{N}^{0,1}$
and a Schwarz inequality:
\bey
	\lefteqn{\Big|\Tr_e \g^{err}_K(t)\cO_\e\Big|}
\label{eq:errest1}\\
	&\leq&  C_\cO
	\Tr_{e+ph}\Bigg\{  \nu \sum_{N=0}^{K-1} \langle N \rangle^3
	 \sum_{n=0}^N 
	\cE_{n,N}^0 (t)\Gamma_0\Big[  \cE_{n,N}^0 (t)\Big]^* 
	 + \nu^{-1} \sum_{N=K}^{N_0-1} \langle N \rangle^3
	 \sum_{n=0}^N 
	\cE_{n,N}^0 (t)\Gamma_0\Big[  \cE_{n,N}^0 (t)\Big]^*	
	\nonumber\\
	&& +\nu^{-1} t N_0^2 \sum_{n=0}^{N_0} \int_0^t \rd s\;
	 \cD_{n,N_0}^0 (s)\Gamma_0
	\Big[  \cD_{n,N_0}^0 (s)\Big]^*
	\! +\nu^{-1} t N_0^2 \sum_{N=3}^{N_0}\sum_{n=2}^{N} \int_0^t \rd s\;
	  \cD_{n,N}^1 (s)\Gamma_0
	\Big[  \cD_{n,N}^1 (s)\Big]^*\!
	\Bigg\} . \nonumber
\eey

\bigskip

The main lemma is the following:

\begin{lemma}\label{lemma:main}
Let $0\leq a < 1$ and  $n \leq N$ with the same parity.
For simplicity, we let $M:= {n+N\over 2}$.
For the fully expanded objects:
\be	
	\limsup_{L\to\infty}\Tr_{e+ph}\Bigg( \cE_{n,N}^{0}(t) \Gamma_0 
	\Big[ \cE_{n,N}^{0}(t)\Big]^*\Bigg)
	\leq {(C_a\lambda^2t)^N\over \big[ M!\big]^a} + 
	(C\lambda^2t)^N {n!\over t^{1/2}} (\log^*t)^4
\label{norecestsmall}\ee
(this will be used for $N < K$);
\be	
	\limsup_{L\to\infty}
	\Tr_{e+ph}\Bigg( \cE_{n,N}^{0}(t) \Gamma_0 
	\Big[ \cE_{n,N}^{0}(t)\Big]^*\Bigg)
	\leq {(C_a\lambda^2t)^N\over \big[M!\big]^a}
	+(C\lambda^2t)^N {n!\over t^6} (\log^*t)^{n+10}
\label{norecest}\ee
(this will be used for $K\leq N < N_0$).
For the amputated objects, $N\ge 3$,
\be
	\limsup_{L\to\infty}
	\Tr_{e+ph}\Bigg( \cD_{n,N}^{0}(t) \Gamma_0 
	\Big[ \cD_{n,N}^{0}(t)\Big]^*\Bigg)
	\leq  {1\over t}\;
	{(C_a\lambda^2t)^N\over \big[M!\big]^a}
	+ {1\over t}\; (C\lambda^2t)^N {n!\over t^6} (\log^*t)^{n+10}
	\chi(N\ge 7)\; ;
\label{norecampest}
\ee
\be
	\limsup_{L\to\infty}\Tr_{e+ph}\Bigg( \cD^{1}_{n,N}(t) \Gamma_0 
	\Big[  \cD^{1}_{n,N}(t) \Big]^*\Bigg)
	\leq {1\over t}\;
	 (C\lambda^2t)^N \Bigg[ {1\over t^2} (\log^*t)^6
	 +{n!\over t^6} (\log^*t)^{n+10}
	\chi(N\ge 7)\Bigg]\;.
\label{1recampestor}\ee
We introduced the notation $\log^* t : = \max\{ 1, \log t\}$
and $\chi$ is the characteristic function. 
\end{lemma}

{\it Remark 1.} Each estimate has two parts. The first terms
include the contribution of the  direct pairing term
({\it "ladder"} diagram) with possible immediate recollisions
({\it "one-loop renormalization"}). The second terms contain the non-classical
indirect terms (so-called {\it "crossing"} terms).

{\it Remark 2.} From the indirect terms in (\ref{norecestsmall})
 we gain only $t^{1/2}$
instead of $t$, but the power of $\log^* t$ is bounded.
In (\ref{norecest}) we gain full $t$-powers 
but we lose in the power of $\log ^*t$. With more careful
estimates it is possible to remove the logarithms and still gain
full $t$-powers but we do not need it.

%{\it Remark 3.}
% The estimate (\ref{norecest}) would be enough in a weaker form:
%as
%\be
%	\leq {(C_a\lambda^2t)^N\over \big[M!\big]^a}
%	+(C\lambda^2t)^N {n!\over t^5} (\log^*t)^{n}
%\label{weaknorecest}
%\ee
%but we will need the stronger estimate for the amputated
%object  (\ref{norecampest}) which is proven exactly as 
%(\ref{norecest}).
% The weaker form corresponds to using $\kappa =5$ 
%crossings (or so-called
%"peaks", see Definition \ref{def:peak}), 
%while (\ref{norecest}) represents $\kappa=6$
%crossings (for the  $\om(\bk)=(const.)$ case).
%For the general $\om\neq (const.)$
%  one uses $\kappa=7$ or $\kappa=8$ crossings,
%respectively, whether the weaker (\ref{weaknorecest}) or
%the stronger (\ref{norecest}) 
% estimate is proven.

{\it Remark 3.} The basic idea in these estimates is that we do not
gain factorials  {\it together with} crossing  or recollision
gains. This is only a technical convenience.
In fact, instead of (\ref{1recampestor}), it is possible to prove
\be
	\limsup_{L\to\infty}\Tr_{e+ph}\Bigg( \cD^{1}_{n,N}(t) \Gamma_0 
	\Big[  \cD^{1}_{n,N}(t) \Big]^*\Bigg)
	\leq {1\over t}\; 
	 {(C_a\lambda^2t)^N\over\big[ M!\big]^a}
	\; {n!\over t^2} 
\label{1recampest}\ee
which would be equally sufficient to estimate the recollision terms.
In this estimate we would gain a factorial together
with a recollision gain. We follow the first path 
since it is somewhat shorter.

\bigskip

%Once we have these estimates, we can  finish the 
%estimate (\ref{eq:errest1}) for $6\leq K\leq N_0$
%\bey
%	\lefteqn{\limsup_{L\to\infty} \Big|\Tr_e \g^{err}_K(t)\cO_\e\Big|}
%	\nonumber\\
%	&\leq& C_\cO
%	\Bigg\{ \nu
%	\sum_{N=0}^{K-1} \langle N\rangle^3
%	\sum_{n=0}^N \Bigg[{(C_a\lambda^2t)^N\over
%	 \big[M!\big]^a} + 
%	(C\lambda^2t)^N {n!\over t^{1/2}} (\log^*t)^4 \Bigg]
%	\nonumber\\
%	&&+ \nu^{-1} 
%	\sum_{N=K}^{N_0-1} \langle N\rangle^3 \sum_{n=0}^N  \Bigg[ 
%	 {(C_a\lambda^2t)^N\over  \big[M!\big]^a}
%	+(C\lambda^2t)^N {n!\over t^6} (\log^*t)^{n+10}\Bigg]
%	\nonumber\\
%	&&
%	+ \nu^{-1}t N_0^2\sum_{n=0}^{N_0} \int_0^t {\rd s\over s} 
%	 (C_a\lambda^2s)^{N_0}\Bigg( {1\over 
%	\big[  M_0!\big]^a}
%	+{n!\over s^6}  (\log^*s)^{n+10} \Bigg) 
%	\nonumber\\
%	&&+  \nu^{-1}t N_0^2\sum_{N=3}^{N_0}
%	 \sum_{n=2}^{N} \int_0^t {\rd s\over s} 
%	(C\lambda^2s)^{N}\Bigg( {(\log^*s)^6\over s^2}
%	+ {n!\over s^6} (\log^*s)^{n+10}
%	\chi(N\ge 7)
%	\Bigg)	\Bigg\}\;,
%\label{eq:errest2}
%\eey
%where we introduced $M_0: = (N_0+n)/2$ analogously to $M=(N+n)/2$.

We use  Lemma  \ref{lemma:main} to estimate the terms in 
(\ref{eq:errest1}).
We choose  ${1\over 1.01}< a <1 $ and
\be
	N_0: = {2.2 \; \log t \over 
	\log \log t}\; ,
\label{nzerochoice}
\ee
hence
$$
	t^{2.1}\ll N_0!\ll t^{2.2}\;,
	\quad (\log^*t)^{N_0} =  t^{2.2}\;,
	 \quad C^{N_0} \ll  t^{\delta}
$$
for any $\delta>0$, $t\gg 1$.

First we let $\e\to 0$ then $K\to \infty$
and finally $\nu\to0$ in (\ref{eq:errest1}).
Recall that $\lambda^2 =\e$ and $t= T/\e$, i.e., $\lambda^2 t = T$.
Easy calculation shows that
\be
	\lim_{\nu\to0}\limsup_{K\to\infty} \limsup_{\e\to0} 
 	\limsup_{L\to\infty} \Big|\Tr_e \g^{err}_K(t)\cO_\e \Big| =0\;,
\label{errzero}
\ee
which proves (\ref{errz}).

\section{Formulas}
\setcounter{equation}{0}

In this section is we derive 
expressions for the terms
in Lemma \ref{lemma:main}.

\subsection{Definition and estimates of basic functions}

We define the following functions for $\bp,\bk\in \bR^d$, 
$\sigma\in \{ \pm \}$, $\eta\neq 0$, $\alpha, s\in \bR$
\bey
	 M(\bk, \s) : &=& |Q(\bk)|^2 \Big( \cN(\bk ) + {\s +1\over 2}\Big)\;,
\label{def:M}\\
	\Phi_\s ( \bp, \bk): &=& e(\bk+\bp) + \s\om(\bk)\;,
\label{def:Phi}\\
	\Theta(s, \bp, \Om): &=&
	\sum_{\s\in \{\pm\}}
	\int e^{-is [\Phi_\s ( \bp, \bk)
	+\Om]} M(\bk, \s) \rd\bk \;,
\label{def:Theta}\\
	\Upsilon_\eta(\alpha, \bp):&=&
	\sum_{\s\in \{ \pm \}}
	\int{M(\bk, \s) \rd \bk\over \a -\Phi_\s ( \bp, \bk)
	+ i\eta}\;  .
\label{def:Up}
\eey
 We also define
\be
	M^*(\bk): = \max_{\sigma \in \{\pm\} } \max_{j=0,1,\ldots, 2d}
	 |\nabla^j_\bk M(\bk, \s)|
\label{def:Mstar}
\ee
and from (\ref{Qdec}) and  the properties of $\cN$ from
Section \ref{sec:assump} we have
\be
	|M^*(\bk)|\leq C\langle \bk \rangle^{-4d-24} \;.
\label{Mstardec}
\ee
The following lemma estimates the functions $\Theta$ and $\Upsilon$:

\begin{lemma}\label{lemma:Thetabound}
Under the conditions in Section \ref{sec:assump} and (\ref{Mstardec})
we have
\be
	\sup_{\bp, \Om}
	| \Theta(s, \bp, \Om)|\leq {C\over \langle s \rangle^{d/2}}\;;
\label{Thetaest}\ee
\be
	\sup_{\bp,\a,\eta} |\Upsilon_\eta(\alpha, \bp)|\leq C\;,
	\qquad 
	\sup_{\eta} |\Upsilon_\eta(\alpha, \bp)|
	\leq {C \langle \bp \rangle^d\over \langle \a - e(\bp)\rangle}\;;
\label{Upsilonest}\ee
\be
	\sup_{\bp,  \a, \eta}\Bigg( 
	\Big|\nabla_\bp \Upsilon_\eta (\a, \bp)\Big|+
	\Big|\partial_\a \Upsilon_\eta (\a, \bp)\Big|
	+\Big|\partial_\eta \Upsilon_\eta (\a, \bp)\Big|\Bigg)
	\leq C\eta^{-1/2}\;.
\label{derup}
\ee
For $\om = (const.)$ we also have
\be
	\sup_{\bp,  \a, \eta} |
	\nabla_\bp^{\ell}\Upsilon_\eta(\alpha, \bp)|
	\leq C\;, \qquad \sup_{\eta} |
	\nabla_\bp^{\ell}\Upsilon_\eta(\alpha, \bp)|
	\leq {C \langle \bp \rangle^d\over \langle \a - e(\bp)\rangle}\;,
	  \qquad \ell\leq d\;.
\label{Upsder}
\ee
Moreover, the limit 
\be
	\Upsilon_{0+}(\a, \bp): = \lim_{\eta\to0+0}
	\Upsilon_{\eta}(\a, \bp)
\label{Upslim}
\ee
exists, and
\be
	\mbox{Im} \; \Upsilon_{0+}(\a, \bp)= 
	-\pi   \sum_{\sigma\in \{\pm\}}
	 \int |Q(k)|^2  \Big( \cN(\bk)+ {\sigma+1\over 2}\Big)\;
	\delta\Big(\a - \Phi_\sigma(\bp,\bk)\Big) 
	\rd\bk\; .
\label{Upszero}
\ee

\end{lemma}

{\it Proof.} Apart from the proof of (\ref{Upszero}), we fix $\sigma$ and
for simplicity we omit it from the notation, i.e., 
 we assume that $\Theta$ and $\Upsilon$ are defined without
the summation over $\s$.
The inequality (\ref{Thetaest}) for $|s|\leq 1$ is trivial.
For large $|s|$ it
follows from the stationary phase formula
\be
	\Big|\int e^{-is\Phi(\bp,\bk)}
	 M(\bk) \rd \bk\Big| \leq {C \over s^{d/2}}
	\Big( \| \langle \nabla \rangle^d M \|_{L^2} + \|\langle\nabla
	\rangle^d M\|_{L^1}\Big)
\label{statph}
\ee
The prototypes of (\ref{statph}) are the linear and  quadratic cases;
$\Phi(\bp, \bk)=\bu\cdot \bk$ or
$\Phi(\bp, \bk)=\bk^2$. The linear phase factor is trivial:
\be
	\Big|\int e^{-is\bu\cdot\bk} M(\bk)\rd\bk\Big| \leq
	{C\over \langle s|\bu|\rangle^{d}} \|\langle\nabla
	\rangle^d M\|_{L^1} \; . 
\label{lin}
\ee
For the quadratic case, by
 standard estimate on the Schr\"odinger evolution kernel
\be
	\Big|\int e^{-is\bk^2}
	 M(\bk) \rd \bk\Big| \leq {C\over s^{d/2}} \| \wh M \|_{L^1}
\label{quad}
\ee
and $\|  \wh M \|_{L^1} \leq C\| \langle \nabla \rangle^d M \|_{L^2}$.
Similar estimates are valid for other purely quadratic phase factors.

The proof of (\ref{statph}) for a
general phase  function uses a partition of unity separating
the isolated critical point. In a small neighborhood of the
 nondegerate critical
point, a smooth change of variables transforms the phase 
into a purely quadratic function and (\ref{quad}) applies.
Away from the critical point the gradient of $\Phi$
is separated away from zero, i.e., (\ref{lin}) applies
with $|\bu|\ge C>0$. The size of the neighborhood and
the Jacobians of the change of variable transformations
are controlled uniformly in $\bp$ using (\ref{hessom}).
The details of this
standard technique are left to the reader (see also Sec. VIII.2
of \cite{St}).

\medskip

For the first estimate in (\ref{Upsilonest}) we use the formula
\be
	\int {M(\bk)\rd\bk\over \a -\Phi(\bp,\bk)+i\eta}
	= i\int_0^\infty \rd s \; e^{is(\a + i\eta)}
	\int e^{-is\Phi(\bp,\bk)} M(\bk) \rd \bk
\label{Upsfor}
\ee
for any $\eta>0$. We estimate $\int e^{-is\Phi(\bp,\bk)} M(\bk) \rd \bk$
from (\ref{Thetaest})
 and use that $d\ge 3$ hence $\langle s\rangle^{-d/2}$ is
integrable.
This inequality is a   extension  of Lemma 3.10
of \cite{EY2}. 

To obtain the factor decaying in $\langle \a - e(\bp)\rangle$, we
insert a smooth cutoff function $\bk\mapsto \theta_\bp(\bk)$
supported on the set $\{ \bk \; : \; |\a - \Phi(\bp,\bk)|\leq 2\}$, so that
$1-\theta_\bp$ is supported on $\{ \bk\; : \; |\a - \Phi(\bp,\bk)|\ge 1\}$.
Using (\ref{eq:econd}) and (\ref{ombound}),
 we see that $| \nabla^\ell \theta_\bp(\bk)|
\leq C_\ell (\langle\bp\rangle^\ell+ \langle \bk\rangle^\ell)$
 for any $\ell$. On the support
of $\theta_\bp$ we repeat the argument above and use (\ref{statph}) with
 $\Big\| \langle\nabla \rangle^d (M\theta_\bp)\Big\|_{L^2} \leq
C\langle\bp\rangle^d$. On the support of $1-\theta_\bp$ the denominator
can be estimated as
$$
	{1\over |\a - \Phi(\bp,\bk)+ i\eta|}
	\leq {C\over\langle \a - e(\bp+\bk)\pm \om(\bk)\rangle }
	\leq {C \langle \bk \rangle \Big( \langle \bp \rangle
	+\langle \bk \rangle \Big) \over\langle \a - e(\bp)\rangle}\;,
$$
again by (\ref{eq:econd}), and then the $\bk$ integration is finite.

\medskip

We will  sketch
the estimate  of the first term in (\ref{derup}), the other two are similar.
We have from (\ref{Upsfor})
$$
	\Big|\nabla_\bp \Upsilon_\eta (\a, \bp) \Big|\leq
	 \int_0^\infty \;  s \; e^{-\eta s}\Bigg| 
	\int e^{-is\Phi(\bp,\bk)} \Big[\nabla_\bp \Phi(\bp,\bk)\Big]
	M(\bk) \rd\bk\Bigg| \; \rd s\;.
$$
Using stationary phase (\ref{statph}), the smoothness and decay
properties of $\Phi$ and $M$, we estimate the term in
absolute value by $C\langle s \rangle^{-d/2}$, which gives
(\ref{derup}) for $d\ge 3$. 

\medskip

If $\om = (const.)$, then after a change of variables
\be
	\nabla_\bp^\ell\Upsilon_\eta(\alpha, \bp) =
	\int {\nabla^\ell M (\bk-\bp) \; \rd\bk\over
	\a - e(\bk) \pm \om +i\eta}\;.
\label{chang}
\ee
Applying the same argument as in the proof of (\ref{Upsilonest}),
using that $\nabla^\ell M$ is $d$-times differentiable
with decaying derivatives, we obtain (\ref{Upsder}).

Finally, the limit (\ref{Upslim}) exists 
by the  estimate of the last term in (\ref{derup}). The formula
(\ref{Upszero}) is a straighforward calculation using the
smoothness of $e(\bk), \om(\bk), \cN(\bk), Q(\bk)$.
$\;\;\Box$

\subsection{Graphical representation of collision histories}

We will write $\cE^0_{n,N}$ (\ref{def:E0})
in momentum representation; $\bp_j$ stand for
electron momenta, and $\bk_j$ denote phonon momenta
(see Section \ref{sec:Duh} later for more details).
The history of the electron-phonon collisions will be represented by a
graph, where the electron lines are solid and the phonon lines
are dashed.
 The bullets represent collisions. We fix an orientation of the edges
of the graph (Fig.~1). The momentum label on each edge expresses
the momentum flow in the direction of the oriented edge.

At each collision
the adjacent three momenta are subject to momentum conservation.
Fig.~1 represents a history with $N$ collisions with
initial momentum $\bp_N$ and final momentum $\bp_0$.

\bigskip\bigskip
\centerline{\epsffile{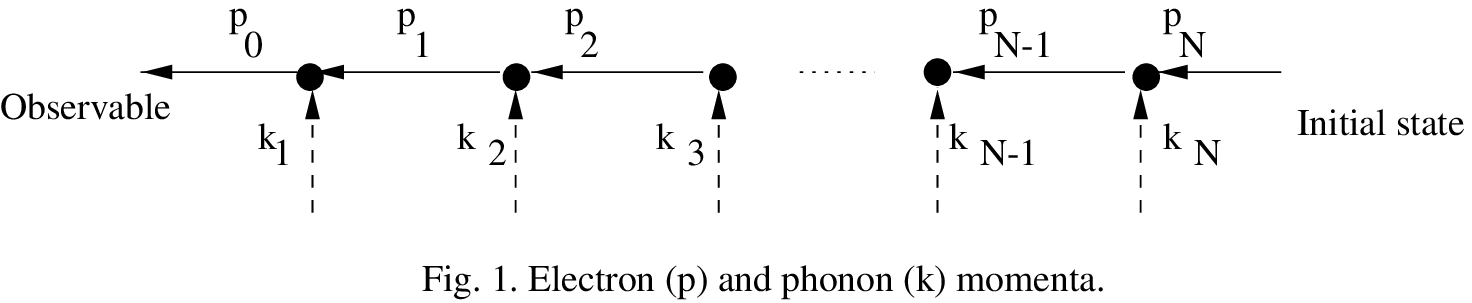}}
\bigskip

When we compute $\Tr_{ph} \cE^0_{n,N} \Gamma_0 \Big[\cE^0_{n,N}\Big]^*$
or similar quadratic expressions,
 we need another copy of the electron history and we will
pair the $2N$ phonon momenta lines using the analogue of the Wick theorem.
 This will be represented
by another copy of the graph on Fig.~1 but all the arrows reversed.
  The corresponding momenta
will be distinguished by tilde. {\it Pairing} means joining the phonon
lines and identifying the associated phonon momenta. 
More precisely,
pairing of $\bk_a$ and $\bk_b$ means identification $\bk_a=-\bk_b$ and
similarly for two tilde variables. If $\bk_a$ and $\tbk_b$
are paired, then the identification is $\bk_a=\tbk_b$.
The graph consisting of two copies of Fig.1 with a pairing
of the phonon lines will be called the {\it pairing graph}.
The orientation is neglected on the figures.

 Immediate reabsorption occurs
when neighboring phonon momenta are paired and these momenta will
 be integrated
out separately. Therefore we will mainly focus on the pairing structure
of the $n$ external phonon lines.

\begin{definition}\label{def:skeleton}
A pairing line in the graph is called an {\bf internal} 
(immediate recollision) line
if it pairs $(\bk_a, \bk_{a+1})$ or $(\tbk_a, \tbk_{a+1})$ for some
$a\leq N-1$. Otherwise it is called an {\bf external} (genuine pairing) line.
The {\bf skeleton}
 of a graph is defined by removing all internal
lines together with their vertices (Fig.2).
\end{definition}

\bigskip\bigskip
\centerline{\epsffile{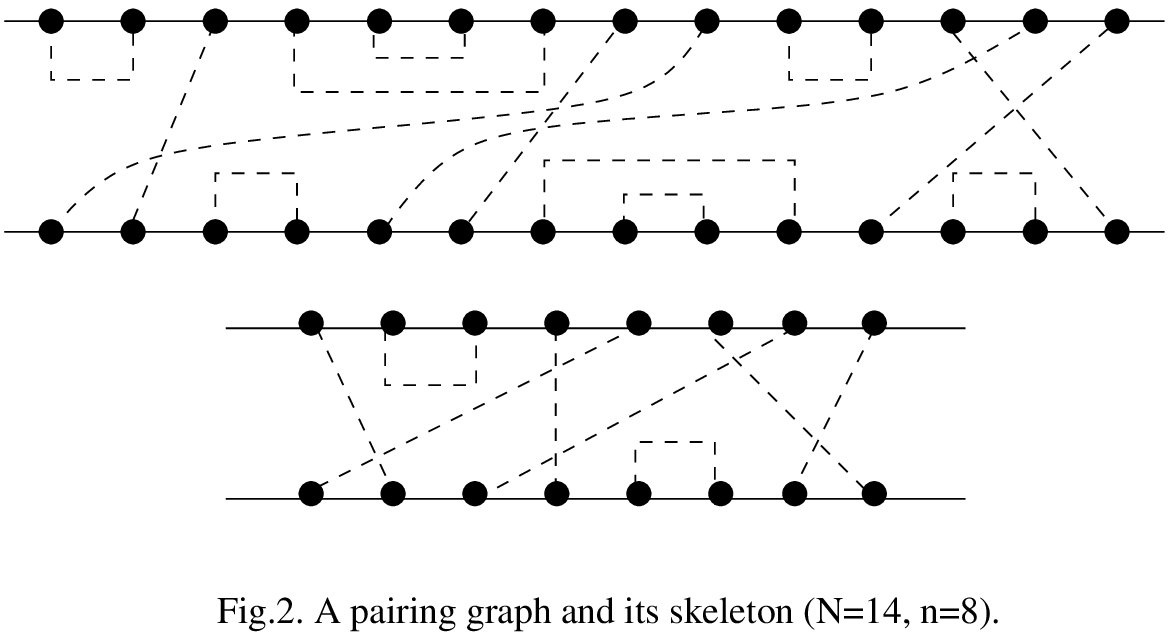}}
\bigskip

For most of our estimates only the skeleton will play a role and
 only the identity of the external lines
should be kept in the notation.
Hence we will relabel the indices of the phonon momenta
(Fig.~3). Moreover, due to momentum conservation,
 the electron momenta between successive
immediate recollisions are the same so they can also be relabelled.

\bigskip\bigskip
\centerline{\epsffile{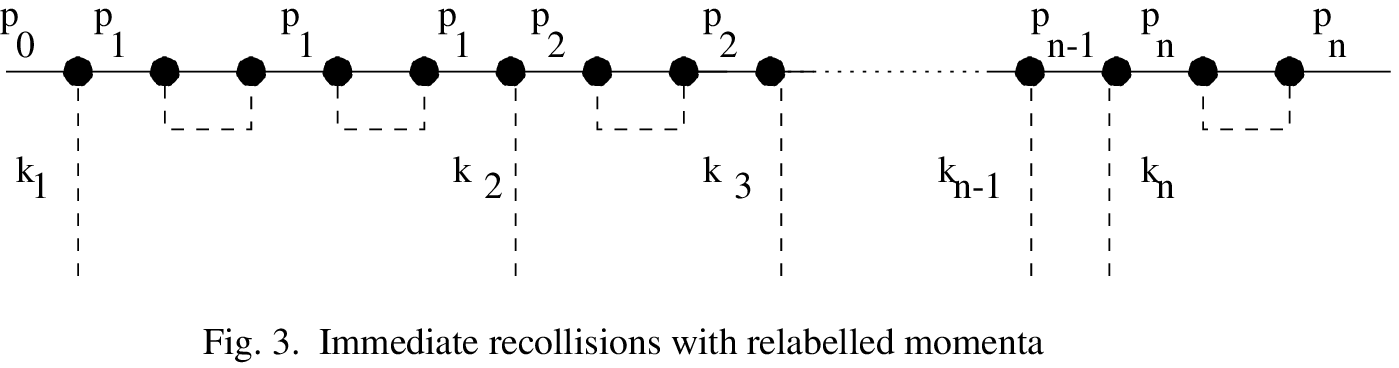}}
\bigskip

The size of the contribution of a specific pairing to 
 $\Tr_{ph} \cE^0_{n,N} \Gamma_0 \Big[\cE^0_{n,N}\Big]^*$
is determined by the skeleton of the pairing. 
There is no recollision line, i.e., all external lines
join an "upper" bullet with a "lower" one.
The main
(physical) contribution comes from the so-called {\bf direct} pairing
(or "ladder graph"). 
 A pairing and the corresponding graph
is called {\bf indirect} or {\bf crossing} if some $(\bk_a, \tbk_b)$ and
$(\bk_{a'}, \tbk_{b'})$ are paired with $a<b$, $a'>b'$.
This notion is independent of relabelling.
Fig.4  shows a direct and an indirect graph.

\bigskip\bigskip
\centerline{\epsffile{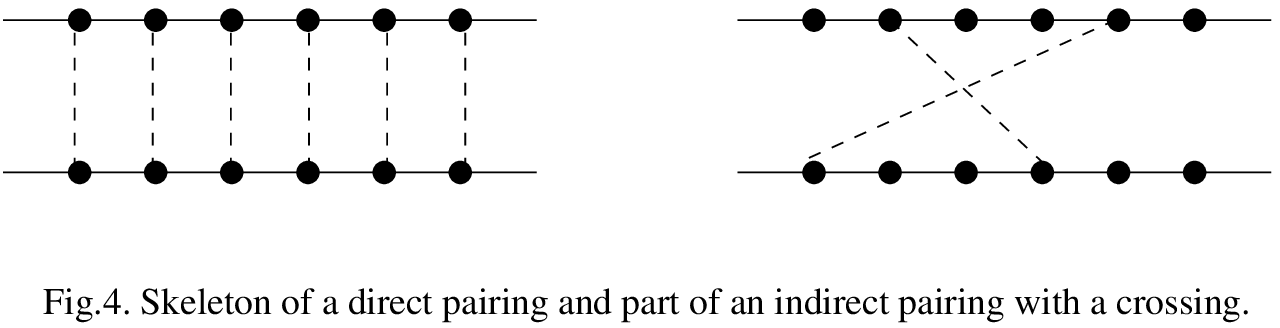}}
\bigskip

A graph is said to contain a {\it genuine reabsorption} 
if some $\bk_a$ is paired with $\bk_b$, $|a-b|\ge 2$,
or the same happens with some 
 $\tbk_a$ and  $\tbk_b$. 
This concept is defined
according to the original labelling, but notice that it
depends only on one copy of the electron history.
 Fig.5 shows graphs with genuine reabsorption. Such graphs represent
the trace of recollision terms, 
$\Tr \; \cD_{n,N}^1\Gamma_0\Big[ \cD_{n,N}^1\Big]^* $ (\ref{def:D1}).

Notice that neighboring momenta may be paired in a skeleton
if they have corresponded to a genuine recollision in the original graph
and all the momenta in between formed immediate pairs
(Fig.2).

\bigskip\bigskip
\centerline{\epsffile{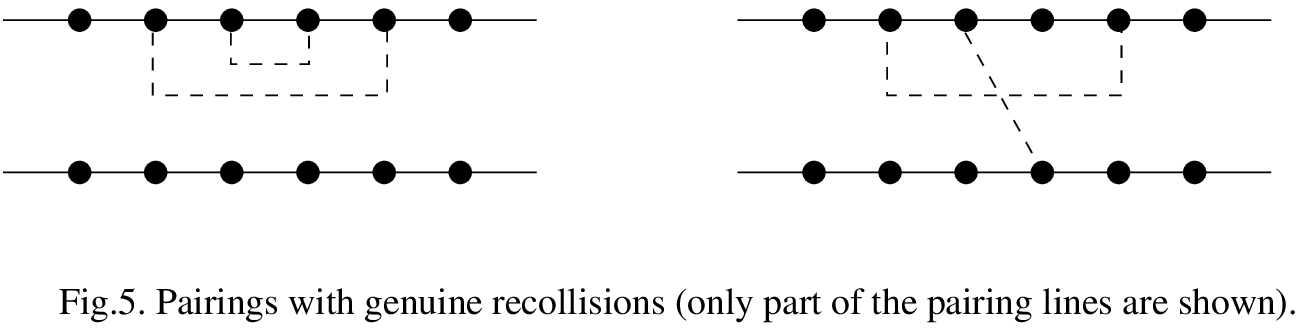}}
\bigskip

\subsection{The main representation formula}\label{sec:mainrep}

Let $\Pi_n$ denote the set of permutations on $\{ 1, 2, \ldots, n\}$.

\begin{proposition}\label{prop:formula}
For any $n\leq N$, $N-n$ even, we have
\be
	\limsup_{L\to\infty}\Bigg|
	\Tr_{e+ph}\Bigg( \cE^{0}_{n,N}(t) \Gamma_0 
	\Big[ \cE^{0}_{n,N}(t)\Big]^*\Bigg)
	-\sum_{\um,\utm \in \cM(n,N)}
	\sum_{\pi\in \Pi_n}
	C_{\um,\utm,\pi}(t)\Bigg| =0
\label{sumcpi1}\ee
with 
\be
	C_{\um,\utm,\pi}(t):=
	 \lambda^{2N}\sum_{\s_j \in \{ \pm\}\atop j=1,\ldots, n}
	\int \rd\nu_\pi(\bp_n,\tbp_n, \ubk, \utbk, \usi)
	\; Y_{\um,\utm,\pi}(t; \bp_n, \tbp_n,
	 \ubk, \utbk, \usi)\;,
\label{CY}
\ee
where we define the measure
\be
	\rd\nu_\pi(\bp_n,\tbp_n, \ubk, \utbk, \usi): =
	\Big(\prod_{j=1}^n M(\bk_j, \s_j) \delta(\bk_j -\tbk_{\pi(j)})
	\rd\bk_j\rd\tbk_j \Big) 
	 \rd \bp_n\rd \tbp_n \delta (\bp_n-\tbp_n)
	\wh\g_e(\bp_n, \tbp_n)
\label{def:nu}
\ee
on $\Lambda^*\times \Lambda^*\times (\Lambda^*)^n\times (\Lambda^*)^n\times
\{ \pm \}\subset
\bR^d\times\bR^d\times (\bR^d)^n\times(\bR^d)^n\times \{ \pm\}^n$,
and we let
\bey
	Y_{\um,\utm,\pi}(t; \bp_n, \tbp_n,\ubk, \utbk, \usi):&=& 
	  \int_0^{t*} [\rd s_b]_{b=0}^N
	 \prod_{j=0}^n\Bigg[ 
	  \prod_{b\in  I_j^c}
	e^{-is_b [e(\bp_j)+
	\Om_j]}
	 \prod_{b\in I_j}\Theta(s_b, \bp_j, \Om_j) \Bigg]
\label{1Y}\\
	&&\times\int_0^{t*} [\rd \ts_\tb]_{\tb=0}^N
	 \prod_{j=0}^n  \Bigg[\prod_{\tb\in  \tI_j^c}
	e^{i\ts_\tb [e(\tbp_j)+
	\wt\Om_j]} 
	 \prod_{\tb\in \tI_j}\overline{\Theta}
	(\ts_\tb, \tbp_j, \wt\Om_j) \Bigg]\nonumber\\
	&=& e^{2t\eta}\int_{-\infty}^\infty \rd\a\; e^{-it\a} \prod_{j=0}^n
	 R_j^{m_j+1}
	\Big[ \Upsilon_\eta(\alpha-\Om_j, \bp_j) \Big]^{m_j}
	\label{2Y}\\
	&&\times\int_{-\infty}^\infty \rd\ta\;  e^{it\ta} \prod_{j=0}^n
	 \tR_j^{m_j+1}
	\Big[ \overline{\Upsilon}_\eta(\ta- \wt\Om_j, \tbp_j) 
	\Big]^{\tm_j} \; ,
\nonumber
\eey
where $I_j, \wt I_j$ etc. depend on $\um,\utm$ (see  Definition \ref{def:J})
and we define
\be
	R_j: = R_j(\a, \bp_j, \Om_j, \eta)
	 = {1\over \a -e(\bp_j)-
	\Om_j + i\eta}\;,
\label{def:R}
\ee
$$
	 \tR_j: = R_j(\ta, \tbp_j, \wt\Om_j, \eta)
	 ={1\over \ta -e(\tbp_j)-
	\wt\Om_j - i\eta} \; .
$$
In these formulas, $\bp_j$'s are functions of $\bp_n$ and 
$\ubk=(\bk_1, \ldots, \bk_n)$, and
similarly for the tilde variables:
\be
	\bp_j:= \bp_n +\sum_{\ell=j+1}^n \bk_\ell,
	\qquad \tbp_j:= \tbp_n +\sum_{\ell=j+1}^n \tbk_\ell,
\label{def:pj}
\ee
and
\be
	\Om_j = \Om_j(\ubk, \usi):= \sum_{\ell=j+1}^n \s_\ell \om (\bk_\ell),
	\qquad \wt\Om_j: = \Om_j(\utbk, \usi\circ\pi^{-1})=
	\sum_{\ell=j+1}^n \s_{\pi^{-1}(\ell)} \om (\tbk_\ell)
\label{def:Om}
\ee
with $\usi=(\s_1, \ldots, \s_n)$.
\end{proposition}

{\it Remark.} This proposition shows that only those pairings
are relevant that respect the no-recollision rule. In particular,
every pairing can be identified with a permutation
on the (relabelled) indices of the external  lines.
In the sequel we use this identification freely.

\medskip

The following subsections contain
 the  proof of Proposition \ref{prop:formula}.

\subsection{Duhamel formula in momentum space}\label{sec:Duh}

We express $\cE_{n,N}^0$ (\ref{def:E0}) in momentum space.
Given $\ubk:= (\bk_1, \bk_2, \ldots, \bk_N)$ and $\bp_N$, we
 define variables $\bp_0, \bp_1,  \ldots, \bp_{N-1}$ as follows
(Fig.1)
\be
	\bp_j = \bp_N + \sum_{\ell=j+1}^N \bk_\ell, 
	\qquad j =0, 1, \ldots, N-1\;.
\label{def:pjN}
\ee
Sometimes these relations will be expressed as
$$
	\int \Big( \prod_{j=0}^{N-1} \rd \bp_j\Big)
	\Delta(\ubp, \ubk)
$$
with
\be
	\Delta(\ubp, \ubk): 
	= \prod_{j=0}^{N-1}
	 \delta\Big( \bp_j - \bp_N - \sum_{\ell=j+1}^N \bk_\ell\Big)\;,
\label{def:Delta}
\ee
but mostly we  consider $\bp_0, \bp_1, \ldots, \bp_{N-1}$
as functions of $\bk_1, \ldots, \bk_N$ and $\bp_N$.

For expressions that are quadratic in $\cE_{n,N}^0$, e.g. (\ref{sumcpi1}),
 we  need another set of variables
distinguished by tilde. Given 
$\utbk:= (\tbk_1, \tbk_2, \ldots, \tbk_N)$ and $\tbp_N$, we 
 define, for $j=0, 1,2, \ldots, N-1$,
\be
	\tbp_j := \tbp_N + \sum_{\ell=j+1}^N \tbk_\ell\;.
\label{def:tpjN}
\ee

Note that the $(\bk_1, \bk_2, \ldots, \bk_n)$, 
and $(\bp_0, \bp_1, \ldots, \bp_n)$ variables in Proposition
 \ref{prop:formula}
will be only  a subset of $( \bk_1, \ldots, \bk_N)$ and $(\bp_0,\ldots,
\bp_N)$ after relabelling them (see later).

Using this notation, we can rewrite the kernel 
$\cB(t, \ubk, N; \bp_0, \bp_N)$
of the operator $\cB(t, \ubk, N)$ (\ref{def:cB}) in the
Fourier space  of $\cH_e$
\bey
	\lefteqn{\cB (t, \ubk, N; \bp_0, \bp_N):}
\label{def:cBker}\\
	&=& \lambda^N\int \Big( \prod_{j=1}^{N-1} \rd \bp_j\Big)
	\Delta(\ubp, \ubk)
	\int_0^{t*} [\rd s_j]_0^N \; e^{-is_0[e(\bp_0)+H_{ph}]}
	 \Big(\prod_{j=1}^N Q(\bk_j) b_{\bk_j} e^{-is_j[e(\bp_j)+H_{ph}]}
	 \Big) \;.
\nonumber
\eey
This is an operator acting on $\cH_{ph}$.
Notice that $\bp_0$ is not integrated out, i.e.,  there is
a delta function $\delta \Big( \bp_0 - \bp_N -\sum_{\ell=1}^N \bk_\ell\Big)$
on the right hand side showing
that $\ubk, \bp_0, \bp_N$ are  not independent.
Equivalently,  we can  write
\be
	\cB \Big(t, \ubk, N; \bp_N + \sum_{j=1}^N \bk_j, \bp_N\Big):
	= \lambda^N
	\int_0^{t*} [\rd s_j]_0^N \; e^{-is_0[e(\bp_0)+H_{ph}]}
	 \Big(\prod_{j=1}^N Q(\bk_j) b_{\bk_j} e^{-is_j[e(\bp_j)+H_{ph}]}\Big)
\label{def:cBkersub}
\ee
if we consider $\bp_j$ (for $j=0,1, \ldots, N-1)$ 
as functions of  $\bk_1, \ldots, \bk_N$ and $\bp_N$ given by
(\ref{def:pjN}).

We collect the terms depending on the phonon operators.
Recall that
$$
	e^{-isH_{ph}}c_\bk e^{isH_{ph}} = e^{is\om(\bk)}c_\bk ,
	\qquad
	e^{-isH_{ph}}c_\bk^\dagger e^{isH_{ph}} = 
	e^{-is\om(\bk)}c_\bk^\dagger\; .
$$
Define
$$
	b_\bk(s) : = e^{-isH_{ph}}b_\bk e^{isH_{ph}}
	= e^{-is\om(\bk)}c_\bk^\dagger - e^{is\om(\bk)}c_{-\bk}\;,
$$
then  its adjoint is
$b_\bk^*(s)  =-b_{-\bk}(s)$ and 
\be
	[ b_{\bk}(\tau), b_\bm(s)]
	=\delta(\bk+\bm)\Big[ e^{-i(\tau-s)\om(\bk)} -  e^{i(\tau-s)\om(\bk)}
	\Big]\; .
\label{btaucomm}
\ee
Let
\be
	G^\# (\bu, \tau ):=\Big\{ e^{-i\tau \om(\bu)} \cN(\bu) 
	+ e^{i\tau \om(\bu)} (\cN(\bu) +1)\Big\}\; ,
\label{cov}
\ee
then
\be
	\Tr_{ph}\gamma_{ph} b_\bu(\tau)b_\bv(s) =
	\Tr_{ph}\gamma_{ph} b_\bu^*(\tau)b_\bv^*(s)
	= -G^\# (\bu, \tau -s) \delta(\bu+\bv)\; ,
\label{bb}
\ee
$$
	\Tr_{ph}\gamma_{ph} b_\bu(\tau)b_\bv^*(s) =
	\Tr_{ph}\gamma_{ph} b_\bu^*(\tau)b_\bv(s)
	=G^\# (\bu, \tau -s) \delta(\bu-\bv) \; .
$$

We define
\be
	\tau_j := s_0+ s_1+\ldots + s_{j-1}\;,
	\qquad \tilde\tau_j:=
	\ts_0+ \ts_1 + \ldots + \ts_{j-1}
\label{def:tau}
\ee 
for $j\ge 1$ and $\tau_0=\ttau_0:=0$ and let
$\utau: =(\tau_1, \ldots, \tau_N)$, $\uttau:=(\ttau_1, \ldots ,
\ttau_N)$.

The terms involving phonon operators
in (\ref{def:cBkersub}) are:
\be
	 e^{-is_0H_{ph}}\Bigg(\prod_{j=1}^N b_{\bk_j}
	e^{-is_jH_{ph}}	\Bigg)
	 = \Big(\prod_{j=1}^N 
	b_{\bk_j}(\tau_j)\Big)e^{-itH_{ph}}\;
\label{eq:collphon}
\ee
recalling the convention (\ref{opprod}). We define 
\bey
	T(\ubk, \utbk, \utau, \uttau):&=& 
	\Tr_{ph}\Bigg[ \Big(\prod_{j=1}^N b_{\bk_j}(\tau_j)\Big)
	e^{-itH_{ph}}\g_{ph} e^{itH_{ph}}  
	\Big(\prod_{j=N}^1 b^*_{\tbk_j}(\ttau_j)\Big)
	\Bigg]
\label{def:T}\\
	&=&\Tr_{ph}\Bigg[\gamma_{ph}b^*_{\tbk_N}(\ttau_N)\ldots
	b^*_{\tbk_2}(\ttau_2) b^*_{\tbk_1}(\ttau_1) b_{\bk_1} (\tau_1)
	 b_{\bk_2}(\tau_2) \ldots
	 b_{\bk_N}(\tau_N)
	 \Bigg] \; . \nonumber
\eey

%This gives the  formula
%\be
%	\cB \Big(t, \ubk, N; \bp_N + \sum_{j=1}^N \bk_j, \bp_N\Big)
%	=  \lambda^N
%	\int_0^{t*} [\rd s_j]_0^N \; e^{-is_0e(\bp_0)}
%	 \Big(\prod_{j=1}^N Q(\bk_j) e^{-is_je(\bp_j)}\Big)
%	\Big(\prod_{j=1}^N b_{\bk_j}(\tau_j)\Big)e^{-itH_{ph}}\;.
%\label{eq:Bker}
%\ee

Combining (\ref{def:E0}), (\ref{def:cBker}), (\ref{eq:collphon})
and (\ref{def:T}) we 
obtain  an integral formula
for the fully expanded term with no reabsorptions:
\bey
	\lefteqn{\Tr_{e+ph}\Bigg( \cE^{0}_{n,N}(t) \Gamma_0 
	\Big[ \cE^{0}_{n,N}(t)\Big]^*\Bigg)}\label{eq:ceexp}\\ 
	&=& \lambda^{2N}\int^{\#(n,N)}\rd\ubk\rd\utbk
	  \int_0^{t*} [\rd s_j]_0^N  [\rd \ts_j]_0^N
	T(\ubk, \utbk, \utau, \uttau)
	 \int \rd \bp_N\rd \tbp_N \delta (\bp_N-\tbp_N)
	\wh\g_e(\bp_N, \tbp_N)\nonumber\\
	&&\times\Bigg[ \int \Big(\prod_{j=0}^{N-1} \rd \bp_j\Big) 
	\Delta(\ubp, \ubk)  e^{-is_0e(\bp_0)}
	\Big(\prod_{j=1}^{N} e^{-is_je(\bp_j)} Q(\bk_j)\Big)
	\Bigg]
	\nonumber\\
	&&\times
	\Bigg[\int \Big(\prod_{j=0}^{N-1} \rd \tbp_j\Big) 
	\Delta(\utbp, \utbk)
	e^{i\ts_0e(\tbp_0)}
	 \Big(\prod_{j=1}^N e^{i\ts_je(\tbp_j)} Q(\tbk_j)\Big)
	\Bigg] \; . \nonumber
\eey
Here we used that $Q(\bk)$ is real and that $\delta(\bp_0-\tbp_0)$
from $\Tr_e$ is equivalent to $\delta(\bp_N-\tbp_N)$.

\subsection{Computing the phonon trace and the thermodynamic limit}
\label{sec:thermotrace}

We use Wick's theorem and the definition of $G^\#$ (\ref{cov})
to compute the phonon trace (\ref{def:T}) appearing in (\ref{eq:ceexp}). 
We have to consider all pairings
within the indices of the set $\{\ubk,\utbk\}:=
\{ \bk_1, \ldots , \bk_N, \tbk_1, \ldots , \tbk_N\}$.
We recall the definition of $\int^{\# (n,N)}$ from (\ref{def:schint}),
and similarly we have
$$
	\int^{\#(n,N)} \rd\utbk
	=\sum_{\utm\in \cM(n,N)} 
	\int^{(\utm)} \rd\utbk\;.
$$
We also recall that $\um \in \cM(n, N)$ is equivalent to  a subsequence
$\mu$, which is viewed as a monotonic map $\mu: \{ 1, \ldots, n\}
\mapsto\{1, \ldots, N\}$.
The sets $I, I_j, J$ depend on $\um$ and we use
tilde for the sets $\tI:= I(\utm)$, $\tI_j: = I_j(\utm)$
and $\tJ: = J(\utm)$.

Certain pairs are already prepared by imposing the
immediate reabsorptions in the measures via $\um, \utm$.
We say that a Wick-pairing of $\{\ubk,\utbk\}$
{\it respects} a given
pair $\um,\utm\in \cM(n,N)$, if
$b$ is paired with $b+1$ for all $b\in J$
and similarly for $\tb\in\tJ$ (see (\ref{def:Xi})).

If a Wick-pairing respects $(\um,\utm)$,
then only the  indices from $I$ and $\tI$
will be paired freely. Moreover, since the momenta with
indices from $I$ cannot be paired (see (\ref{def:schintmu})),
 and so are the momenta
within $\tI$, the pairing  occurs between the momenta
$\bk_{b}$ ($b\in I$) and $\tbk_{\tb}$ ($\tb\in \tI$).
Hence any Wick-pairing that respects $(\um,\utm)$
can be uniquely identified
with a map $\pi^* \in \cP_n:=\tmu \Pi_n \mu^{-1}$ between the
sets $I$ and $\tI$. Here $\mu,\tmu$ are the maps
associated with $\um,\utm$, and
$\cP_n=\tmu\Pi_n\mu^{-1}$ is the
set of  maps of the form $\tmu\circ \pi \circ\mu^{-1}: I\mapsto\tI$
with $\pi\in \Pi_n$.

The following lemma states that only those Wick-pairings
are relevant that respect a  given $(\um,\utm)$, the rest are 
negligible in the thermodynamic limit.

\begin{lemma}\label{lemma:thermo}
Let $\um, \utm\in \cM(n,N)$
and let $\utau, \uttau$
be fixed times. For any function $F$ of the momenta $\ubk,\utbk$
and times $\utau, \uttau$ we have
\bey
        \lefteqn{\int^{(\um)} \rd\ubk
 	\int^{(\utm)} \rd\utbk \; 
	F(\ubk, \utbk,\utau,\uttau)
	T(\ubk, \utbk, \utau, \uttau) 
	=  \int \rd\ubk
 	\int \rd\utbk\;
	 F(\ubk, \utbk,\utau,\uttau)}
\nonumber\\
	&&\times\sum_{\pi^*\in \cP_n}
	 \Bigg[ \prod_{b\in I(\um)} \delta\Big(\bk_{b}-
	\tbk_{\pi^*(b)}\Big) G^\#(\bk_{b}, \ttau_{\pi^*(b)}
	- \tau_{b}) \Bigg]
	\;
	\overline{\cX}(\ubk, \utau, \um)\; \cX(\utbk, \uttau, \utm)+ 
	O\Big( {\| F\|_\infty\over |\Lambda|}\Big)\; ,\nonumber
\eey
where
\be
	\cX(\ubk, \utau, \um): = 
	 \prod_{b\in J(\um)}
	\delta(\bk_{b}+\bk_{b+1}) G^\#(\bk_b, \tau_{b+1}-
	\tau_b)
\label{def:cX}
\ee
expresses the trace of the immediate reabsorptions.
Notice that the delta functions in $\cX$ coincide with the
characteristic functions of $\Xi$ built into the measure $\int^{(\um)}$
(\ref{def:Xi}).
The error depends only on  $N$ and on the maximum
of the function $\cN$.
\end{lemma}

{\it Proof.} 
Consider a factor $\chi(\bk_{b} 
+\bk_{b+1})$ in the definition of $\Xi (\ubk,\um)$.
Suppose that the Wick pairing did not pair $\bk_b$ 
with $\bk_{b+1}$. Then one can integrate out all the other
$\bk,\tbk$ variables in such order that eventually
all delta functions disappear and we are
left with an integration
\be
	\Bigg|\int \rd\bk_{b}\rd\bk_{b+1} \chi(\bk_{b} 
	+\bk_{b+1}) H(\bk_{b}, \bk_{b+1},\utau, \uttau) \Bigg|
	\leq {\| H \|_\infty\over |\Lambda|}\; ,
\label{Hlambda}\ee
where $H$ is the result of all the other integrations
and clearly $\|H\|_\infty\leq C(N)\| F \|_\infty$. 

Therefore only those Wick-pairings are relevant
that respect the prepared immediate reabsorptions
built into the measures via $\um,\utm$ and they are
described by a map $\pi^*\in \cP_n$.
For such pairing one can remove all further restrictions
from the measure (imposed by $(\um), (\utm)$ superscripts
in (\ref{def:schintmu}))
since $\Xi (\ubk,\um)$ is imposed by the delta functions in $\cX$,
while the complement of the condition,
 that $\bk_b\neq - \bk_{b'}$ for $b\neq b'\in I$
 (see (\ref{def:schintmu})), is small
by a volume factor using an estimate similar to (\ref{Hlambda}).
$\;\;\Box$

\bigskip

{\it Remark:} Similar argument is valid for amputated and
recollision
terms, $\cD_{n,N}^{0,1}$.
 Since we always take the thermodynamic limit $L\to\infty$
first, we can always assume that the prepared immediate reabsorptions
are respected by the Wick pairings, modulo negligible errors.

\bigskip

Using this Lemma and the fact that the expression in
the big curly bracket in (\ref{eq:ceexp}) has
a volume independent bound, we  obtain from (\ref{eq:ceexp})
%$$
%	\Tr_{e+ph}\Bigg( \cE^{0}_{n,N}(t) \Gamma_0 
%	\Big[ \cE^{0}_{n,N}(t)\Big]^*\Bigg)
%	= \lambda^{2N}\sum_{\um,\utm \in \cM(n,N)}
%	\int^{(\um)}\Big(\prod_{j=1}^N \bk_j \Big)
%	 \int^{(\utm)}\Big(\prod_{j=1}^N \tbk_j \Big) 
%$$
%\be
%	\times \int_0^{t*} [\rd s_j]_0^N  \int_0^{t*} [\rd \ts_j]_0^N
%	\sum_{\pi^*\in \cP_n}
%	 \Bigg[ \prod_{b\in I} \delta\Big(\bk_{b}-
%	\tbk_{\pi^*(b)}\Big) G^\#(\bk_{b}, \ttau_{\pi^*(b)}
%	- \tau_{b}) \Bigg]
%	\; \cX(\utbk, \uttau, \utm)\overline{\cX}(\ubk, \utau, \um)
%\label{eq:1ceexp}
%%\ee
%$$
%	\times\Bigg\{ \int \rd \bp_N\rd \tbp_N \delta (\bp_0-\tbp_0)
%	\wh\g_e(\bp_N, \tbp_N)
%	\Bigg[ \int \Big(\prod_{j=0}^{N-1} \rd \bp_j\Big) \Delta(\ubp, \ubk)
%	  e^{-is_0e(\bp_0)}
%	\Big(\prod_{j=1}^{N} e^{-is_je(\bp_j)} Q(\bk_j)\Big)
%	\Bigg]
%$$
%$$
%	\times
%	\Bigg[\int \Big(\prod_{j=0}^{N-1} \rd \tbp_j\Big) \Delta(\utbp, \utbk)
%	e^{i\ts_0\tbp_0^2/2}
%	 \Big(\prod_{j=1}^N e^{i\ts_j\tbp_j^2/2} Q(\tbk_j)\Big)
%	\Bigg] \Bigg\} + O(|\Lambda|^{-1})
%$$
%and the constant in the last error term depends on every parameter.

%\subsection{Proof of the representation formula}\label{sec:profrep}

\be
	\Tr_{e+ph}\Bigg( \cE^{0}_{n,N}(t) \Gamma_0 
	\Big[ \cE^{0}_{n,N}(t)\Big]^*\Bigg)
	=\sum_{\um,\utm \in \cM(n,N)}
	\sum_{\pi\in \Pi_n}
	C_{\um,\utm,\pi}(t) + O(|\Lambda|^{-1})\;,
\label{sumcpi}
\ee
where for any $\um,\utm\in \cM(n, N)$
and for any $\pi\in \Pi_n$ we define
$\pi^*:= \tmu\circ\pi\circ\mu^{-1}$  and
\bey
	\lefteqn{C_{\um,\utm,\pi}(t):= 
	\lambda^{2N}\sum_{\s_b \in \{ \pm\}\atop b\in I\cup J\cup \tJ }
	\int\rd\ubk
	 \; \rd\utbk \;
	 \prod_{b\in I} \delta\Big(\bk_{b}-
	\tbk_{\pi^*(b)}\Big)\Bigg[ \prod_{b\in J}
	\delta(\bk_{b}+\bk_{b+1})
	 \prod_{\tb\in \tJ}
	\delta(\tbk_\tb +\tbk_{\tb+1}) \Bigg]}
\nonumber\\
	&& \times\int \!\rd \bp_N\rd \tbp_N \delta (\bp_N-\tbp_N)
	\wh\g_e(\bp_N, \tbp_N)
	 \Big( \prod_{b\in I\cup J}\!\! M(\bk_b, \s_b)
	\prod_{\tb\in \tJ} \! M(\tbk_\tb, \s_\tb)\Big)
\label{def:cpi}\\
	&&\times\int_0^{t*} [\rd s_j]_0^N \int_0^{t*} [\rd \ts_j]_0^N
	 \Big(\prod_{b\in I} e^{i\s_b (\ttau_{\pi^*(b)}-\tau_b) 
	\om(\bk_b)} 
	\Big)\Bigg[  \prod_{b\in J}
	e^{-i\s_b s_b\om(\bk_b)} 
	\prod_{\tb\in \tJ} e^{i\s_\tb \ts_\tb \om (\tbk_\tb)}
	\Bigg]
\nonumber\\
	&&\times
	\Bigg[ \int \Big(\prod_{j=0}^{N-1} \rd \bp_j\Big) \Delta(\ubp, \ubk)
	 \Big(\prod_{j=0}^N e^{-is_je(\bp_j)} \Big)
	\Bigg]
	\Bigg[\int \Big(\prod_{j=0}^{N-1} \rd \tbp_j\Big) \Delta(\utbp, \utbk)
	 \Big(\prod_{j=0}^N e^{i\ts_je(\tbp_j)}\Big)
	\Bigg] \; . \nonumber
\eey
 We  used  $\tau_{b+1}-\tau_b=s_b$ and
the definition of $M$ (\ref{def:M}) 
to  combine the $G^\#$ terms with the $Q(\bk)$ factors:
\be
	 Q(\br)^2 G^\#(\br, \zeta)
	= \sum_{\s =\pm 1} e^{i\s \zeta \om(\br)} M(\br, \s)\;.
\label{def:Gph}
\ee

\subsection{Proof of the representation formula}\label{sec:profrep}

Recall that $\{ 0, 1,\ldots, N\}$ is a disjoint union
$J\cup J^c$. Let $b\mapsto\chi_b$ be the characteristic
function of $J$, i.e., $\chi_b=1$ if $b\in J$ and $\chi_b=0$
otherwise. Similarly, $\tchi_b$ is the 
characteristic function of $\tJ$.

Using the definition of $\tau_j, \ttau_j$, we can rewrite
the phonon phase factors in (\ref{def:cpi})
$$
	\prod_{b\in I}  e^{-i\s_b \tau_b \om(\bk_b)}
	 \prod_{b\in J}
	e^{-i\s_b s_b\om(\bk_b)}
	= \prod_{j=0}^n \Big( \prod_{b\in I_j\cup I_j^c}
	e^{-is_b [\Om_j(\bk,\usi)+\chi_b\s_b\om(\bk_b)]} \Big)
$$
and
$$	
	\prod_{b\in I}  e^{i\s_b \ttau_{\pi^*(b)} \om(\bk_b)}
	\prod_{\tb\in \tJ} e^{i\s_\tb \ts_\tb \om (\tbk_\tb)}
	=\prod_{j=0}^n \Big( \prod_{\tb\in \tI_j\cup\tI_j^c}
	e^{i\ts_\tb [\wt\Om_j(\tbk,\usi\circ(\pi^*)^{-1})
	+\tchi_\tb\s_\tb\om(\tbk_\tb)]} 
	\Big)
$$
with 
\be
	\Om_j:= \Om_j(\bk, \usi): = \sum_{b\in I\;, \;  b>\mu(j)}
	 \sigma_b \om (\bk_b)
\label{def:Omdef}
\ee
$$
 	\wt\Om_j: =\wt\Om_j (\tbk, \usi\circ(\pi^*)^{-1}): = 
	\sum_{\tb\in \tI\; , \;\tb>\tmu(j)} 
	\s_{(\pi^*)^{-1}(\tb)}\omega(\tbk_\tb)
$$
(recall that $\mu, \tmu$ are associated with $\um,\utm$).
Here $\usi: = \{ \s_b\; : \; b\in I\}$ and usually
 we will omit the arguments.

Hence we have for $\um,\utm\in \cM(n, N)$ that
\bey
	C_{\um,\utm,\pi}(t):&=& 
%	\lambda^{2N}\sum_{\s_b \in \{ \pm\}\atop b\in I\cup J\cup \tJ }
%	\int\Big(\prod_{j=1}^N \rd\bk_j \Big)
%	 \int\Big(\prod_{j=1}^N \rd\tbk_j \Big) 
%	 \prod_{b\in I} \delta\Big(\bk_{b}-
%	\tbk_{\pi^*(b)}\Big)
%$$
%$$
%	\times\Bigg[ \prod_{b\in J}
%	\delta(\bk_{b}+\bk_{b+1})
%	 \prod_{\tb\in \tJ}
%	\delta(\tbk_\tb +\tbk_{\tb+1}) \Bigg]
%	 \int \rd \bp_N\rd \tbp_N \delta (\bp_N-\tbp_N)
%	\wh\g_e(\bp_N, \tbp_N)
%	 \Big( \prod_{b\in I\cup J} M(\bk_b, \s_b)
%	\prod_{\tb\in \tJ} M(\tbk_\tb, \s_\tb)\Big)
%$$
%\be
%	\times\int_0^{t*} [\rd s_j]_0^N
%	\Bigg[ \int \Big(\prod_{j=0}^{N-1} \rd \bp_j\Big) \Delta(\ubp, \ubk)
%	 \prod_{j=0}^n \Big( \prod_{b\in I_j\cup I_j^c}
%	e^{-is_b [\bp_b^2/2+\Om_j(\bk,\usi)+\chi_b\s_b\om(\bk_b)]}\Big)
%	\Bigg]
%\label{eq:cpi2}
%\ee
%$$
%	\times
%	 \int_0^{t*} [\rd \ts_j]_0^N
%	\Bigg[\int \Big(\prod_{j=0}^{N-1} \rd \tbp_j\Big) \Delta(\utbp, \utbk)
%	 \prod_{j=0}^n \Big( \prod_{\tb\in \tI_j\cup\tI_j^c}
%	e^{i\ts_\tb [\tbp_\tb^2/2+\wt\Om_j+\tchi_\tb
%	\s_\tb\om(\tbk_\tb)]} 
%	\Big)\Bigg]
%$$
%$$
	\lambda^{2N}\sum_{\s_b \in \{ \pm\}\atop b\in I\cup J\cup \tJ }
	\int\Big(\prod_{b\in I\cup J}\rd\bk_b \Big)
	\Big(\prod_{\tb\in \tI\cup\tJ}\rd\tbk_\tb \Big) 
	 \prod_{b\in I} \delta\Big(\bk_{b}-
	\tbk_{\pi^*(b)}\Big) 
\label{eq:cpi3}\\
	&& \times\int \rd \bp_N\rd \tbp_N \delta (\bp_N-\tbp_N)
	\wh\g_e(\bp_N, \tbp_N)
	 \Big( \prod_{b\in I\cup J} M(\bk_b, \s_b)
	\prod_{\tb\in \tJ} M(\tbk_\tb, \s_\tb)\Big)
\nonumber\\
	&&\times\int_0^{t*} [\rd s_j]_0^N
	\Bigg[ 
	 \prod_{j=0}^n  \prod_{b\in I_j\cup I_j^c}
	e^{-is_b [e(\bp_{\mu(j)}+\chi_b\bk_b)+
	\Om_j+\chi_b\s_b\om(\bk_b)]} \Bigg]
\nonumber\\
	&&\times
	 \int_0^{t*} [\rd \ts_j]_0^N
	\Bigg[
	 \prod_{j=0}^n  \prod_{\tb\in \tI_j\cup\tI_j^c}
	e^{i\ts_\tb [e(\tbp_{\tmu(j)}+\tchi_\tb\tbk_\tb)
	+\wt\Om_j+\tchi_\tb\s_\tb
	\om(\tbk_\tb)]} \Bigg]\;.
\nonumber
\eey
In this formula we considered $\bp_j, \tbp_j$ as functions
of $\bp_N=\tbp_N$ and $\bk$'s or $\tbk$'s, respectively (see (\ref{def:pjN})),
and we used that 
\bey
	\bp_b =& \bp_{\mu(j)}+\bk_b & \mbox{for} \qquad b\in I_j\; ,
\nonumber\\
	\bp_b =& \bp_{\mu(j)}  & \mbox{for} \qquad b\in I_j^c\;,
\nonumber
\eey
and similarly for tildes. We also freely integrated out
$\bk_{b+1}, \tbk_{\tb+1}$ for $b\in J$, $\tb\in \tJ$ since they
do not appear any more in the formulas.

\bigskip

Let $\bp_j^*:=\bp_{\mu(j)}$, $\tbp_j^*:=\bp_{\tmu(j)}$, 
$\bk_j^*:=\bk_{\mu(j)}$, $\tbk_j^* :=\tbk_{\tmu(j)}$, 
$\s_j^*:=\s_{\mu(j)}$ for
$j=1,\ldots, n$ and $\bp_0^*: = \bp_0$, $\tbp_0^* : = \tbp_0$.
These will be the relabelled momenta (see Figs.~1-3.), temporarily
distinguished by star.
Then by (\ref{def:pjN})
we have $\bp_n^*= \bp_N$, $\tbp_n^*=\tbp_N$ and
\be
	\bp_j^* = \bp_n^* + \sum_{\ell=j+1}^n \bk_\ell^*, \qquad
	\tbp_j^* = \tbp_n^* + \sum_{\ell=j+1}^n \tbk_\ell^*
\label{def:pjstar}
\ee
for $j=0,1,\ldots, n$.
Also notice that 
$$
	\Om_j =  \sum_{\ell=j+1}^n \s_\ell^*\om(\bk_\ell^*),
	\qquad \wt\Om_j = \sum_{\ell=j+1}^n \s_{\pi^{-1}(\ell)}^* 
	\om(\tbk_\ell^*)\;.
$$
Clearly $\Om_j, \wt\Om_j$ depend only on the star  variables, i.e.,
with a slight abuse of notation we continue to denote them as
\be
	\Om_j = \Om_j(\ubk^*, \usi^*)\;,
	\qquad \wt\Om_j=
	\wt\Om_j(\utbk^*,
	 \usi^*\circ \pi^{-1}) \; .
\label{Omdefabuse}
\ee

With these variables and after separating the terms
with immediate recollisions, 
we obtain from (\ref{eq:cpi3}) that
\be 
	C_{\um,\utm,\pi}(t):=
	 \lambda^{2N}\sum_{\s_j^* \in \{ \pm\}\atop j=1,\ldots, n}
	\int \rd\nu_\pi(\bp_n^*, \tbp_n^*, \ubk^*, \utbk^*, \usi^*)
	\; Y_{\um,\utm,\pi}
	(t; \bp_n^*, \tbp_n^*,
	 \ubk^*, \utbk^*, \usi^*)
\label{eq:cpi4}
\ee
with
\bey
	\lefteqn{Y_{\um,\utm,\pi}(t; \bp_n^*, \tbp_n^*,\ubk^*, \utbk^*, \usi^*)
	: = \sum_{\s_b\in \{ \pm\}\atop b\in J\cup \tJ}
	\int \Big(\prod_{b\in J}M(\bk_b, \s_b) \rd \bk_b\Big)
	\Big(\prod_{\tb\in \tJ} M(\tbk_\tb, \s_\tb) \rd \tbk_\tb\Big)}
	\qquad\qquad\qquad
\label{def:Y}\\
	&&\times\int_0^{t*} [\rd s_b]_{b=0}^N
	 \prod_{j=0}^n \Bigg[ 
	 \prod_{b\in  I_j^c}
	e^{-is_b [e(\bp_j^*)+\Om_j]} 
	 \prod_{b\in I_j}
	e^{-is_b [e(\bp_j^*+\bk_b)+
	\Om_j+\s_b\om(\bk_b)]} \Bigg]
\nonumber\\
	&&\times\int_0^{t*} [\rd \ts_\tb]_{\tb=0}^N
	 \prod_{j=0}^n \Bigg[ 
	 \prod_{\tb\in  \tI_j^c}
	e^{i\ts_\tb [e(\tbp_j^*)+\wt\Om_j]} 
	 \prod_{\tb\in \tI_j}
	e^{i\ts_\tb [e(\tbp_j^*+\tbk_\tb)+
	\wt\Om_j+\s_\tb\om(\tbk_\tb)]} \Bigg]\;,
\nonumber
\eey
where $\ubk^*$ stands for the sequence $\bk_1^*, \ldots, \bk_n^*$ and
similarly for the other variables.  We recall that 
$\bp_j^*$, $\tbp_j^*$, $\Om_j$, $\wt\Om_j$
 are functions of $\pi$, $\usi^*$, $\ubk^*$, $\utbk^*$
and $\bp_n^*$ ($\um, \utm$ are considered given).

\bigskip

There are two different expressions for $Y_{\um,\utm,\pi}$
after integrating 
 out the internal momenta for all immediate recollisions.
Recalling the definition of $\Theta$ (\ref{def:Theta}),
we have
\bey
	Y_{\um,\utm,\pi}(t; \bp_n^*, \tbp_n^*,\ubk^*, \utbk^*, \usi^*)&=&
	  \int_0^{t*} [\rd s_b]_{b=0}^N
	 \prod_{j=0}^n\Bigg[ 
	  \prod_{b\in  I_j^c}
	e^{-is_b [e(\bp_j^*)+
	\Om_j]}
	 \prod_{b\in I_j}\Theta(s_b, \bp_j^*, \Om_j) \Bigg]
\label{eq:Y2}\\
	&&\times\int_0^{t*} [\rd \ts_\tb]_{\tb=0}^N
	 \prod_{j=0}^n  \Bigg[\prod_{\tb\in  \tI_j^c}
	e^{i\ts_\tb [e(\tbp_j^*)+
	\wt\Om_j]} 
	 \prod_{\tb\in \tI_j}\overline{\Theta}
	(\ts_\tb, \tbp_j^*, \wt\Om_j) \Bigg]\;.
\nonumber
\eey
Notice that all non-star variables have disappeared.

\bigskip

The other expression for $Y_{\um,\utm,\pi}$ relies on performing
all time integrals by using
$$
	\delta\Big( t-\sum_{b=0}^N s_b\Big)
	= \int_{-\infty}^{\infty}
	 e^{-i\alpha \Big( t-\sum_{b=0}^N s_b\Big)}
	\rd\alpha\; ,
$$
and we can regularize the $\rd s$ integrations by shifting
$e(\bp_j) \to e(\bp_j) -i\eta$.

The result is
\be
	\int_0^{t*} [\rd s_b]_{b=0}^N
	 \prod_{j=0}^n \Bigg[ 
	 \prod_{b\in  I_j^c}
	e^{-is_b [e(\bp_j^*)+\Om_j]} 
	 \prod_{b\in I_j}
	e^{-is_b [e(\bp_j^*+\bk_b)+
	\Om_j+\s_b\om(\bk_b)]} \Bigg]
\label{eq:timeint}
\ee
$$
	= e^{t\eta}
	\int_{-\infty}^\infty \rd\a e^{-it\a} \prod_{j=0}^n
	\Bigg[  \Bigg({1\over \a -e(\bp_j^*)-
	\Om_j + i\eta}\Bigg)^{m_j+1}
	\prod_{b\in I_j}
	{1\over \a -e(\bp_j^*+\bk_b)-
	\Om_j-\s_b\om(\bk_b) + i\eta}\Bigg]\;,
$$
and we will always choose $\eta:=t^{-1}$.
We can use a similar identity for the $\rd \ts_j$ integrals.

Again, we can integrate out $\bk_b$, $b\in J$, and similarly 
for the tilde variables. Recalling the definition of $\Upsilon$
(\ref{def:Up}) we obtain from (\ref{eq:Y2})
\bey
	Y_{\um,\utm,\pi}
%	(t; \bp_n^*, \tbp_n^*,\ubk^*, \utbk^*, \usi^*)
	&=&
	e^{2t\eta}\int_{-\infty}^\infty \rd\a e^{-it\a} \prod_{j=0}^n
	  \Bigg({1\over \a -e(\bp_j^*)-
	\Om_j + i\eta}\Bigg)^{m_j+1}
	\Big[ \Upsilon_\eta(\alpha- \Om_j, \bp_j^*) \Big]^{m_j}
\nonumber\\
	&&\times\int_{-\infty}^\infty \rd\ta  e^{it\ta} \prod_{j=0}^n
	  \Bigg({1\over \ta -e(\tbp_j^*)-
	\wt\Om_j - i\eta}\Bigg)^{\tm_j+1}
	\Big[ \overline{\Upsilon}_\eta(\ta-\wt\Om_j, \tbp_j^*) 
	\Big]^{\tm_j}
\label{eq:Y3}
\eey
and again, all non-star variables have disappeared.

Finally, Proposition \ref{prop:formula} follows from
(\ref{sumcpi}), (\ref{eq:cpi4}), (\ref{eq:Y2}) and
(\ref{eq:Y3}) if we remove the stars from each variable.
 $\;\;\;\Box$.

\section{Apriori bound}\label{sec:triv}
\setcounter{equation}{0}

We start with
 the following result that is
weaker than (\ref{norecestsmall}) but we need it
to estimate certain pairings later.
 
\begin{lemma}\label{lemma:trivbound}
 For any $0\leq a < 1$, $n\leq N$, $\um,\utm\in \cM(n,N)$,
$\pi\in \Pi_n$ we have
\be
	|C_{\um,\utm,\pi}(t)|\leq {(C_a\lambda^2 t)^N\over
	 \big[ M!\big]^a}
\label{cpiest}
\ee
(recall that $M:= (n+N)/2$),
hence
\be
	\limsup_{L\to\infty}\Tr_{e+ph}\Bigg( \cE^{0}_{n,N}(t) \Gamma_0 
	\Big[ \cE^{0}_{n,N}(t)\Big]^*\Bigg) 
	\leq n! \times  {(C_a\lambda^2 t)^N\over
	 \big[ M!\big]^a} \; .
\label{roughtrace}\ee
\end{lemma}

{\it Remark.} The estimate (\ref{roughtrace}) is summable
over $n$ and $N$ only for short macroscopic time $T=\lambda^2t\leq T_0$.

\medskip

{\it Proof.}  The second inequality (\ref{roughtrace})
easily follows from (\ref{cpiest}) and (\ref{sumcpi1}) since 
the number of pairs $\um,\utm\in\cM(n,N)$
 is bounded by
\be
	{N \choose n}^2 \leq C^N\;.
\label{combest} 
\ee

The bound (\ref{cpiest}) is trivial if $N=0$. Otherwise
 we split $|Y| = |Y|^{a/2}|Y|^{1-a/2}$
in (\ref{CY}) and we apply supremum bound in the first term.
Using (\ref{1Y}) and (\ref{Thetaest}), we see
that
$$
	|Y_{\um,\utm,\pi}(t; \bp_n, \tbp_n,\ubk, \utbk, \usi)|\leq
	C^N  \int_0^{t*} [\rd s_b]_{b=0}^N
	 \prod_{b\in J}{1\over \langle s_b \rangle^{d/2}}
	\int_0^{t*} [\rd \ts_\tb]_{\tb=0}^N
	 \prod_{\tb\in \tJ}{1\over \langle \ts_\tb \rangle^{d/2}}\;.
$$
An easy calculation shows that ($d\ge 3$)
\be
	\int_0^{t*} [\rd s_b]_{b=0}^N
	 \prod_{b\in J}{1\over \langle s_b \rangle^{d/2}}
	\leq C^N {t^{|J^c|-1}\over \big( |J^c|-1\big)!}
	= C^N {t^{M}\over M!}
\label{befYa}
\ee
(recall that $M:=(n+N)/2 = n+|\um|$),
hence
\be
	|Y_{\um,\utm,\pi}(t; \bp_n, \tbp_n,\ubk, \utbk, \usi)|^{a/2}\leq
	C^N\Bigg( {t^{M}\over M!} \Bigg)^a\;.
\label{Ya}
\ee
For the $|Y|^{1-a/2}$ part we use a Schwarz inequality to separate
the tilde variables. The result from (\ref{CY}), (\ref{2Y}),
(\ref{Upsilonest}) and
(\ref{Ya}) is
\bey
	|C_{\um,\utm,\pi}(t)|&\leq&
	C^N\lambda^{2N}\Bigg( {t^{M}\over M!} \Bigg)^a
	\sup_{\usi}
	\int \rd\nu^*_\pi (\bp_n, \tbp_n, \ubk, \utbk)
\label{eq:cpi6}\\
	&&\times \Bigg\{ \Big[  \int_{-\infty}^\infty
	 \rd\a\prod_{j=0}^n
	|R_j|^{m_j+1}\Big]^{2-a} + \Big[ \int_{-\infty}^\infty
	 \rd\ta  \prod_{j=0}^n
	|\tR_j|^{\tm_j+1}\Big]^{2-a} \Bigg\}
\nonumber
\eey
with $\eta:=t^{-1}$ and with a slight modification of the measure
$\rd\nu_\pi$:
\be
	 \rd\nu^*_\pi (\bp_n, \tbp_n, \ubk, \utbk):=
	\Big(\prod_{j=1}^n   M^*(\bk_j)\rd\bk_j\rd\tbk_j \Big)
	\Big(\prod_{j=1}^n \delta(\bk_j -\tbk_{\pi(j)})\Big)
	 \rd \bp_n\rd \tbp_n \delta (\bp_n-\tbp_n)
	\wh\g_e(\bp_n, \tbp_n) 
\label{def:nustar}
\ee
and we  recall the definition of $M^*$ from
(\ref{def:Mstar}).
The estimate of the two terms in the last line of (\ref{eq:cpi6})
 are identical, so we consider only the
first one. 

The case $n=0$, $m_0\ge 1$ is trivial by integrating out $\rd\a$,
and collecting $t^{m_0(2-a)} = t^{\um(2-a)}$ using
$$
	\sup_{\bp_0} \int_{-\infty}^\infty 
	{\rd\a\over |\a - e(\bp_0) +i\eta|^{m_0+1}}
	\leq (Ct)^{m_0}\;.
$$
This gives (\ref{cpiest})
together with the $t^{Ma}=t^{\um a}$ factor from (\ref{Ya}). 

{F}rom now on we assume that $n\ge 1$.
We can immediately integrate out  $\tbk_1, \tbk_2, \ldots,
\tbk_n$ and  $\tbp_n$.  
Moreover, we use the trivial estimate
%\be
%	 \Bigg|{1\over \a -\bp_j^2/2-
%	\Om_j + i\eta}\Bigg|^{m_j+1} \leq t^{m_j}
%	 \Bigg|{1\over \a -\bp_j^2/2-
%	\Om_j + i\eta}\Bigg|
%
\be
	|R_j|^{m_j+1} \leq t^{m_j} |R_j|
\label{ttriv}\ee
for all $j$. So we  obtain
\be
	|C_{\um ,\utm,\pi}(t)|\leq
	C^N\lambda^{2N}\Bigg({t^{M}\over M!} \Bigg)^a t^{(2-a)|\um|}
	\sup_{\usi}
	\int \rd\mu(\ubk, \bp_n) \Bigg[ \int_{-\infty}^\infty
	 \rd\a \prod_{j=0}^n
	|R_j|\Bigg]^{2-a}\;,
%	 \Bigg[ \int \rd\a   \prod_{j=0}^n
%	  {1\over \Big|\a -\bp_j^2/2-
%	\Om_j + i\eta\Big|}\Bigg]^{2-a}
\label{schw}\ee
where
\begin{equation}
	 \rd\mu (\ubk,\bp_n) : = \wh\gamma_e(\bp_n, \bp_n) \Big( 
	\prod_{j=1}^n M^*( \bk_j) \rd \bk_j \Big)  \rd \bp_n
\label{mudef}
\end{equation}
is a positive measure.
The rest of the proof is  similar to the proof of (3.16)
in \cite{EY2}.

Fix all momentum variables, $\ubp, \ubk$'s,  for a moment and let
\begin{equation}
	G(\alpha) 
	: = \prod_{j=0}^{n-2} |R_j| \; .
%	{1\over \Big|\alpha - \bp_j^2/2
%	-\Om_j+ i\eta\Big|} \;.
\label{Gdef}
\end{equation}
For any given  $\bp_n, \bk_n, \sigma_n$ we denote by $\varrho$
the probability measure on $\bR$
$$
	\varrho (\rd\alpha) = \varrho (\rd\alpha; \bp_n, \bk_n)
	 : = {\rd\alpha\over Z( \bp_n, \bk_n)
	\Big| \alpha - A + i\eta\Big|\Big| \alpha - B + i\eta\Big|}
$$
with $A:= e(\bp_n)$, $B: = e(\bp_{n-1})+\s_n\om(\bk_n)=
e(\bp_{n} + \bk_n) + \sigma_n \omega(\bk_n)$ 
and
\begin{equation}
	Z=Z( \bp_n, \bk_n): = \int_{-\infty}^\infty
	{\rd\alpha\over | \alpha - A + i\eta|| \alpha - B + i\eta|}
\label{Zdef}
\end{equation}
(we neglect $\sigma_n$ dependence in the notation).
We have a simple estimate for $Z$:
\be
	Z\leq
	 {C\over |A-B + i\eta|}\Bigg[ 1 + \log_+ \Big|{A-B\over \eta}\Big|
	\Bigg]\;.
\label{Zest}\ee
Rewrite the $\rd\alpha$ integration in (\ref{schw}) as
$$
	 \int_{-\infty}^\infty \rd\alpha  \;
	\prod_{j=0}^n |R_j|
%	 \Bigg|{1\over  \alpha - \bp_j^2/2-\Om_j+ i\eta }\Bigg|
	= Z\int_{-\infty}^\infty \varrho(\rd\alpha) G(\alpha)
$$
and use H\"older (recall that $2-a> 1$)
\begin{equation}
	 \Big(  \int_{-\infty}^\infty \rd\alpha  \;
	\prod_{j=0}^n |R_j|
	 \Big)^{2-a}
%	 =\Bigg( Z
%	\int_{-\infty}^\infty \varrho(\rd\alpha) 
%	G(\alpha) \Bigg)^{2-a}
	 \leq Z^{2-a} \int_{-\infty}^\infty
	  \varrho(\rd\alpha) G(\alpha)^{2-a}\;.
\label{holder}
\end{equation}

We then  obtain
\be
	|C_{\um,\utm,\pi}(t)|\leq
	(C\lambda^2)^{N}\Bigg({t^{M}\over M!} \Bigg)^a t^{(2-a)|\um|}
	\sup_{\usi}
 	\int \rd\bp_{n}\rd\bk_n \big[ Z( \bp_n, \bk_n)\big]^{2-a}
	\wh\gamma_e (\bp_n, \bp_n)M^*(\bk_n) 
\label{succ}\ee
$$
	\times  \int_{-\infty}^\infty\varrho(\rd\a; \bp_n, \bk_n)
	\int {M^*(\bk_{n-1})\rd\bk_{n-1}
	\over \Big| \alpha - e(\bp_{n-2}) + i\eta
	-\sum_{\ell=n-1}^n \s_\ell\omega (\bk_\ell)\Big|^{2-a}}
$$
$$
	\times\int {M^*(\bk_{n-2})\rd\bk_{n-2}
	\over \Big| \alpha - e(\bp_{n-3}) + i\eta
	-\sum_{\ell=n-2}^n \s_\ell\omega (\bk_\ell)\Big|^{2-a}}
	\times\ldots\times \int {M^*(\bk_{1})\rd\bk_{1}
	\over \Big| \alpha - e(\bp_{0}) + i\eta
	- \sum_{\ell=1}^n \s_\ell\omega (\bk_\ell)\Big|^{2-a}}\;.
$$
We have
\begin{equation}
	\sup_\alpha\sup_{\usi} \sup_{\bp_n, \bk_{j+1}, \bk_{j+2}, \ldots,
	\bk_n}
	\int {M^*(\bk_{j})\rd\bk_{j}
	\over \Big| \alpha - e(\bp_{j-1}) + i\eta
	-\sum_{\ell=j}^n \s_\ell\omega (\bk_\ell)\Big|^{2-a}}
	\leq C_a t^{1-a}
\label{succest}
\ee
with $\eta: = t^{-1}$. Recalling $\bp_{j-1} = \bp_j + \bk_j$,  
the prototype of this inequality for $b>1$ is
\be
	\sup_{\bp,  \theta}\int { M^*(\bk) \rd\bk
	 \over | \theta - \Phi_\pm(\bp, \bk) +i\eta|^{b}}
	 \leq C_b t^{b-1}\; .
\label{eq:prototype}
\ee

There will be several similar inequalities in the rest of the paper.
They follow from the assumptions in Section \ref{sec:assump} by elementary
arguments. Their proofs use the same idea and the details
 will be omitted. Here we only sketch the proof of (\ref{eq:prototype}).

\medskip

{\it Proof of (\ref{eq:prototype}).} We consider a
tiling of $\bR^d$ by identical cubes $\{ Q_j \}$
 of size $\wt\varrho$. We recall the quantity $\wt\varrho$
from (\ref{layer}).
On each fixed cube $Q=Q_j$ we resolve the singularity on an exponential scale.
We define the sets
\be
	S_\ell = S_\ell(\bp): = Q\cap \Big\{ \bk\; : \; 2^{-\ell-1}
	\wt\varrho < 
	|\Phi_\pm(\bp, \bk)-\theta|\leq 2^{-\ell}\wt\varrho\Big\}
\label{def:Sell}\ee
for $1\leq \ell <\ell_0:= [\log^* t]$, $\ell\in \bN$.
We let
$$
	S_{\ell_0}: =  Q\cap \Big\{ \bk\; : \; 
	|\Phi_\pm(\bp, \bk)-\theta|\leq 2^{-\ell_0}\wt\varrho\Big\}
$$
and $S_0:= Q\setminus \bigcup_\ell S_\ell$.
The integrand on $S_\ell$, $\ell =0, 1,\ldots, \ell_0$
 is bounded by $2^{b\ell} C \langle\mbox{dist}(Q,0)\rangle^{-4d-24}$
using the decay of $M^*(\bk)$. The volume of $S_\ell$ is 
of order $2^{-\ell}$ by (\ref{layer}). We can sum up these estimates for $\ell$
to obtain $C_Qt^{b-1}$. Notice that for $b=0$ one obtains
$C_Q\log^*t$. Finally, the strong decay of the coefficients 
$C_Q\sim \langle\mbox{dist}(Q,0)\rangle^{-4d-24}$ 
allows to sum up the contributions
from each cube. 
Notice that only the following property of $M^*$ was used in this proof
\be
	\sum_j \| M^* \|_{L^\infty(Q_j)} <\infty \; . \qquad \Box
\label{onlyMstar}
\ee

%This inequality can be checked by  using the same partition of unity as
%in the proof of (\ref{statph}). After a suitable change of
%variables the denominator in (\ref{eq:prototype}) is transformed
%into either a linear or a quadratic function. For
%such functions the proof is an explicit calculation.

\bigskip

We continue the estimate of (\ref{succ}).
Using (\ref{succest}), we integrate out $\bk_1, \bk_2, \ldots, \bk_{n-1}$
in this order, then we integrate out $\rd\a$ using that $\varrho(\rd\a)$
is a probability measure:
\be
	|C_{\um,\utm,\pi}(t)|\leq
	C_a^N\lambda^{2N}\Bigg({t^{M}\over M!} \Bigg)^a t^{(2-a)|\um| 
	+(1-a)(n-1)} 
 	\int \rd\bp_{n}\rd\bk_n \big[ Z( \bp_n, \bk_n)\big]^{2-a}
	\wh\gamma_e (\bp_n, \bp_n)M^*(\bk_n) \;.
\label{trivfin}
\ee
We then use the estimate (\ref{Zest}) and integrate out $\rd\bk_n$
to collect one more factor $ C_a t^{1-a}$ using (\ref{eq:prototype}). 
At the end, to do the
$\rd\bp_n$ integration, we use that $\int\wh\gamma_e(\bp, \bp)\rd\bp =
\mbox{Tr}\; \gamma_e =1$. The total power of $t$ is 
$Ma + (2-a)|\um| + (1-a)n = n+2|\um|=N$ using $M=n+|\um|$.
This completes the proof of Lemma 
\ref{lemma:trivbound}.
$\,\,\,\Box$

\section{Fully expanded terms with small $n$ and no recollision}
\label{sec:norecestsmall}
\setcounter{equation}{0}

Here we prove (\ref{norecestsmall}). 
We use the representation (\ref{sumcpi}).
We can assume that $n\ge 4$, otherwise the apriori bound
(\ref{roughtrace}) applies.

\begin{definition}\label{def:strongcross}
A pairing  $\pi\in \Pi_n$ is called {\bf crossing} if $\pi\neq \mbox{id}$.
In particular, there exists
$a< b$,   with $\pi(b) < \pi(a)$.
 In this case the pair of indices $(a, b)$
is called {\bf crossing pair}.
\end{definition}

{\it Remark.} This definition differs from the one given
in \cite{EY2} (Definition 2.5). Here only pairing lines
between tilde and non-tilde variables can form a crossing pair.

\bigskip

The only non-crossing pairing is the so-called {\it direct} pairing, $\pi=id$,
 it is  estimated by (\ref{cpiest}) and it gives  the first term
on the right hand side of (\ref{norecestsmall}).

\begin{lemma}\label{lemma:onecross} Let $n\leq N$, $\um,\utm\in \cM(n,N)$
and let $\pi\in \Pi_n$ be a crossing pairing. Then
\be
	\limsup_{L\to\infty}
	|C_{\um,\utm,\pi}(t)|\leq  t^{-1/2}(C\lambda^2 t)^N (\log^* t)^4.
\label{1crossest}
\ee
\end{lemma}

{F}rom this lemma and the combinatorial bound  (\ref{combest})
 the estimate (\ref{norecestsmall}) follows
immediately.

\medskip

{\it Proof of Lemma \ref{lemma:onecross}}.
If $(a, b)$ is  a crossing pair, then for
fixed $b$ there could be many different $a$'s which
form a crossing pair with $b$. Consider the smallest
one, i.e.,  $\pi(b) < \pi(a)$, but
for all $c<a$, $\pi(c)< \pi(b)$. This property will be
called the {\it minimality} of the crossing pair.

{F}rom  (\ref{CY}), (\ref{2Y}), (\ref{Upsilonest}) and the bound
(\ref{ttriv}) we have
\be
	|C_{\um, \utm,\pi}(t)| \leq (C\lambda)^{2N}t^{2|\um|}
	\sup_{\usi} \int \rd\nu_\pi^*(\bp_n, \tbp_n, \ubk, \utbk)
	\;\int_{-\infty}^\infty \rd\a  \prod_{j=0}^n  |R_j|
	\int_{-\infty}^\infty \rd \ta \prod_{j=0}^n |\tR_j|\;.
\label{eq:cpinew}
\ee 
Here we considered $\bp_j, \tbp_j$ as functions
of $\bp_n=\tbp_n$ and $\ubk$ or $\utbk$, respectively,
according to (\ref{def:pj}) and  recall the definition
of $R_j, \tR_j$ from (\ref{def:R}).

We need the following estimate to take care of the $\a, \ta$ 
integrations. For any fixed index $\ell$
\be
	\langle \bp_\ell \rangle \leq C^n\Bigg( 
	\prod_{j=1}^n \langle \bk_j\rangle\Bigg)
	\langle \bp_n\rangle
\label{telescope}
\ee
and the same estimate is true for $\langle \tbp_\ell \rangle$.
This easily follows from $\langle a + b \rangle \leq C
\langle a  \rangle\langle  b \rangle$.

We also define
\be
	L(\bk) : = \langle \bk\rangle^{12} M^*(\bk)
\label{def:L}
\ee
and we use that 
\be
	L(\bk) \leq C \langle \bk \rangle^{-4d-12}
\label{Ldec}
\ee
by (\ref{Mstardec}).

Let $A:= \{ a-1, a,  n\}$ and $\tA: =\{ \pi(b)-1, \pi(b),  n\}$.
Notice that these are sets of three distinct elements.
 We then perform a  Schwarz estimate in (\ref{eq:cpinew}) to obtain
\bey
	|C_{\um, \utm,\pi}(t)| &\leq& (C\lambda)^{2N}t^{2|\um|}
	\sup_{\usi}
	\int\rd\nu_\pi^*(\bp_n, \tbp_n, \ubk, \utbk) 
	\int_{-\infty}^\infty \rd\a \rd\ta 
\label{fourschw}\\
	&&\times 
	\Bigg[  t^{-1/2}\prod_{j\not \in \tA} |\tR_j|^2
	+  t^{1/2}\prod_{j\not\in A} |R_j|^2\Bigg] 
	 \prod_{j \in A} |R_j| \prod_{j\in \wt A} |\tR_j|
	 =: (I.) + (II.) \; , 
\nonumber
\eey
%$$
%	\times \Bigg[  t^{-1/2}\prod_{j\not \in \tA}
%	 {1\over \Big|\ta -  \tbp_j^2/2
%	+ i\eta - \Om_j (\tbk, \usi_\pi)\Big|^2} 
%	+ t^{1/2}\prod_{j\not\in A}
%	  {1\over \Big|\a -  \bp_j^2/2
%	+ i\eta - \Om_j (\bk, \usi)\Big|^2}\Bigg]
%$$
%$$
%	\times \prod_{j = a-1, a}{1\over \Big|\a -  \bp_j^2/2
%	+ i\eta  - \Om_j (\bk, \usi)\Big|} \prod_{j=\pi(b)-1, \pi(b)}
% 	{1\over \Big|\ta -  \tbp_j^2/2
%	+ i\eta -\Om_j (\tbk, \usi_\pi) \Big|} =: (I.) + (II.)
%\label{fourschw}
%\ee
where the decomposition $ (I.) + (II.)$ is according
to the summation in the big square bracket.

\bigskip

To estimate the term $(I.)$,
we use $\tbp_n$ and $\tbk_j$, $j=1, \ldots, n$, as
variables and we simply estimate the  terms $R_a, R_{a-1}$
by $Ct$ each.
Moreover, we can gain an extra factor $\langle \a \rangle$
at the expense of 
$\langle \bp_n\rangle^2 \prod_{j=1}^n \langle \bk_j\rangle^4$.
Using (\ref{telescope}) and (\ref{ombound}):
\be
	|R_a|
	\leq {Ct \over \langle \a - e(\bp_a) - \Om_a(\bk, \usi)\rangle}
	\leq Ct { \langle \bp_a\rangle^2 \langle  \Om_a(\bk, \usi) \rangle
	\over  \langle \a \rangle}
	\leq  	  C^n t{ \langle \bp_n\rangle ^2
	\over  \langle \a \rangle}
	\prod_{j=1}^n \langle \bk_j\rangle^4\; .
\label{prot:tel}
\ee
We also need the estimate
\be
	\prod_{j=1}^n \langle \tbk_j\rangle^{-2}
	\leq C^n
	{\langle \tbp_n \rangle^2 \over  \langle \tbp_{\pi(b)} \rangle^2}
\label{insert}
\ee
to insert a decay in $\tbp_{\pi(b)}$ and we also insert an
explicit extra $\prod_j \langle \tbk_j \rangle^{-2}$ decay. 
 Finally we estimate $M^*(\bk)$ by
$L(\bk) \langle \bk \rangle^{-8}$, which makes up
for the $\prod_j \langle \bk_j\rangle^8$ factors
used in these estimates.

Then we can integrate
out all $\bk_j$ and $\bp_n$ freely, set $\bp_n=\tbp_n$ in $R_n$ and we are left with
\bey
	(I.) &\leq&  t^{-1/2} t^2(C\lambda)^{2N}t^{2|\um|}
	\sup_{\usi} 
	 \int\Big(\prod_{j=1}^nL(\tbk_j) \rd\tbk_j \Big) 
	\int_{-\infty}^\infty
	 {\rd\a \rd\ta\over \langle \a \rangle}
	\int \rd \tbp_n \langle \tbp_n\rangle^4
	\wh\g_e(\tbp_n, \tbp_n)
\nonumber\\
	&&\times |R_n| \; |\tR_n| \; |\tR_{\pi(b)-1}|
	\; |\tR_{\pi(b)}|
	\Bigg( \prod_{j\not \in \tA} |\wt R_j|^2 \Bigg)
	 {1\over  \langle \tbp_{\pi(b)} \rangle^2}
	\prod_{j=1}^n {1\over \langle\tbk_j\rangle^2} \;.
\label{eq:I}
\eey

Since $\tR_j$ depends on $\tbk_{j+1}, \tbk_{j+2},\ldots, \tbk_n$ and
$\tbp_n$, the variables
 $\tbk_1, \tbk_2 ,\ldots, \tbk_{n-1}, \tbk_n$ can be integrated out 
successively in this order. 
The integrals of $\tbk_j$ with $j\neq \pi(b), \pi(b)+1$
successively eliminate the factors $\tR_{j-1}$ that depend on
 $\tbp_{j-1} = \tbp_j+ \tbk_j$, and
they each give $Ct$ by 
\be
	\sup_{\bp, \theta}\int { L(\bk)
	\rd \bk\over \Big| \theta - \Phi_\pm(\bp, \bk)
	+ i\eta \Big|^2} \leq Ct
\label{prot:Ct}
\ee
(see (\ref{eq:prototype}))
with $\bk=\tbk_j$, $\bp=\tbp_j$ and $\theta =\ta-\wt\Om_j$,
where we recall (\ref{Omdefabuse}) that
$$
	\wt\Om_j 
%	:=\Om_j (\tbk, \usi\circ\pi^{-1})
	 = \sum_{m=j+1}^n \s_{\pi^{-1}(m)} \om(\tbk_m)
%	= \sum_{\ell: \pi(\ell)\ge j+1} \s_\ell \om(\bk_\ell)
$$
and that $\tbp_j= \tbp_n + \sum_{m=j+1}^n \tbk_m$.
We recall that the function $L(\bk)$ also satisfies
(\ref{onlyMstar}).

The  integrals of $\tbk_{\pi(b)}$ and $\tbk_{\pi(b)+1}$
eliminate the factors  $\tR_{\pi(b)-1}$ and $\tR_{\pi(b)}$
%(denominators with $\tbp_{\pi(b)-1}$ and $\tbp_{\pi(b)}$)
and they each give $C\log^*t$  by the estimate
\be
	\sup_{\bp, \theta}\int { L(\bk)
	\rd \bk\over \Big| \theta - 
	\Phi_\pm(\bp, \bk)
	+ i\eta \Big|} \leq C\log^*t \; .
\label{prot:Clog}
\ee
 Moreover, 
from the $\tbk_{\pi(b)}$ integral we gain an
additional $\langle \ta \rangle$:
$$
	\int {L(\tbk_{\pi(b)})\rd \tbk_{\pi(b)}
	\over \Big| \ta -  e(\tbp_{\pi(b)} + \tbk_{\pi(b)})
	+ i\eta \pm \om (\tbk_{\pi(b)})-
	\wt\Om_{\pi(b)}\Big|
	\;  \langle \tbp_{\pi(b)} \rangle^2}
	\leq {C^n(\log^* t)  \prod_j \langle \bk_j\rangle^2
	\over \langle \ta \rangle}\;.
$$
The prototype of this inequality is
\be
	\sup_{\bp}\int {L(\bk)\rd \bk
	\over \Big| \theta - \Phi_\pm(\bp , \bk)
	+ i\eta  \Big|
	\;  \langle \bp \rangle^2}
	\leq {C\log^* t \over \langle\theta\rangle}
\label{prot:Clogalpha}
\ee
with $\theta =\ta-\wt\Om_{\pi(b)}$.
The proofs of (\ref{prot:Clog}) and (\ref{prot:Clogalpha})
follow the same route as (\ref{eq:prototype}) and we omit them.
To gain  $\langle \ta \rangle$, we  use (\ref{ombound})
\be
	{1\over \langle \ta- \wt\Om \rangle}	
	\leq {C\langle \wt\Om \rangle \over \langle \ta \rangle}	
	\leq {C^n \prod_j \langle \bk_j\rangle^2 \over \langle \ta \rangle}\;.
\label{cOm}
\ee
We insert these estimates into (\ref{eq:I});
 we have collected $(Ct)^{n-2}( C\log^*t)^2$
so far and we also gained $\langle \ta\rangle$.

Finally we integrate out $\a, \ta$ and $\tbp_n$:
we obtain
\bey
	\lefteqn{\int_{-\infty}^\infty
	 {\rd \a \rd \ta \over \langle \a \rangle \langle \ta \rangle}
	\int { \wh\g_e(\tbp_n,\tbp_n)\langle \tbp_n \rangle^4\rd\tbp_n\over
	| \a - e(\tbp_n) + i\eta|\; | \ta - e(\tbp_n) + i\eta|}}\qquad\qquad
	\qquad\qquad\qquad
\label{int:alpha3}\\
	&\leq& (C\log^*t)^2
	 \int \wh\g_e(\tbp_n,\tbp_n)\langle \tbp_n \rangle^4\rd\tbp_n
	\leq (C\log^*t)^2  
\nonumber
\eey
using (\ref{gammadecay}).
Altogether the first term in (\ref{fourschw}) gives
$$
	(I.)\leq t^{-1/2}(C\lambda^2 t)^N (\log^* t)^4.
$$
\bigskip

We now estimate the second term  $(II.)$ in (\ref{fourschw}) after integrating
out $\tbk_j$'s and $\tbp_n$
\bey
	(II.) &=&  t^{1/2} (C\lambda^2)^N t^{2|\um|}
	\sup_{\usi}
	\int\Big(\prod_{j=1}^n 
	  M^*(\bk_j) \rd\bk_j \Big) \rd \bp_n \wh\g_e(\bp_n, \bp_n)
\label{eq:II}\\
	&&\times \int_{-\infty}^\infty \rd\a \rd\ta
	 \prod_{j\not\in A}  |R_j|^2  \prod_{j\in A}  |R_j| 
	 \prod_{j\in \wt A}  |\tR_j| \;.
\nonumber\eey
Recall that $R_j$ depends on $\bp_j$'s that
 are functions of the variables $\bk_j$ and $\bp_n$.
We express everything in terms of $\bk_j, \bp_n$ variables
by (\ref{def:pj}). Similarly,
$$
	\tbp_j = \bp_n + \sum_{m\; : \; \pi(m) >j} \bk_{m}
$$
  by the pairing $\bk_j = \tbk_{\pi(j)}$.
In particular we have
\bey
	\bp_{a-1} =  \bp_a + \bk_a = \bv + \bk_a + \bk_b \qquad
	&\mbox{with}&
	\quad	\bv := \bp_n + \sum_{j\ge a+1\atop j\neq b} \bk_j\;,
\nonumber\\
	\tbp_{\pi(b)} = \bu + \bk_a \qquad
	&\mbox{with}&
	\quad
	\bu : = \bp_n + \sum_{j \; : \; \pi(j)>\pi(b)\atop j\neq a} \bk_j\;,
\nonumber\\
	\wt\Om_{\pi(b)} = \wt\Om^* +\sigma_a \om(\bk_a)
	\qquad &\mbox{with}&\quad
	\wt\Om^*  : = \sum_{j \; : \; \pi(j)>\pi(b)\atop j\neq a}\sigma_j
	\om( \bk_j)\;.
\nonumber
\eey
Notice that $\bv$, $\bu$ and $\wt\Om^*$ are independent 
of $\bk_1, \ldots, \bk_a$ and $\bk_b$. This is clear for $\bv$,
and it follows  for $\bu$ and
 $\wt\Om^*$ from the minimality of $(a,b)$.

\bigskip\bigskip
\centerline{\epsffile{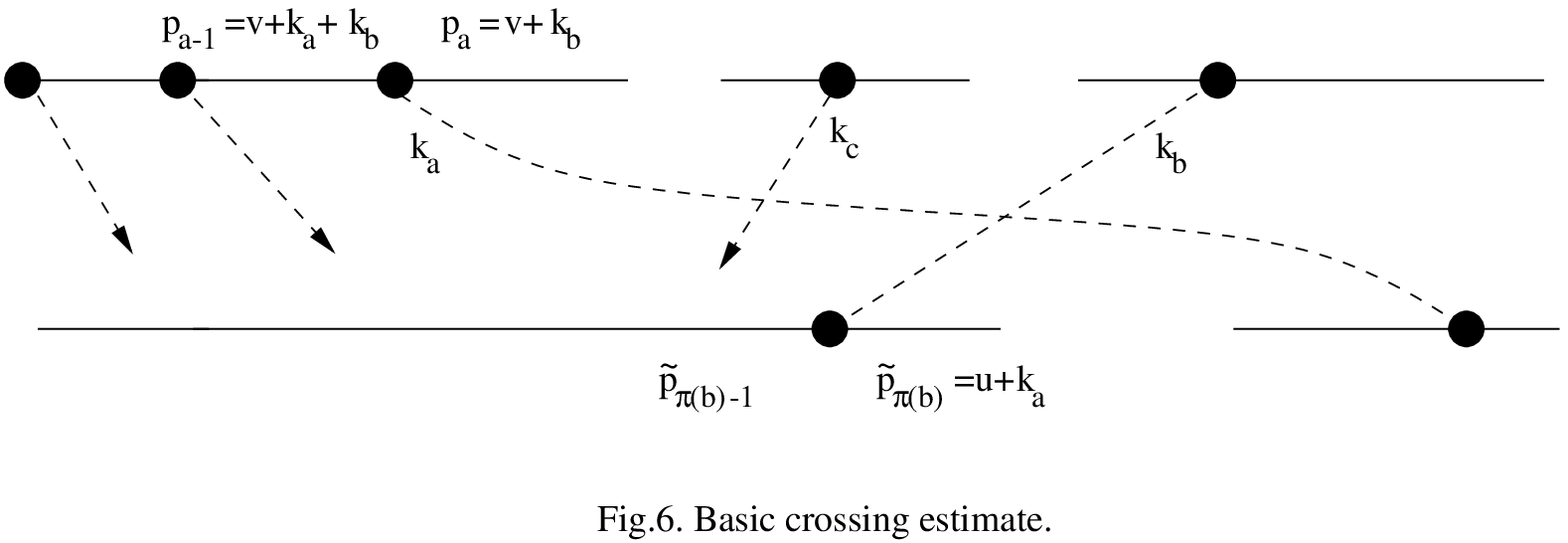}}
\bigskip

Now we start estimating (\ref{eq:II}).
First we estimate $\tR_{\pi(b)-1}$ by
$Ct$ and we also gain a factor $\langle \ta\rangle^{-1}$ using 
\be
	|\tR_{\pi(b)-1}|={1\over \Big|\ta -  e(\tbp_{\pi(b)-1})+i\eta
	-\wt\Om_{\pi(b)-1}\Big|}\leq C^nt {\langle \bp_n\rangle^2\over
	\langle \ta\rangle} \prod_{j=1}^n \langle \bk_j\rangle^{4}
\label{tagain}
\ee
similarly to
(\ref{prot:tel}). We also need
$$
	\prod_{j=1}^n \langle \bk_j\rangle^{-2}\leq
	C^n { \langle \bp_n\rangle^2\over \langle \bp_a\rangle^2}
$$
analogously to (\ref{insert}). Hence we have
\bey
	(II.) &\leq&   t^{3/2} (C\lambda^2)^N t^{2|\um|}
	\sup_{\usi}\int\Big(\prod_{j=1}^n 
	  L(\bk_j) \rd\bk_j \Big)\int_{-\infty}^\infty
	 {\rd\a \rd\ta\over \langle \ta\rangle}
\label{IIestcont}\\
	&&\times 
	\int \rd \bp_n \langle \bp_n\rangle^4\wh\g_e(\bp_n, \bp_n)
	|\tR_{\pi(b)}| \; |\tR_n|
	\Bigg( \prod_{j\not\in A}  |R_j|^2  \prod_{j\in A}  |R_j|\Bigg)
	{1\over  \langle \bp_a\rangle^2} \prod_{j=1}^n {1\over \langle\bk_j
	\rangle^2} \; .
\nonumber
\eey

We then integrate out $\bk_1, \ldots, \bk_{a-1}$ using (\ref{prot:Ct})
and collecting $(Ct)^{a-1}$.
This eliminates $R_0, R_1, \ldots, R_{a-2}$. Notice
that $\wt R_{\pi(b)}$
is independent of these variables.

Next we integrate out $\bk_a$. This occurs only in  $R_{a-1}$ and
$\tR_{\pi(b)}$
%denominators with $\bp_{a-1}$ and $\tbp_{\pi(b)}$
\be
	\int  |R_{a-1}| \; |\tR_{\pi(b)}| \langle \bp_a\rangle^{-2}
	L(\bk_a)\rd\bk_a \leq  {(C
	\log^*t)^2 \over |\bp_a - \bu|\langle \a -\Om_a\rangle }\;.
\label{eq:2den}\ee
Here we used the inequality
\be
	\sup_{\ttheta}\int {L(\bk)\rd\bk\over \Big|\theta -  \Phi_\pm(\bp, \bk)
	+ i\eta \Big| \;
	 \Big|\ttheta - \Phi_\pm(\bu , \bk)
	+ i\eta  \Big|\langle \bp\rangle^2 }
	\leq {(C\log^*t)^2 \over |\bp - \bu| \langle \theta \rangle }
\label{prot:twoden}
\ee
with $\bk= \bk_a$, $\bp= \bp_a$, $\theta= \a - \Om_a$, $\ttheta
 = \ta - \wt\Om^*$.
\medskip

{\it Proof of (\ref{prot:twoden}).} We use the transversality condition
(\ref{trans}) and a similar resolution
of singularities as in the proof of (\ref{eq:prototype}).
Again, we consider a tiling by cubes of size $\sim \wt\varrho$.
 We fix a cube $Q$
 and we recall the definition of the sets $S_\ell(\bp)$
(\ref{def:Sell}). 
On the set $S_\ell(\bp) \cap \wt S_{\wt \ell}(\bu)$ the integrand
is bounded by  $2^{\ell + \wt\ell} C\langle\mbox{dist}(Q,0)\rangle^{-4d-12}
\langle \bp\rangle^{-2}$, 
the volume is bounded
by $2^{-(\ell + \wt\ell)} C|\bp-\bu|^{-1}$.
To ensure the decay in $\theta$ for $\bk\in S_\ell(\bp)$, $\ell\ge1$,
we use $\langle \bp\rangle^{-2}\leq  
\langle\mbox{dist}(Q,0)\rangle^2 \langle\theta\rangle^{-1}$
by (\ref{eq:econd}) and (\ref{ombound}). On $S_0(\bp)$ we have
$$
	{1\over  \Big|\theta -  \Phi_\pm(\bp, \bk)
	+ i\eta \Big| \langle \bp\rangle^2} \leq
	{ \langle\mbox{dist}(Q,0)\rangle^2 \over \langle\theta\rangle}\; .
$$
 After summing
these bounds for $\ell, \wt\ell = 0, \ldots, [\log^* t]$, then
summing up the result for all $Q$, we obtain
(\ref{prot:twoden}). $\;\;\; \Box$

\bigskip 

Continuing (\ref{eq:2den}),
from $\langle \theta\rangle^{-1}=\langle \a-\Om_a\rangle^{-1}$
we also gain a factor $\langle \a \rangle^{-1}$
similarly to (\ref{cOm}) using the 
 decaying factor $\prod_j \langle \bk_j \rangle^{-2}$
in (\ref{IIestcont}).
Hence we have
\bey
	(II.) &\leq&   t^{3/2} (Ct)^{a-1}(C\log^*t)^2 (C\lambda^2)^N
	t^{2|\um|}
	\sup_{\usi}
	\int\Big(\prod_{j=a+1}^n 
	  L(\bk_j) \rd\bk_j \Big)
\label{eq:IIest}
\\
	&&\times \int_{-\infty}^\infty
	 {\rd\a \rd\ta\over \langle \a\rangle \langle \ta\rangle}
	\int \rd \bp_n \langle \bp_n\rangle^4\wh\g_e(\bp_n, \bp_n) \;
	|R_a|\; |R_n| \; |\tR_n| \;\Bigg( \prod_{j=a+1}^{n-1} |R_j|^2\Bigg)
	 {1\over |\bp_a - \bu|} \;.
\nonumber
\eey
% 	{1\over \Big|\a -  \bp_n^2/2+ i\eta \Big|}
%	 {1\over \Big|\ta -  \bp_n^2/2	+ i\eta \Big|}
%	\times \prod_{j=a+1}^{n-1}  {1\over \Big|\a -  \bp_j^2/2
%	+ i\eta - \Om_j\Big|^2}
%	 \times {1\over \Big|\a -  \bp_a^2/2
%	+ i\eta  - \Om_a\Big|}{1\over |\bp_a - \bu|}
%$$

Let $c\ge a+1$ be the smallest index such that $\pi(c)\leq \pi(b)$
(see Fig. 6).
Such index exists, if not else then $c=b$. By the definition
of $\bv$, $\bu$, we have
$$
	\bp_a - \bu = \bv + \bk_b - \bu =: \bk_c + \bw
$$
where $\bw$ depends only on $\bk_{c+1}, \ldots, \bk_n, \bp_n$
by the minimality of $c$. Simply $\bk_c$ is the momentum with the
smallest
index which appears in the difference
$$
	\bv -\bu =  \sum_{j\ge a+1\atop j\neq b} \bk_j
	-\sum_{j \; : \; \pi(j)>\pi(b)\atop j\neq a} \bk_j,
$$
i.e., the first momentum which appears in $\bv$ but not in $\bu$;
or if the smallest such index is bigger than $b$, then $c:=b$.

Now we can integrate out $\bk_{a+1}$ in (\ref{eq:IIest})
to eliminate $|R_a|$.
This gives only a $C\log^*t$ factor by (\ref{prot:Clog})
if $c> a+1$, and  by
$$
	\int {L(\bk_{a+1})\rd\bk_{a+1}\over
	\Big|\a - e(\bp_{a+1} + \bk_{a+1}) + i\eta
	\pm \om(\bk_{a+1}) - \Om_{a+1}\Big|
	\; | \bk_{a+1} + \bw|}\leq C\log^*t
$$
if $c=a+1$. The prototype of this inequality is
\be
	\sup_{\bp,\bw, \theta}\int {L(\bk)\rd\bk\over
	\Big|\theta - \Phi_\pm(\bp, \bk) + i\eta \Big|
	\; | \bk + \bw|}\leq C\log^*t \; 
\label{prot:plus}
\ee
and it is proven similarly as (\ref{eq:prototype}).
Here the point singularity around $\bw$ also has to be
resolved on an exponential scale; we find that
within each cube $Q$ the integrand is bounded
by $2^{\ell + \wt\ell}$ except for a set of
measure $\sim 2^{-\ell - (d-1)\wt\ell}$ using (\ref{layer}).
The details are omitted.

\medskip

If $c=a+1$, then all the remaining factors $|R_j|^2$, $j=a+1, a+2, \ldots, 
n-1$,
 can be integrated out successively. The order of the integration
 is $\rd\bk_{a+2},\ldots, \rd\bk_n$. This gives
$(Ct)^{n-a-1}$, which makes altogether a $t$-power $N-1/2$,
taking the $ t^{3/2} (Ct)^{a-1} t^{2|\um|}$ prefactor into account
and recalling that $n+2|\um|=N$.

If $c>a+1$, then we integrate out $\bk_{a+2} \ldots, \bk_{c-1}$
in this order
(it is void if $c=a+2$).
These integrations eliminate $|R_j|^2$, $j=a+1, \ldots, c-2$,
they do not affect the denominator $|\bp_a - \bu|= |\bk_c+\bw|$, and
they give $(Ct)^{c-a-2}$ by (\ref{prot:Ct}).
Now we perform the $\bk_c$ integral and eliminate $R_{c-1}$:
$$
	\int {L(\bk_c)\rd  \bk_c\over
	\Big| \a - e(\bp_c +\bk_c) +i\eta
	\pm\om(\bk_c) - \Om_c\Big|^2\; | \bk_c + \bw|}
	\leq Ct
$$
uniformly in $\bw$ by
\be
	\sup_{\bp, \bw,\theta}\int {L(\bk)\rd  \bk\over
	\Big| \theta - \Phi_\pm(\bp,\bk) +i\eta\Big|^2\; | \bk + \bw|}
	\leq Ct \; .
\label{prot:Ctdenom}
\ee
This inequality is obtained exactly as (\ref{prot:plus}).

Then we do the $\bk_{c+1},\ldots, \bk_n$ integrations
using (\ref{prot:Ct}), eliminating all $|R_j|^2$ factors in (\ref{eq:IIest}).
This also collects
$$
	t^{3/2} (Ct)^{a-1}(C\log^* t)^3 (Ct)^{c-a-2} Ct (Ct)^{n-c}t^{2|\um|} = 
	(Ct)^{N-1/2}(\log^* t)^3\;.
$$

Finally, we finish the estimate of (\ref{eq:IIest})
 with the $\a, \ta, \bp_n$  integrations
as in (\ref{int:alpha3}). This completes the proof of
Lemma \ref{lemma:onecross}. $\;\;\;\Box$

\bigskip

\section{Fully expanded terms with big $n$ and no recollision}
\label{sec:norecbign}
\setcounter{equation}{0}

Here we prove (\ref{norecest}). The basic idea is that we consider
a few disjoint crossing pairs (see Definition \ref{def:strongcross}),
which are well ordered (to make successive integration possible)
and we plan to gain $t^{-1}$ from each.

The key is the right definition of "well-ordered" 
crossing pairs.   The main difficulty lies in the
fact that it is not true that "more crossings" give
better bound. In particular, the contribution of the fully crossing
pairing, $\pi(j)=(n+1)-j$, is of order $t^{-d+1}$
 independently on the number of crossing pairs.

So we need a  more refined definition, which actually
excludes the full crossing, and lets crossing pairs alternate with
noncrossing ones.

\begin{definition}\label{def:peak} Fix a pairing $\pi\in \Pi_n$.
 A number $a$ is called {\bf peak}   
if $\pi(a-1)< \pi(a)> \pi(a+1)$  (Fig. 7) and it is called
{\bf valley} if $\pi(a-1)> \pi(a)< \pi(a+1)$.
 The endpoints $a=1$ and $a=n$ are not considered peak
or a valley. The valleys and peaks clearly alternate.
We define a complete ordering on the set of peaks according
to the value of their height $\pi(a)$.
\end{definition}

\bigskip\bigskip
\centerline{\epsffile{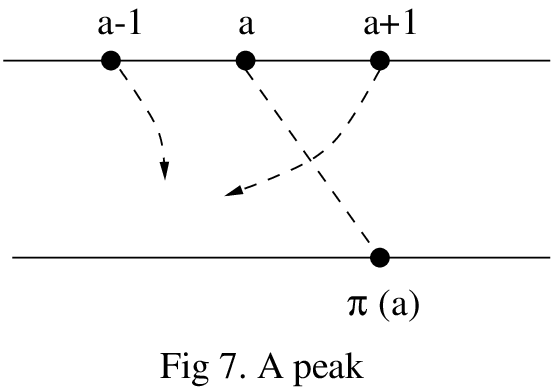}}
\bigskip

The number of peaks measures the complexity of the pairing.
The proof of the following lemma is given in the Appendix
\ref{app:comb}.
\begin{lemma}\label{lemma:number}
The number of pairings in $\Pi_n$ with no more than $K$ peaks is at most 
$n^{4K+3}(2K+2)^n$.
\end{lemma}

We want to gain a factor $t^{-1}$ from each peak. The idea is
that the factor $\wt R_{\pi(a)-1}$ (together with $R_{a-1}, R_a$)
 will be eliminated by integrating
out $\bp_{a-1}, \bp_{a}$ at the expense of only $\log^*t$ factors
instead of the trivial estimate $|\wt R_{\pi(a)-1}|\leq t$.
In order to estimate this integral, we make sure
that only these three denominators depend on  $\bp_{a-1}, \bp_{a}$
since we cannot estimate integrals with many denominators.
This requires a certain combinatorial structure of the peaks.

There is a difference whether $\om$ is constant or not.
The $\om =(const.)$ case is simpler to integrate out
and we will be able to gain a factor $t^{-1}$ from each
element of a {\it monotonic} peak sequence.

In the $\om\neq (const.)$ case the propagators $R_j, \wt R_j$
 have more complicated momentum
dependence (see Appendix \ref{sec:indnon} for more details), 
hence we will need 
a stronger combinatorial structure for the  peaks to be able
to perform the integrations. The appropriate
 structure ({\it "monotonic staircase"})
is introduced in the Appendix \ref{sec:gencomb}.

We explain the proof of the $\om =(const.)$ case in details.
The necessary modifications for the nonconstant case are
sketched in the Appendix \ref{sec:gencomb}, \ref{sec:indnon}.

\subsection{Combinatorics for the case of the constant $\omega$}

 The following theorem is a
well known  Ramsey-type theorem in combinatorics:

\begin{lemma}\label{lemma:ramsey1}
Any sequence of different numbers of $\a \beta + 1$ elements
either contains an increasing subsequence of $(\a +1)$ elements
or it contains a decreasing subsequence of $(\beta +1)$ elements.
$\;\;\Box$
\end{lemma}

\noindent
We will prove the following estimate.
\begin{proposition}\label{prop:omconst}
 Let $n\leq N$, $\um,\utm\in \cM(n,N)$.
Suppose that $\om $ is constant and that $\pi\in \Pi_n$
has either an increasing sequence of $\kappa$ peaks or
a decreasing sequence of $\kappa$ peaks. Then
$$
	|C_{\um,\utm,\pi}(t)| \leq t^{-\kappa} (C\lambda^2 t)^N 
	(\log^* t)^{n+\kappa+2}\;.
$$
\end{proposition}

{F}rom these statements (\ref{norecest}) follows for
the $\om$ constant case. Choose $\kappa=6$, then
 all pairings which have a monotonic sequence
of at least 6
 peaks can be included into the second term in  (\ref{norecest}).
Choosing $\a = \beta =5$ and $K= 5\cdot 5 +1 =26$, it is
clear from Lemmas \ref{lemma:number} and  \ref{lemma:ramsey1}
that with the exception of at most
$54^nn^{4\cdot 26 + 3}\leq C^n$ pairings,
we are in the situation of Proposition \ref{prop:omconst}.
The exceptional pairings are estimated by the apriori
bound (\ref{cpiest})  to give the first term
in (\ref{norecest}). 

\bigskip

{\it Remark:}
In general, we get $t^{-\kappa}$ with the exception of $(C\kappa)^{2n}$
pairings.

\subsection{Estimate for the constant $\om$ case.}\label{sec:constom}

{\it Proof of  Proposition \ref{prop:omconst}.}
Let $a_1 < a_2 < \ldots < a_\kappa$ be the locations of
the monotonic peak-sequence.
We start with the expression (\ref{eq:cpinew}).
We estimate all the $\tR_j$'s trivially by $Ct$
except $\tR_{\pi(a_m)-1}$, with $m=1, 2, \ldots, \kappa$,
and except $\tR_n$. This gives a factor $(Ct)^{n-\kappa}$.
In fact, with the help of (\ref{tagain}), we gain a $\langle \ta \rangle^{-1}$
factor from one of these estimates, at the expense of
collecting $\langle \bp_n\rangle^2 \prod_j \langle \bk_j\rangle^4$
factors. We also insert a factor
\be
	 1 \leq C^n {1\over \langle \bp_{a_1}\rangle^2}
	 \Big( \langle \bp_{n}\rangle^2
	\prod_j \langle \bk_j\rangle^2 \Big)\;,
\label{insert1} 
\ee
this will help to secure a decay in $\langle\a\rangle$.
All these factors will be incorporated into the integration measures
as before at the expense of changing $M^*(\bk)$ to $L(\bk)= M^*(\bk)
\langle \bk\rangle^{12}$.

Figs.~8--9 show the leftover $\tR$'s
represented by their electron momenta $\tbp_{\pi(a_m)-1}$
(bold lines) in case of an increasing and in a decreasing
 sequence of peaks, respectively.
 We want to gain a factor of $t$ from each bold line.

\bigskip\bigskip
\centerline{\epsffile{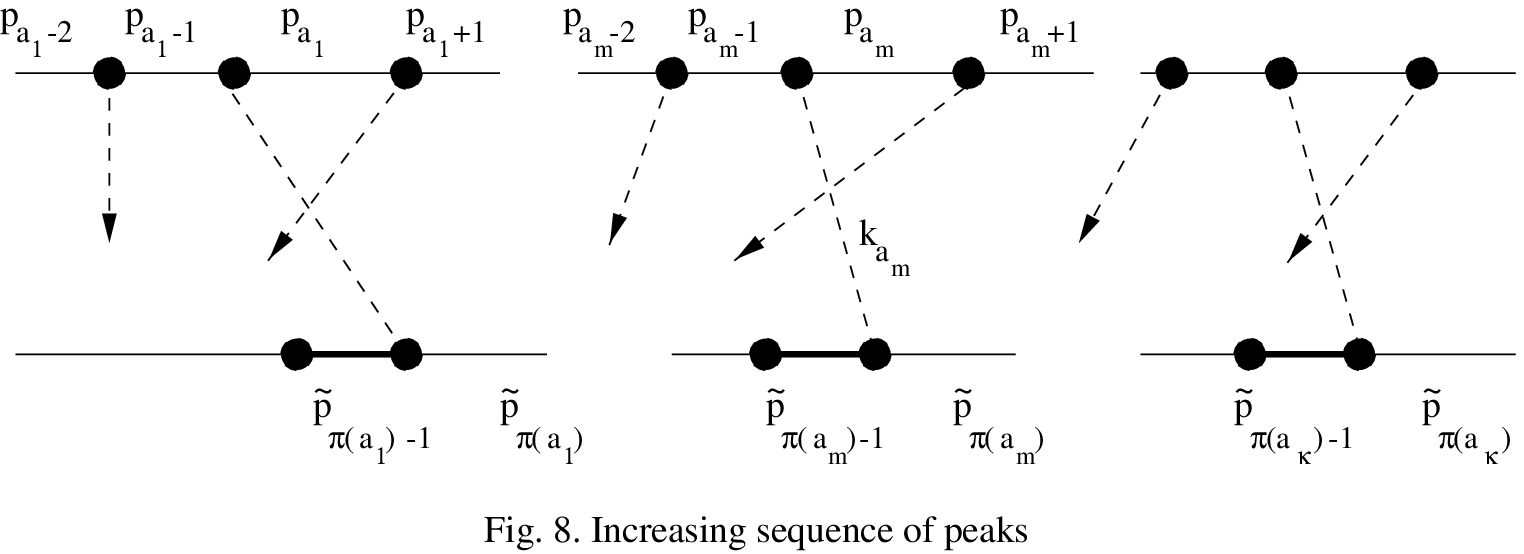}}
\bigskip

\bigskip\bigskip
\centerline{\epsffile{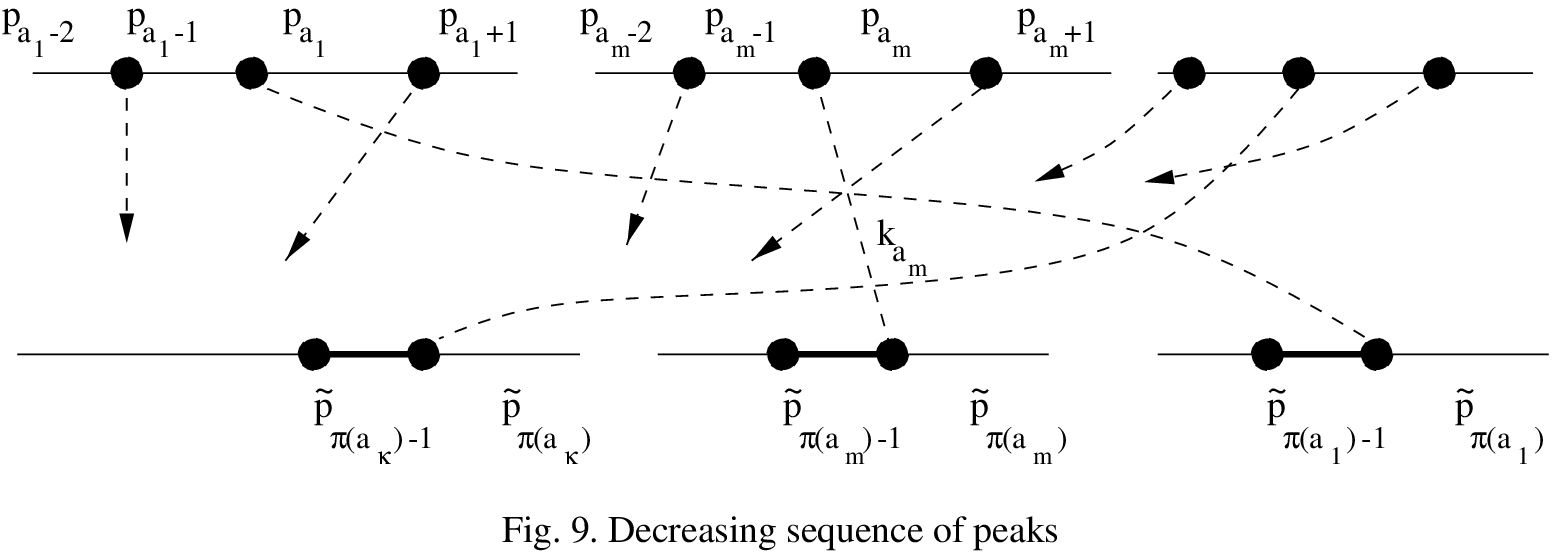}}
\bigskip

Now  we express everything in terms of $\bp_j$, $j=0, 1, \ldots, n$.
Recall that
$$
	\tR_{\pi(a_m)-1}=
	{1\over \Big| \ta - e\Big(\tbp_{\pi(a_m)} + \bp_{a_m-1} -
	\bp_{a_m}\Big) + i\eta  - \wt\Om_{\pi(a_m)-1}\Big|}
$$
where now
$$
	\wt\Om_{\pi(a_m)-1}
	 = \Big(\sum_{j\; : \; \pi(j)\ge \pi(a_m)} \s_j\Big)\om\;,
$$
i.e., it is independent of the electron momenta. Similar
formula is valid for $\Om_{a_m}$.

The important thing is that $\tbp_{\pi(a_m)}$ is independent of
$\bp_{a_j-1}$ and $\bp_{a_j}$ for all $j\leq m$ 
in the case of monotonically  increasing peaks, and it
is  independent of
$\bp_{a_j-1}$ and $\bp_{a_j}$ for all $j\ge m$ 
in the case of monotonically  decreasing peaks.
This can be seen from the expression
$$
	\tbp_{\pi(a_m)} = \bp_n +
	\sum_{j\; : \; \pi(j)>\pi(a_m)}
	(\bp_{j-1}- \bp_{j})
$$
and from the structure of the peaks.

Hence we can integrate out $\bp_{a_m}$ and $\bp_{a_m-1}$,
 $m=1,2,\ldots, \kappa$,
variables in the following order
$$
	\bp_{a_1-1}, \bp_{a_1}, \bp_{a_2-1}, \bp_{a_2},
	\ldots,  \bp_{a_\kappa-1}, \bp_{a_\kappa},
$$
if we have monotonically increasing peaks; and
in the order
\be
	 \bp_{a_\kappa-1}, \bp_{a_\kappa},
	  \bp_{a_{\kappa-1}-1}, \bp_{a_{\kappa-1}},
	\ldots,\bp_{a_2-1}, \bp_{a_2},
	\bp_{a_1-1}, \bp_{a_1}
\label{peaks}
\ee
if we have monotonically decreasing peaks. 
Notice that the  $\rd\bp_{a_m}$ and $\rd\bp_{a_m-1}$
integrations involve only the factors $R_{a_m-1}, R_{a_m}$
and $\tR_{\pi(a_m)-1}$ assuming that  all other $p_{a_j-1}, p_{a_j}$
have already been integrated out for $j<m$ in case of increasing peaks, and
for $j>m$ in case of decreasing peaks.

The result of each
integration is a $C\log^*t$ factor by using the following
inequality:
\bey
	\int {L(\bp_{a_m-2}-\bp_{a_m-1}) L(\bp_{a_m-1} - \bp_{a_m})
	L(\bp_{a_m} -\bp_{a_m+1}) \rd \bp_{a_m-1}\rd\bp_{a_m}\over
	\langle \bp_{a_m}\rangle^2
	\Big| \a - e(\bp_{a_m-1}) + i\eta - \Om_{a_m-1}\Big|
	\; \Big|  \a - e(\bp_{a_m}) + i\eta - \Om_{a_m}\Big|}&&
\label{int:pam}
\\
	\times {1\over  \Big| \ta - e\Big(\tbp_{\pi(a_m)} + \bp_{a_m-1} -
	\bp_{a_m}\Big) + i\eta  - \wt\Om_{\pi(a_m)-1} \Big|}
	&\leq& {(C\log^* t)^3 \langle n\om\rangle\over  \langle\a\rangle
	\langle \bp_{a_m-2} - \bp_{a_m+1}\rangle^{d+1}}\;.
\nonumber
\eey
The factor $\langle \bp_{a_m}\rangle^{-2}$ on the left hand side
and the factor $ \langle\a\rangle^{-1}$ on the right hand side are
present only for $m=1$.
The prototype of (\ref{int:pam}) is the following estimate
\be
	\sup_{\bp, \bq, \br, \theta_1,  \ttheta}
	\int { \langle \bp - \bu\rangle^{-d-1}
	\langle \bu - \bv \rangle^{-d-1} \langle \bv - \bq\rangle^{-d-1}
	\rd\bu \rd \bv\over  \langle \bv \rangle^2
	\Big| \theta_1- e(\bu) + i\eta \Big|
	\; \Big|  \theta_2 - e(\bv) + i\eta \Big| \; 
	\Big| \ttheta - e(\br + \bu-\bv)+ i\eta \Big|}
	\leq {(C\log^* t)^3 \over\langle \theta_2\rangle}
\label{prot:pam}
\ee
with the choice $\bp= \bp_{a_m-2}$, $\bu = \bp_{a_m-1}$,
$\bv = \bp_{a_m}$, $\bq = \bp_{a_m+1}$ and
$\br=\tbp_{\pi(a_m)}$. Again, if there is no factor $\langle \bv \rangle^{-2}$
on the left, then there is no factor
 $\langle \theta_2\rangle^{-1}$ on the right.
Since $\theta_2=\a - \Om_{a_m}$, we can estimate
$$
 	{1\over \langle \theta_2\rangle}\leq {C\langle n\om\rangle
	\over \langle\a\rangle}\;.
$$

We used that $\br=\tbp_{\pi(a_m)}$
is independent of $\bu = \bp_{a_m-1}$ and $\bv = \bp_{a_m}$
so that we could take the supremum over $\br=\tbp_{\pi(a_m)}$ outside
of the integration.
We also kept a momentum decay relation from the three $L$-factors
using that $L(\bk)\leq 
 C\langle \bk\rangle^{-d-1}\langle \bk\rangle^{-d-1}$ and
$$
	{1\over \langle\bp_{a_m-2}-\bp_{a_m-1} \rangle
	 \langle\bp_{a_m-1}-\bp_{a_m} \rangle
	\langle\bp_{a_m}-\bp_{a_m+1} \rangle}
	\leq {C\over \langle\bp_{a_m-2}-\bp_{a_m+1} \rangle}\;.
$$
Only part of the decay in $L(\bk)$ was needed in (\ref{prot:pam}).

For the proof of (\ref{prot:pam}) we use (\ref{prot:twoden})
to perform the $\rd \bv$ integration, then (\ref{prot:plus})
for the $\rd\bu$ integration.

\bigskip

After having eliminated all remaining $\tR_j$ factors,
 the rest  can be done by
successively integrating out 
$$
	\bp_0, \bp_1, \ldots, 
	\bp_{a_1-2},\wh\bp_{a_1-1},\wh\bp_{a_1},\bp_{a_1+1} \ldots,
	\bp_{a_2-2},\wh\bp_{a_2-1},\wh\bp_{a_2},\bp_{a_2+1} \ldots,
	\bp_{n-1}
$$
in this order
(hat denotes missing variable).
Notice that we saved a successive momentum decay for the remaining
variables so we can use (see (\ref{prot:Clog}))
$$
	\sup_{c, \bq}\int {\langle \bp -\bq \rangle^{-d-1} \rd \bp
	\over | c - e(\bp) + i\eta  |}\leq C\log^*t
$$
with $c=\a - \Om_{a_j}$, $\bp= \bp_j$, $\bq = \bp_{j+1}$
for all $j \not \in \{ a_m-2, a_m-1, a_m\; : \; m=1, 2, \ldots, \kappa\}$,
and simply $\bq$ becomes $\bp_{j+3}$ for $j=a_m-2$. We collect
$(C\log^*t)^{n-2\kappa}$.

At the end we are left with an integration identical to
(\ref{int:alpha3}).
This completes the proof of
Proposition \ref{prop:omconst}.$\;\;\;\Box$

\section{Amputated term without reabsorption}
\label{sec:ampnorec}
\setcounter{equation}{0}

Here we indicate the proof of (\ref{norecampest}) which
is very similar to that of (\ref{norecest}). 
We explain the necessary modifications in the corresponding
formulas leading to (\ref{norecest}). One has
to check only Sections \ref{sec:mainrep}, \ref{sec:triv} and 
\ref{sec:norecbign}.

Our starting point is  Proposition \ref{prop:formula}. Almost the
same representation is valid for the amputated term
$\Tr \cD^0_{n,N}\Gamma_0 \Big[\cD^0_{n,N}\Big]^*$
except that in the set $\cM(n, M)$ we additionally
require $m_0=0$, $\mu(1)=1$,
i.e., $I_0 = \emptyset$ and we also set $I_0^c:=\emptyset$.
In particular the index $0$ is not part of the set $J\cup J^c$
and $|J^c|: = n+|\um|$ is reduced by one.
This expresses the fact that there is no propagator $e^{is_0e(\bp_0)}$
associated with the momentum $\bp_0$.
The definition of $Y_{\um,\utm, \pi}$ (\ref{1Y}), (\ref{2Y}) is changed
 slightly: the products over $j$ start from $j=1$ and there
is no $s_0, \ts_0$.

The additional gain $1/t$ is due to the fact that
the factors $R_0$ and $\tR_0$ are missing and that the set $J^c$
has smaller cardinality. 

In the proof of Lemma \ref{lemma:trivbound} one can easily see
that there is one less factor of $t^{a}$ in
 (\ref{Ya}) due to the decreased size
of $|J^c|$ in (\ref{befYa}). The other $t^{1-a}$ gain comes
from the fact that the last integrand in (\ref{succ})
does not have a denominator, so the $\rd\bk_1$ integration
is bounded by a constant.

In the proof of Propositions \ref{prop:omconst}
the gain comes from the fact that $\tR_0$ is missing
hence it does not have to be estimated by $Ct$ at the very beginning.
Notice that $\tR_0$ has indeed been estimated by $Ct$ since 
$\pi(a_m)\ge 2$ for all peaks.
For nonconstant $\om$ (see Appendix \ref{sec:staircase})
the argument is similar: in  the proof of Proposition
 \ref{prop:omnonconst} the first factor $\tS_0$ is missing
in (\ref{eq:cpinew3}). If $\pi(a_1)=1$ and there are no $\tS_j$ 
factors and $\tbeta$ integration, then we gain from the missing 
$\tR_0$ term.
This completes the proof of  (\ref{norecampest}). $\;\;\;\Box$

\section{One reabsorption}\label{sec:onereabs}
\setcounter{equation}{0}

In this section we prove (\ref{1recampestor}). Recall that the measure
$\int^{*(n,N)} \prod\rd \bk_j$ in the definition of $\cD_{n,N}^1$
 (see (\ref{def:starint}) and (\ref{def:D1})) contains
a double summation over measures $\int^{*(\um, a)} \prod\rd \bk_j$,
 where $\um\in \cM_a(n,N)$.
Hence we can write
$$
	\cD_{n,N}^1(t) = \sum_{a=2}^N\sum_{\um\in \cM_a(n,N)}
	\cD_{n,N}^{\um, a}(t)
$$
with
\be
	\cD_{n,N}^{\um, a}(t) = \int^{*(\um,a)} \Big(\prod_{j=1}^N
	\rd\bk_j\Big)\cA(t,\ubk, N)\;.
\label{def:duma}
\ee
After a Schwarz inequality we can symmetrize
$$
	\Tr_{e+ph} \Bigg( \cD_{n,N}^1\Gamma_0 \Big[\cD_{n,N}^1\Big]^*\Bigg)
	\leq C^N N\sum_{a=2}^N\sum_{\um\in \cM_a(n,N)}
	\Tr_{e+ph} \Bigg( \cD_{n,N}^{\um, a} \Gamma_0 \Big[ \cD_{n,N}^{\um, a}
	\Big]^*\Bigg)
$$
using that the cardinality of $\cM_a(n,N)$ is bounded by $C^N$.

Therefore it is sufficient to show that
\be
	\limsup_{L\to\infty}
	\Tr_{e+ph}
	 \Bigg( \cD_{n,N}^{\um, a} \Gamma_0 \Big[ \cD_{n,N}^{\um, a}
	\Big]^*\Bigg)\leq {1\over t}(C\lambda^2 t)^N
	\Bigg[ {(\log^*t)^6\over t^2} + {n!\over t^6} (\log^* t)^{n+10}
	\chi(N\ge 7)\Bigg] 
\label{suffrec}
\ee
for any $2\leq a\leq N$ and $\um\in \cM_a(n,N)$.
{F}rom now on we fix $a$ and $\um$ and recall that
$I_a = I\setminus\{ 1, \mu(a)\}$. Let $I_a^c:= \{ 1, \ldots, N\}\setminus
I_a$.

We separate the measure of the reabsorpting phonon momenta 
in (\ref{def:duma}):
$$
	 \int^{*(\um,a)}\rd\ubk
	=  \int^{\#(\um,a)} 
	\Big(\prod_{b\in I_a}\rd\bk_b\Big)
	\int  \rd \bk_1\rd \bk_{\mu(a)}
	\Big(\prod_{b\in J}\rd\bk_b\rd\bk_{b+1}
	\Big) F_{a}( [\bk]_{I_a}, [\bk]_{I_a^c})
$$
with 
\be
	  \int^{\#(\um,a)}\Big(\prod_{b\in I_a}\rd\bk_b\Big)
	: =  \int\Big(\prod_{b\in I_a} \rd\bk_b\Big)
	 \prod_{b\neq b'\in I_a}
	\Big(1- \chi(\bk_b+\bk_{b'})\Big)
\label{ma}
\ee
and
$$
	F_{a}( [\bk]_{I_a}, [\bk]_{I_a^c}):=
	\chi(\bk_1 + \bk_{\mu(a)})\Xi(\ubk,\um)
	\prod_{b\not\in  I_a}
	\Big(1- \chi(\bk_b+\bk_{1})\Big)
	\Big(1- \chi(\bk_b+\bk_{\mu(a)})\Big)\;, 
$$
where in the notation we split the variables
 $\ubk=\{ \bk_1, \ldots, \bk_N\}$ into two sets: $\{ \ubk\} =
[\bk]_{I_a}\cup [\bk]_{I_a^c}$ and 
for any $S\subset \{ 1, 2, \ldots, N\}$ we let
$[\bk]_{S}: =
\Big\{  \bk_b \; : \; b\in S\Big\}\;.$
We let
\be
	\cF \Big( [\bk]_{I_a}, \bp_N\Big)
	: = \int \rd\bk_1 \rd \bk_{\mu(a)}
	\Big(\prod_{b\in J}\rd\bk_b\rd\bk_{b+1}\Big)
	F_{a}( [\bk]_{I_a}, [\bk]_{I_a^c})
	\cA \Big(t, \ubk, N;  \bp_N + \sum_{j=1}^N \bk_j , \bp_N\Big)
\label{def:cF}
\ee
be an operator in the phonon space, parametrized by $[\bk]_{I_a}$ and
$\bp_N$.
Here $\cA \Big(t, \ubk, N; \bp_0, \bp_N\Big)$ is the kernel
of the operator $\cA (t, \ubk, N)$ in Fourier space of $\cH_e$
similarly to (\ref{def:cBker}) and (\ref{def:cBkersub}).
Using the delta function $\delta\Big( \bp_0-\bp_N-\sum_{j=1}^N\bk_j\Big)$
we have
\be
	 \cA \Big(t, \ubk, N;  \bp_N + \sum_{j=1}^N \bk_j,  \bp_N\Big):
	= \lambda^N\!\!\int \Big( \prod_{j=1}^{N-1} \rd \bp_j\Big)
	\Delta^*(\ubp, \ubk)
	\int_0^{t*} [\rd s_j]_1^N \; 
	\! \Big(\prod_{j=1}^N Q(\bk_j) b_{\bk_j} 
	e^{-is_j[e(\bp_j)+H_{ph}]}\Big),
\label{def:cAker}
\ee
which is an operator acting on $\cH_{ph}$.
Here we define $\Delta^*$ by the relation
 $\Delta(\ubp, \ubk) = \Delta^*(\ubp, \ubk)
\delta\Big( \bp_0-\bp_N-\sum_{j=1}^N\bk_j\Big)$
(see (\ref{def:Delta})).

The same formulas are valid for the other copy of  $\cD_{n,N}^{\um, a}$
in (\ref{suffrec}); the variables are denoted by tilde. The result
is
\bey
	\Tr_{e+ph}
	 \Bigg( \cD_{n,N}^{\um, a} \Gamma_0 \Big[ \cD_{n,N}^{\um, a}
	\Big]^*\Bigg) &=&\int^{\# (\um, a)} 
	\Big(\prod_{b\in I_a} \rd\bk_b\Big)
 	\int^{\# (\um, a) } 
	\Big(\prod_{b\in I_a} \rd\tbk_b\Big)
	\int \rd\bp_N 
	\wh\g_e(\bp_N, \bp_N)
\nonumber\\
	&&\times \Tr_{ph}\cF \Big( [\bk]_{I_a}, \bp_N\Big)
	 \g_{ph} \Big[\cF \Big( [\tbk]_{ I_a}, \bp_N\Big)\Big]^*\;.
\label{cpi1}
\eey
We again used that the $\delta(\bp_0-\tbp_0)$ coming from taking the
electron trace can be replaced with $\delta(\bp_N-\tbp_N)$
and we integrated out $\tbp_N$.

Similarly to Section \ref{sec:thermotrace},
we again notice that the phonon trace in (\ref{cpi1})
is zero unless there is a complete
pairing between all the involved $2N$ momenta $\ubk, \utbk$.
The pairing must respect
the prepared immediate reabsorptions  and
the reabsorptions between $\bk_1, \bk_{\mu(a)}$ and
 $\tbk_1, \tbk_{\mu(a)}$,
apart from an
error term that is negligible in the thermodynamic limit
(Lemma \ref{lemma:thermo}).
Moreover, there is no pairing between $\bk_b$ and
$\bk_{b'}$, $b,b' \in I_a$
 (and the same for the variables with tilde)
 by the no-reabsorption condition built into the
measure $\int^{\# (\um,a)}\prod \rd\bk_b$ (\ref{ma}).
Hence the possible pairings
are parametrized by a permutation $\pi \in \Pi_n^{a}$, where
$\Pi_n^{a}$ is the set of all permutations
on the set $\{  2, \ldots, \wh a,  \ldots, n\}$.
The map $\pi^*:=\mu \circ\pi\circ\mu^{-1}: I_a\to I_a$
gives the pairing of the indices of those $\bk$'s and $\tbk$'s
that are not prescribed for immediate recollisions
similarly to Section \ref{sec:thermotrace}.
This means that a factor
$$
	\sum_{\pi\in \Pi_n^a}\Bigg(\prod_{j=2\atop j\neq a}^n
	\chi(\bk_{\mu(j)} - \tbk_{\mu(\pi(j))}) \Bigg)
$$
can be freely inserted into (\ref{cpi1})
modulo a negligible error. We obtain
\be
	 \Tr_{e+ph}\Bigg(\cD_{n,N}^{\um, a}(t)
	 \Gamma_0 \Big[ \cD_{n,N}^{\um, a}(t)
	\Big]^*\Bigg)= \sum_{\pi\in \Pi_n^{a}} 
	C_{n,N}^{\um,a}(\pi;t) + O(|\Lambda|^{-1})
\label{deexp}
\ee
with
\bey
	C_{n,N}^{\um,a}(\pi;t):
	&=&\int^{\# (\um, a)} 
	\Big(\prod_{b\in I_a} \rd\bk_b\Big)
 	\int^{\# (\um, a) } 
	\Big(\prod_{b\in I_a} \rd\tbk_b\Big)
	\int \rd\bp_N 
	\wh\g_e(\bp_N, \bp_N)
\label{def:cujpi}\\
	&&\times\Bigg(\prod_{j=2\atop j\neq a}^n
	\chi(\bk_{\mu(j)} - \tbk_{\mu(\pi(j))}) \Bigg)\;
	 \Tr_{ph}\cF \Big( [\bk]_{I_a}, \bp_N\Big)
	 \g_{ph} \Big[\cF \Big( [\tbk]_{ I_a}, \bp_N\Big)\Big]^*\;.
\nonumber
\eey

Peaks and valleys
 of $\pi\in\Pi_n^{a}$ are defined exactly as before (Definition
\ref{def:peak}) and Lemma \ref{lemma:number} remains valid
if $n$ is replaced with $n-2$.
The estimate (\ref{suffrec})  follows from  Lemma \ref{lemma:number}
and the 
lemma below.

\begin{lemma}\label{lemma:smallcom}
Let $N\ge 3$, $a\ge 2$, $n\ge 0$ be integers and $\um\in \cM_a(n,N)$ as before.
For any $\pi \in \Pi_n^{a}$ we have
\be
	C_{n,N}^{\um,a}(\pi;t) \leq
	{1\over t} \cdot  (C\lambda^2 t)^N\cdot {1\over t^2}
	(\log^* t)^6\;.
\label{smallcomest}
\ee
Moreover, if $\pi$ has $\kappa\ge 2$ peaks, then
\be
	C_{n,N}^{\um,a}(\pi;t) \leq
	{1\over t}  \cdot   (C\lambda^2 t)^N \cdot  
	 {1\over t^{\kappa-2}} (\log^* t)^{n+\kappa+2}\;.
\label{largecomest}
\ee
In case of $\om=(const.)$,  the power $\kappa-2$  can be improved to $\kappa$.
\end{lemma}

The proof of  (\ref{smallcomest}) and (\ref{largecomest})
are given in Sections \ref{sec:allpair} and \ref{sec:largepair},
respectively. These are the recollision analogues of the apriori
bound (\ref{cpiest}) in Lemma \ref{lemma:trivbound} and
Proposition \ref{prop:omconst}.

\subsection{General bound for recollision pairings}\label{sec:allpair}

For the proof of (\ref{smallcomest}), we
 start by symmetrizing the momenta in the definition 
(\ref{def:cujpi}). Using a Schwarz inequality within $\Tr_{ph}$
in (\ref{def:cujpi})
\be
	\Bigg| \Tr_{ph}\cF \Big( [\bk]_{I_a}, \bp_N\Big)
	 \g_{ph} \Big[\cF \Big( [\tbk]_{I_a}, \bp_N\Big)\Big]^*\Bigg|
\label{newSchw}
\ee
$$
	\leq {1\over 2} 
	\Tr_{ph}\cF \Big( [\bk]_{I_a}, \bp_N\Big)
	 \g_{ph} \Big[\cF \Big( [\bk]_{I_a}, \bp_N\Big)\Big]^*
	+  {1\over 2} 
	\Tr_{ph}\cF \Big( [\tbk]_{I_a}, \bp_N\Big)
	 \g_{ph} \Big[\cF \Big( [\tbk]_{I_a}, \bp_N\Big)\Big]^*\;.
$$
We deal only with the first term, the second is identical. Hence
\bey
	C_{n,N}^{\um,a}(\pi;t) 
	&\leq&\int^{\# (\um, a)} 
	\Big(\prod_{b\in I_a} \rd\bk_b\Big)
 	\int^{\# (\um, a) } 
	\Big(\prod_{b\in I_a} \rd\tbk_b\Big)
	\int \rd\bp_N 
	\wh\g_e(\bp_N, \bp_N)
\nonumber\\
	&&\times\Bigg(\prod_{j=2\atop j\neq a}^n
	\chi(\bk_{\mu(j)} - \tbk_{\mu(\pi(j))}) \Bigg)\;
	 \Tr_{ph}\cF \Big( [\bk]_{I_a}, \bp_N\Big)
	 \g_{ph} \Big[\cF \Big( [\bk]_{ I_a}, \bp_N\Big)\Big]^*\;,
\nonumber
\eey
and we can freely integrate out all $\tbk_{b}$, $b\in I_a$.
Each $\tbk_{b}$ integration gives a factor ${1\over |\Lambda|}$, so
 after these integrations 
$$
	C_{n,N}^{\um,a}(\pi;t) 
	\leq {1\over |\Lambda|^{n-2}}
	\int^{\# (\um, a)} 
	\Big(\prod_{b\in I_a} \rd\bk_b\Big)
 	\int \rd\bp_N 
	\wh\g_e(\bp_N, \bp_N)\;
	 \Tr_{ph}\cF \Big( [\bk]_{I_a}, \bp_N\Big)
	 \g_{ph} \Big[\cF \Big( [\bk]_{ I_a}, \bp_N\Big)\Big]^*\;.
$$
Notice that this bound is independent of $\pi$.

Now we write out $\cF$ explicitly
using (\ref{def:cF}) and (\ref{def:cAker}).
We obtain
\bey
	\lefteqn{C_{n,N}^{\um,a}(\pi;t) 
	\leq { \lambda^{2N}\over |\Lambda|^{n-2}}			
	\int^{\# (\um, a)} 
	\Big(\prod_{b\in I_a} \rd\bk_b\Big)
 	\int \rd\bp_N 
	\wh\g_e(\bp_N, \bp_N)
	\int \rd\bk_1 \rd \bk_{\mu(a)} \rd\tbk_1 \rd \tbk_{\mu(a)}}
\nonumber\\
	&&\times  \int 
	\Big(\prod_{b\in J}\rd\bk_b\rd\bk_{b+1}\Big)
	\Big(\prod_{b\in J}\rd\tbk_b\rd\tbk_{b+1}\Big)
	F_{a}( [\bk]_{I_a}, [\bk]_{I_a^c})
	F_{a}( [\bk]_{I_a}, [\tbk]_{I_a^c})
\nonumber\\
	&&
	\times \int \Big( \prod_{j=1}^{N-1} \rd \bp_j\Big)
	\Delta^*(\ubp, \ubk)
	\int \Big( \prod_{j=1}^{N-1} \rd \tbp_j\Big)\Delta^*(\utbp, \ubk^*)
	\int_0^{t*} [\rd s_j]_1^N  [\rd \ts_j]_1^N
	\Big( \prod_{j=1}^N Q(\bk_j)
	e^{-is_je(\bp_j)}\Big)
\nonumber\\
	&&
	\times
	\Tr_{ph}\Bigg[ \Big(\prod_{j=1}^N b_{\bk_j}(\tau_j)\Big)
	e^{-itH_{ph}}
	\g_{ph} e^{itH_{ph}}\Big(\prod_{j=1}^N
	 b^*_{\bk_j^*}(\ttau_j)\Big)
	\Bigg]
	\Big( \prod_{j=1}^N Q(\bk_j)
	e^{i\ts_je(\tbp_j)}\Big)\;,
\nonumber
\eey
with $\bk_j^*: = \bk_j$ if $j\in I_a$ and $\bk_j^*:=\tbk_j$
if $j\in I_a^c$, and
we recall the definition of $\tau_j, \ttau_j$ from (\ref{def:tau}).

Again, the pairing in the phonon trace must respect
the  recollisions prepared in the $F_a$ factors
and the prepared $b_{\bk_j}b_{\bk_j}^*$ pairing for $j\in I_a$
(modulo an error $O(|\Lambda|^{-1})$).
These latter pairings yield a factor $|\Lambda|^{n-2}$
from the $n-2$ delta functions $\delta(\bk_j-\bk_j)=|\Lambda|$.

%The result is
%$$
%	C_{n,N}^{\um,a}(\pi;t) 
%	\leq  \lambda^{2N}
%	\int^{\# (\um, a)} 
%	\Big(\prod_{b\in I_a} \rd\bk_b\Big)
% 	\int \rd\bp_N 
%	\wh\g_e(\bp_N, \bp_N)
%$$
%$$
%	\times  \int \rd\bk_1 \rd \bk_{\mu(a)} \rd\tbk_1 \rd \tbk_{\mu(a)}
%	\Big(\prod_{b\in J}\rd\bk_b\rd\bk_{b+1}\Big)
%	\Big(\prod_{b\in J}\rd\tbk_b\rd\tbk_{b+1}\Big)	
%	\Bigg( \prod_{ b\in I_a} G^\#(\bk_b, \ttau_b-\tau_b)\Bigg)
%	\cX(\utbk, \uttau, \um)\overline{\cX}(\ubk, \utau, \um)
%$$
%$$
%	\times \delta(\bk_1 + \bk_{\mu(a)}) 
%	G^\#(\bk_1, \tau_1-\tau_{\mu(a)})
%	\delta(\tbk_1 + \tbk_{\mu(a)}) G^\#(\tbk_1, \ttau_{\mu(a)}-\ttau_1)
%$$
%$$
%	\times\int_0^{t*} [\rd s_j]_1^N \int_0^{t*} [\rd \ts_j]_1^N
%	\Big( \prod_{j=1}^N Q(\bk_j)
%	e^{-is_j\bp_j^2/2}\Big)
%	\Big( \prod_{j=1}^N Q(\bk_j)
%	e^{i\ts_j\tbp_j^2/2}\Big) + O(|\Lambda|^{-1})\;.
%$$

Now we proceed similarly to Section \ref{sec:profrep}
by integrating out the internal momenta $\bk_b, \tbk_b$ for $b\in J$,
 relabelling the external ones and expressing the relabelled
electron momenta by the relabelled phonon momenta (\ref{def:pj}).
We need a slight modification of the expression $\tbp_j$
($j= 1, \ldots, n-1$) compared to (\ref{def:pj})
\be
	\bp_j : = \bp_n 
	+\sum_{m=j+1}^n \bk_m\;, \quad
	\tbp_j : =  \bp_n 
	+ \chi( j \leq a-1) \tbk_a +
	\sum_{m=j+1\atop m \neq a}^n \bk_m \; .
\label{def:pjmod}
\ee
i.e., now these are considered functions of $\bk_2, 
\ldots, \bk_n$, $\tbk_a$ and $\bp_n$. 
Similarly, we slightly modify the definition of
 $\wt\Om_j$, $j=1,2,\ldots , n$,  (compare with (\ref{def:Om})):
\be
	\Om_j:=
	 \sum_{m=j+1}^n \s_m \om(\bk_m)\;,\qquad
	\wt\Om_j:
	=\chi(j\leq a-1) \tsi_a \om (\tbk_a)
	+\sum_{m=j+1\atop m\neq a}^n \s_m \om(\bk_m)\; .
\label{def:Omstar}
\ee
The exact dependence on the $\sigma$'s are irrelevant.
The important property is
that $\Om_j$ depends only on the momenta
$\bk_{j+1}, \bk_{j+2} \ldots, \bk_n$
and the same for $\wt\Om_j$ just $\bk_a$ is replaced by $\tbk_a$.

The result of mimicking the argument in Section \ref{sec:profrep} 
leading to Proposition \ref{prop:formula} is
\be
	 C_{n,N}^{\um,a}(\pi;t) \leq \wt C_{n,N}^{\um,a}(t)
\label{cc}
\ee
with
\bey
	\wt C_{n,N}^{\um,a}(t) : &=&
	 \lambda^{2N}\sum_{\s_2, \ldots, \s_n; \tsi_a} \int \rd\bp_n
	\wh\g_e(\bp_n, \bp_n)
	\int\Big(\prod_{j=2}^n M( \bk_j,\s_j)\rd\bk_j\Big)
\label{eq:cer}\\
	&&\times\int
	  M(\tbk_a,\tsi_a)\rd\tbk_a 
	\;Y_{\um,\um}(t, \bp_n, \bp_n, \ubk, \utbk, \usi \cup\{\tsi_a\})
	+ O(|\Lambda|^{-1})\;,
\nonumber
\eey
where the  definition of $Y$ (\ref{2Y})--(\ref{def:R})
is slightly modified as follows. There is no permutation $\pi$.
 The  products over $j$ in (\ref{1Y})--(\ref{2Y}) start
from $j=1$ and there is no $s_0, \ts_0$. The sets of phonon momenta 
are $\ubk= [\bk]_{I_a}\cup \{ \bk_a\}$ and
$\utbk=[\bk]_{I_a}\cup \{ \tbk_a\}$.
The electron momenta $p_j$'s  are functions
of $\ubk$ as given in (\ref{def:pjmod}), similarly for the tilde
variables. Finally,  we use the
definition of $\Om_j, \wt\Om_j$ 
(\ref{def:Omstar}),
in particular $Y$ depends on $\{ \s_2,  \ldots, \s_n, \tsi_a\}$ 
 which replace the variables $\usi$ in the original definition
of $Y$.

\bigskip

The estimate of  $\wt C_{n,N}^{\um,a}(t)$ is  different for $a\ge 3$
and for $a=2$;
these cases will be discussed separately. The first case
is similar to the crossing estimate
 in Section 3.3. \cite{EY2}. The second case corresponds 
to the nested pairings. We recall the definition from 
Section 3.4 \cite{EY2}:

\begin{definition}
A pairing of the  momenta $\{ \bk_1, \ldots,
\bk_N\}, \{\tbk_1, \ldots, \tbk_N\}$ 
is called {\bf nested}
if there exist indices $j_1<j_2<j_3<j_4$ such that $(\bk_{j_1}, \bk_{j_4})$
and $(\bk_{j_2}, \bk_{j_3})$ are paired, or the same is true
for the tilde variables. Notice that this notion is defined in
the original graph and not in its skeleton.
\end{definition}
 In particular, a nested recollision graph with $a=2$ corresponds to
$j_1:=1$, $j_4:=\mu(2)\ge 4$ and $(j_2, j_3)$ with $j_3:=j_2+1$ being the
indices of any immediate recollisions in between ($1 < j_2 < \mu(2)$
and $j$ is even).

\subsubsection{Pairing 
with a distant recollision: Case $a\ge 3$}\label{sec:ageq3}

We assume that $a\ge 3$. In this case the recollision pairing
line actually crosses at least another pairing line. In \cite{EY2}
we treated this case together with the crossing estimates.
The proof we give here is similar although in this paper we
do not call it crossing pairing.

 To estimate $Y$  in (\ref{eq:cer}), 
we bound all $\Upsilon$'s by a constant 
(\ref{Upsilonest}) and use (\ref{ttriv}). We obtain
\bey
	C_{n,N}^{\um,a}(\pi;t) \leq \wt C_{n,N}^{\um,a}(t) 
	&\leq&  (C\lambda)^{2N} t^{2|\um|}\sup_{\usi, \tsi_a}
	 \int \rd\bp_n 
	\wh\g_e(\bp_n, \bp_n)\int_{-\infty}^\infty \rd\a \rd\ta
\label{eq:cer1}\\
	&& \times
	\int\Big(\prod_{j=2}^n M^*( \bk_j)\rd\bk_j\Big)\;
	 M^*(\tbk_a)\rd\tbk_a 
	 \prod_{j=1}^n	|R_j| \; |\tR_j|\;,
\nonumber
\eey
and we choose $\eta=t^{-1}$ in the definition of $R_j, \tR_j$
(\ref{def:R}).

\bigskip

For $j\ge 1$ we define 
$$
	\bP_j := \bp_n+ \sum_{m=j+1\atop m\neq a}^n \bk_m, \qquad
	\Om^\#_j := \sum_{m=j+1\atop m\neq a}^n\s_{m} \om(\bk_{m})\;.
$$

We start by integrating $\bk_{2}$, this  involves $R_1, \tR_1$,
explicitly (see (\ref{prot:twoden}))
\bey
	\int 
	 {M^*(\bk_2)\rd\bk_2
	 \over \Big| \a -e (\bk_{2} +  \bP_{2} + \bk_a)
	+ i\eta -
	 \Om^\#_2 - \s_2 \om(\bk_2)- \s_a\om(\bk_a)\Big|}&&
\nonumber\\
	\times {1 \over \Big| \ta - e(\bk_{2} +  \bP_{2} + \tbk_a)
	+ i\eta - \Om^\#_2-
	\s_{2} \om(\bk_{2}) - \tsi_a\om(\tbk_a)\Big|}
	&\leq& {(C\log^*t)^2\over  |\bk_a -\tbk_a|}\;.
\nonumber
\eey

\bigskip\bigskip
\centerline{\epsffile{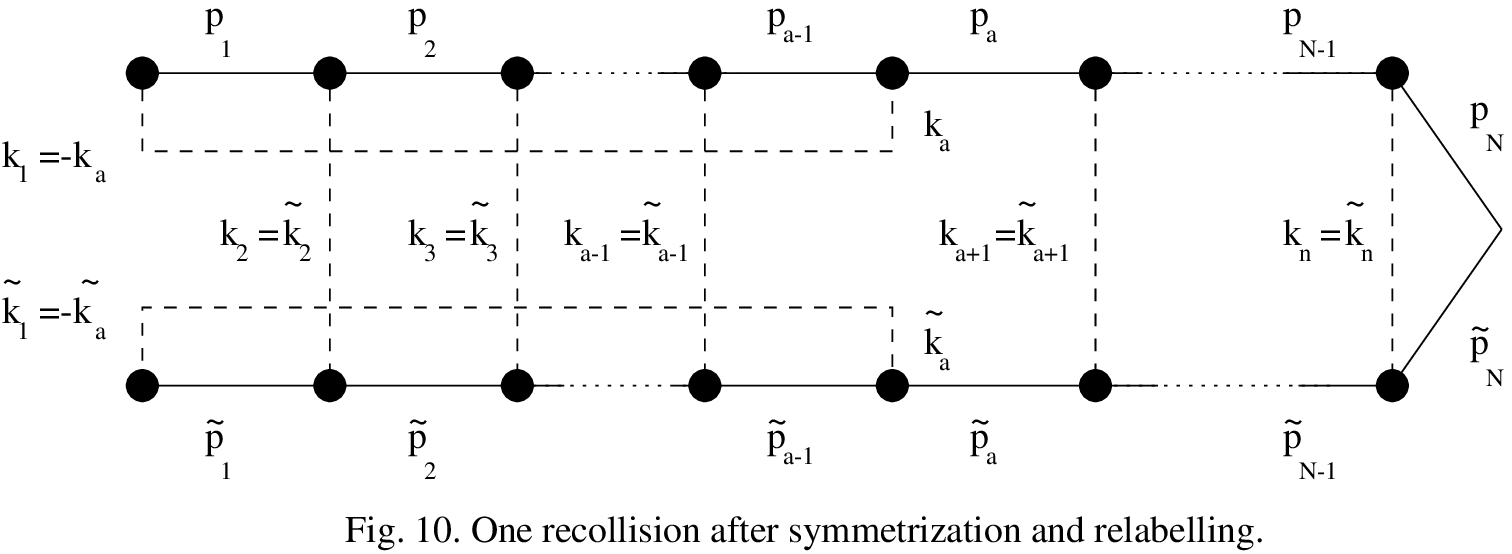}}
\bigskip

Now we integrate out $\bk_{3}, \ldots, \bk_{a-1}$
in this order if $a\ge 4$. At the $j$-th step, $3\leq j \leq a-1$, we
eliminate $R_{j-1},\tR_{j-1}$, using
\bey
	\int {M^*(\bk_j)\rd \bk_j \over\Big|\a - 
	e(  \bk_j + \bP_j + \bk_a)
	+i\eta - \Om^\#_j-\s_j\om(\bk_j)  - \s_a\om(\bk_a)\Big|}&&
\label{eq:middle}
\\
	\times{1\over \Big|\ta - e(  \bk_j + \bP_j + \tbk_a)
	+i\eta ) -  \Om^\#_j -\s_j \om(\bk_j)  - \tsi_a\om(\tbk_a)\Big|}
	& \leq&
	Ct\;,
\nonumber
\eey
and this gives $(Ct)^{a-3}$ (this is valid even if $a=3$).
This estimate follows from (\ref{prot:Ct}) after a Schwarz inequality.

Now we do the $\rd\bk_a, \rd\tbk_a$ integrals  to eliminate
 $R_{a-1}$ and $\tR_{a-1}$:
\bey
	\int 
	{ M^*(\bk_a)\rd \bk_a
	\over \Big| \a - e(\bk_a + \bP_a)+ i\eta
	-\s_a \om(\bk_a) - \Om^\#_a\Big|} &&
\label{int:b}
\\
	\times \int {1\over |\bk_a-\tbk_a|} { M^*(\tbk_a)\rd \tbk_a
	\over \Big| \ta - e( \tbk_a + \bP_a) + i\eta
	-\tsi_a \om(\tbk_a) - \Om^\#_a\Big|}	
	&\leq& { (C\log^* t)^2\over \langle \a -\ta \rangle}\;.
\nonumber\eey
The prototype of the $\rd\tbk_a$ integral is
\be
	\sup_{c,\bq}\int  {1\over |\bk-\bq|}
	{ M^*(\bk)\rd \bk
	\over \Big| c- e( \bk + \bP) + i\eta
	\pm \om(\bk)\Big|}\;  \leq {C\log^*t\over \langle c -e(\bP)\rangle}\;,
\label{eq:prot4}
\ee
which is a straightforward strengthening of (\ref{prot:plus}).
The same estimate is valid without the $ |\bk-\bq|^{-1}$ factor,
and that is the prototype of the $\rd\bk_a$ integration.
Finally, we use
\be
	\sup_{\bP} {1\over  \langle c -e(\bP)\rangle 
	\langle \tc -e(\bP)\rangle}\leq {C\over  \langle c -\tc \rangle}
\label{alphagain}\ee
to obtain the $ \langle \a -\ta \rangle^{-1}$ decay.

\medskip

Now we do $\rd\bk_{a+1}, \rd \bk_{a+2}, \ldots, \rd \bk_{n}$
integrals; each
gives a factor $Ct$ and  altogether we collect $(Ct)^{n-a}$.
 In the $j$-th step ($a+1 \leq j \leq n$) we 
eliminate $R_{j-1},  \tR_{j-1}$ by using a Schwarz estimate
and (\ref{prot:Ct}) as in (\ref{eq:middle}).

Finally we are left with
\be
	\int_{-\infty}^\infty
	 \rd \a \rd \ta {1\over \langle\a -\ta\rangle}\int
	{ \wh\g_e(\bp_n,\bp_n)\rd\bp_n\over
	| \a - e(\bp_n) + i\eta|\; | \ta - e(\bp_n) + i\eta|}
	\leq  (C\log^*t)^2\;.
\label{int:alpha}
\ee
Altogether we have
$$
	(C\log^*t)^2 (Ct)^{a-3} (C\log^*t)^2 (Ct)^{n-a}(C\log^*t)^2
	= {(\log^*t)^6\over t^3} (Ct)^n
$$
for the integrals in (\ref{eq:cer1}),
which finishes the proof of (\ref{smallcomest})
in the $a\ge 3$ case.

\subsubsection{Pairing with a nested recollision: Case $a=2$}
\label{sec:nestconst}

Here we prove (\ref{smallcomest})
for the remaining case $a=2$. 
In this case  $m_1\ge 1$ since we had a genuine recollision
before relabelling. Therefore the original pairing was nested.

We start from (\ref{eq:cer}) and perform explicitly the $\bk_2$
integration. This involves the following 
\be
	A=A(\a, \bp_2, \Om_2,\s_2):=
	\int M(\bk_2,\s_2)  \Big[R_1(\a, \bp_1, \Om_1,\eta)
	\Big]^{m_1+1} \Big[\Upsilon_\eta(\a- \Om_1, \bp_1)
	\Big]^{m_1}\rd\bk_2  \;
\label{def:A}
\ee
using (\ref{2Y}) and
recalling that $\bp_1 = \bp_2 +\bk_2$, $\Om_1=\Om_2 + \s_2\om(\bk_2)$.

\begin{lemma}\label{lemma:A}
With the notation above and for $m_1\ge 1$, $\eta\leq 1$ we have
\be
	|A(\a, \bp_2, \Om_2,\s_2)|\leq {C^{m_1}\langle\bp_2\rangle^d
	 \eta^{-m_1+1/2}
	 \over \langle \alpha - e(\bp_2)-\Om_2\rangle}\;.
\label{Aest}
\ee
\end{lemma}

{\it Proof of Lemma \ref{lemma:A}.}
We present the proof for $\om=(const.)$ case here
and explain the necessary modifications in Appendix \ref{sec:nest}
for the nonconstant case.
The proof of the constant $\om$ case is a simple 
stationary phase calculation,
while the nonconstant case is an integration by parts
similar to Section 3.4
\cite{EY2}.

For the constant $\om$ case we recall (\ref{def:R}), (\ref{def:Up})
and we write 
$$
	A = i^{m_1+1} \int_0^\infty s^{m_1} e^{is(\a -\Om_2 -\s_2\om + i\eta)}
	\int e^{-ise(\bp)}
	 \Bigg[\Upsilon_\eta\Big(
	\a-\Om_2 - \s_2 \om, \bp\Big)
	\Bigg]^{m_1} M(\bp - \bp_2,\s_2)\rd\bp\rd s
$$
after a change of variables. Using (\ref{Upsilonest}), (\ref{Upsder})
and that $M$ has decaying derivatives up to order $d$ we obtain
by stationary phase argument (see (\ref{statph}))
$$
	\Bigg| \int e^{-ise(\bp)}
	 \Bigg[\Upsilon_\eta\Big(
	\a-\Om, \bp\Big)
	\Bigg]^{m_1} M(\bp - \bp_2,\s_2)\rd\bp\Bigg|
	\leq {C^{m_1}\langle\bp_2\rangle^d \eta^{-m_1+1/2}
	 \over \langle s \rangle^{d/2}
	\langle \alpha - e(\bp_2)-\Om_2\rangle}
$$
After performing the $\rd s$ integration, we obtain (\ref{Aest})
 if $d\ge 3$. $\;\;\Box$

\medskip

The same estimate as (\ref{Aest}) is valid for the $\tbk_2$ integration
which can be performed independently. We collected a factor $(Ct)^{2m_1 -1}$
(with $\eta: = t^{-1}$) and we gain a $\langle \a-\ta\rangle^{-1}$
factor, similarly to (\ref{alphagain}). The extra $\langle\bp_2\rangle^{2d}$
factor can be compensated as before, using (\ref{telescope}).

 Then we  estimate
all the other $R_j$ and $\Upsilon$ factors in absolute value
in (\ref{eq:cer}),
we use (\ref{Upsilonest}) and (\ref{ttriv}) for $j\ge2$,
collecting $(Ct)^{2|\um|-2m_1}$. Then we
 follow the end of the  proof in Section \ref{sec:ageq3}
by integrating $\rd\bk_3, \ldots, \rd\bk_n$
and further collecting  $(Ct)^{n-2}$. Hence
the total $t$ power is $N-3$, finishing the proof of
(\ref{smallcomest}) for the $a=2$ case.
 $\;\;\;\Box$

\subsection{Bound for recollision
pairings with many peaks}
\label{sec:largepair}

Here we prove (\ref{largecomest}) of Lemma \ref{lemma:smallcom}.
The argument is similar to the proof of Proposition
\ref{prop:omconst}; in fact
the recollision will not be used.

\bigskip

We start from (\ref{def:cujpi}) and follow the argument
of Section \ref{sec:allpair} except the Schwarz inequality
(\ref{newSchw}). We obtain (compare with (\ref{eq:cer})):
$$
	C^{\um, a}_{n,N}(\pi; t) =\lambda^{2N}
	\sum_{\s_2, \ldots, \s_n; \tsi_a} \int\rd\bp_n\wh\gamma_e(\bp_n,\bp_n)
	\int \Bigg( \prod_{j=2\atop j\neq a}^n M(\bk_j, \s_j)
	\delta(\bk_j-\tbk_{\pi(j)})\rd\bk_j\rd\tbk_j\Bigg)
$$
$$
	\times \int M(\bk_a, \s_a)\rd\bk_a \int M(\tbk_a, \tsi_a)\rd\tbk_a
	\; Y_{\um,\um,\pi}(t, \bp_n, \bp_n, \ubk, \utbk, \usi \cup\{ \tsi_a\})
	+ O(|\Lambda|^{-1})\;,
$$
where the definition of $Y$ (see (\ref{2Y})) is modified as
follows. The two products over $j$ in (\ref{2Y}) start from $j=1$
and there is no $s_0, \ts_0$.
The electron momenta $\bp_j, \tbp_j$, $j\ge 1$, are given
in (\ref{def:pj}) and $\Om_j, \wt\Om_j$ are given in (\ref{def:Om})
with the  modification that $\s_{\pi^{-1}(a)}:=\tsi_a$
in the definition of $\wt\Om_j$.

After the trivial estimates (\ref{Upsilonest}) and (\ref{ttriv}) we
obtain (compare with (\ref{eq:cer1}))
$$
	\Big| C^{\um, a}_{n,N}(\pi; t)\Big|\leq (C\lambda)^{2N}
	t^{2\um} \sup_{\usi, \tsi_a} \int \rd\bp_n \wh\gamma_e(\bp_n,\bp_n)
	\int_{-\infty}^\infty \rd\a \rd\ta
$$
$$
	\times \int \Big(\prod_{j=2\atop j\neq a}^n M^*(\bk_j)
	\delta(\bk_j-\tbk_{\pi(j)})\rd\bk_j\rd\tbk_j\Big)
	\int M^*(\bk_a)\rd\bk_a
	M^*(\tbk_a)\rd\tbk_a \; \prod_{j=1}^n |R_j| \; |\tR_j|
	+ O(|\Lambda|^{-1})\;,
$$
where (\ref{def:R})--(\ref{def:Om}) are in effect.

\bigskip

This estimate is our starting point and it should be compared
with (\ref{eq:cpinew}) which was the starting point of
the proof of Proposition \ref{prop:omconst}.

For $\om=(const.)$
we follow the proof of Proposition \ref{prop:omconst}. 
Let $a_1< a_2< \ldots < a_{\kappa} \in I_a$ be the location
of the monotonic peak-sequence  as before.
We estimate all $|\tR_j|$'s trivially by $Ct$ except $\tR_{\pi(a_m)-1}$,
$m=1, 2,\ldots, \kappa$. Of course the term $\wt R_0$
is missing, which gives the extra $t^{-1}$ factor 
compared to the proof in  Section \ref{sec:constom}.

The fact that $k_1+k_a=\tbk_1+\tbk_a=0$ does not change
the dependence of $\bp_j$ and $\tbp_j$ on $\ubk$, $\utbk$ ($j\ge 1)$.
Since $\bp_0$ and $\tbp_0$ does not appear in $R_j, \tR_j$, $j\ge1$,
the only new momentum constraints due to recollision,
 $\bp_0-\bp_1 + \bp_{a-1}-\bp_a=0$ and
 $\tbp_0-\tbp_1 + \tbp_{a-1}-\tbp_a=0$, play no role.
Hence the argument in Section \ref{sec:constom} remains unchanged;
the key point being that $\tbp_{\pi(a_m)}$ was independent
of $p_{a_m-1}, p_{a_m}$ in (\ref{int:pam}).
 Therefore the $\rd\bp_{a_m}, \rd\bp_{a_m-1}$ integrations
can be done successively; starting from $a_1$ or $a_\kappa$
depending on the monotonicity of the peaks.

For $\om \neq (const.)$ we follow the  
proof of Proposition \ref{prop:omnonconst} in Appendix \ref{sec:indnon};
the proof is unchanged, just the $\rd\bp_0, \rd\tbp_0$ integrations
and the corresponding $R_0, \tR_0$ factors are missing.
$\;\;\;\Box$

\section{Computing the Wigner transform of the main term}\label{sec:wign}
\setcounter{equation}{0}

To identify the weak limit of $W^\e_{\g^{main}_K(t)}(X,V)$, we
test it against a function $J\in \cS(\bR^d\times \bR^d)$.  We fix $K\ge 5$.
We recall the definitions of $\g^{main}_K(t)$ (\ref{def:gammamain}),
$Y_{\um, \utm, \pi}$ (\ref{1Y})-(\ref{2Y}) and
$\wh J_\e(\xi, v)= \e^{-d}\wh J(\xi\e^{-1}, v)$. Then we have
the following proposition which can be proven exactly
as Proposition \ref{prop:formula}.

\begin{proposition}\label{prop:forK} For any fixed $K\ge 5$
and $J\in \cS(\bR^d\times \bR^d)$
$$
	\limsup_{L\to\infty} \Bigg| \;\Big\langle 
	J, W^\e_{\g^{main}_K(t)}\Big\rangle
	- \sum_{N, \tN=0}^{K-1}\sum_{n=0}^{\min\{ N, \tN\}}
	\sum_{\um\in \cM(n, N)}\sum_{\utm\in \cM(n, \tN)}
	\sum_{\pi\in \Pi_n} C^*_{\um, \utm, \pi}(t) \; \Bigg| =0
$$
with $\bar N : = (N+\tN)/2$ and
\be
	 C^*_{\um, \utm, \pi}(t) : =
	\lambda^{2\bar N} \sum_{\s_j\in \{ \pm \}\atop j=1,\ldots, n}
	\int \rd\bar\nu_\pi(\xi,\bv, \ubk, \utbk, \usi)
	\; Y_{\um, \utm, \pi}\Big(t; v+ \sfrac{\xi}{2}, v-  \sfrac{\xi}{2},
	\ubk, \utbk, \usi\Big)\;
\label{Cstar}
\ee
(compare with (\ref{CY})).
 The difference between
$C^*_{\um, \utm, \pi}$ and $C_{\um, \utm, \pi}$ from (\ref{CY})
is twofold. 
We replaced the measure $\rd\nu_\pi(\bp_n, \tbp_n, \ubk, \utbk, \usi)$
(see (\ref{def:nu})) with
\bey
	\rd\bar\nu_\pi(\xi,\bv, \ubk, \utbk, \usi):&=&
	\delta\Big( v_0-v-\sum_{j=1}^n \bk_j\Big)
	 \;\ov{\wh J_\e(\xi, v_0)}
	\; \wh\g_e\Big(v+\sfrac{\xi}{2},v-  \sfrac{\xi}{2}\Big)
\label{def:nubar}
\\
	&&
	\times \Big( \prod_{j=1}^n  \delta(\bk_j-\tbk_{\pi(j)}) 
	 M(\bk_j, \s_j) \rd\bk_j\rd\tbk_j\Big)\; \rd\xi \rd\bv\;,
\nonumber
\eey
and  instead of (\ref{def:pj}) we considered
$\bp_j$, $\tbp_j$  as functions of $v, \xi, \ubk, \utbk$ as follows
\be
	\bp_j = v+ \sfrac{\xi}{2}+\sum_{\ell=j+1}^n \bk_\ell\;,
	\qquad
	\tbp_j = v- \sfrac{\xi}{2}+\sum_{\ell=j+1}^n \tbk_\ell\;.
\label{newpj}
\ee
With these notations,
the definition of $R_j, \tR_j$ (\ref{def:R}) and $\Om_j, \wt\Om_j$
(\ref{def:Om}) are unchanged.  $\;\;\;\Box$
\end{proposition}

\bigskip

The next lemma gives an estimate on all
$ C^*_{\um, \utm, \pi}(t)$, and a stronger bound if $\pi$ is crossing.

\begin{lemma}\label{lemma:crossstar}
 Let $K\ge 5$, $n\leq N \leq K$, $\tn\leq\tN\leq K$,
$\um \in \cM(n, N)$, $\utm\in \cM(n, \tN)$, then for any $\pi\in \Pi_n$
\be
	\limsup_{L\to\infty} | C^*_{\um, \utm, \pi}(t)| 
	\leq {(C_a\lambda^2 t)^{\bar N} \over ( M! \; \tM!)^{a/2}}\;,
\label{trivK}
\ee
where $M:= (N+n)/2$, $\tM:= (\tN+n)/2$ and $0\leq a < 1$. Moreover, if 
$\pi\in \Pi_n$ is a crossing pairing ($\pi\neq\mbox{id}$), then
\be
	\limsup_{L\to\infty} | C^*_{\um, \utm, \pi}(t)| 
	\leq t^{-1/2} (C\lambda^2 t)^{\bar N} (\log^* t)^4\;.
\label{1crossK}
\ee
\end{lemma}

{\it Proof.} For (\ref{trivK}) we follow the proof of the 
 analogous Lemma \ref{lemma:trivbound} starting from (\ref{Cstar}).
We again split $|Y|= |Y|^{a/2}
|Y|^{1-a/2}$. The first factor
is estimated in supremum norm using
  $|Y| \leq C^{\bar N} t^{M+\tM}/( M! \,\tM!)$ (see (\ref{Ya})).
The estimate (\ref{eq:cpi6}) is modified as
$$
	|C_{\um,\utm,\pi}^*(t)|\leq
	(C\lambda)^{N+\tN}\Bigg( {t^{M+\tM}\over M!\; \tM!} \Bigg)^{a/2}
	\sup_{\usi}
	\int\rd\bar\nu_\pi^*(\xi,\bv, \ubk, \utbk)
$$
$$
	\times\Bigg\{ t^{(|\utm|-|\um|)(1-a/2)}  \Big[  \int_{-\infty}^\infty
	 \rd\a\prod_{j=0}^n
	|R_j|^{m_j+1}\Big]^{2-a} + t^{(|\um|-|\utm|)(1-a/2)}
	 \Big[\int_{-\infty}^\infty
	 \rd\ta  \prod_{j=0}^n
	|\tR_j|^{\tm_j+1}\Big]^{2-a} \Bigg\} \;.
$$
with a modification of $\rd\bar\nu_\pi$
\bey
	\rd\bar\nu_\pi^*(\xi,\bv, \ubk, \utbk):&=&
	\delta\Big( v_0-v-\sum_{j=1}^n \bk_j\Big)
	 \;|\wh J_\e(\xi, v_0)|
	\; \Big| \wh\g_e\Big(v+\sfrac{\xi}{2},v-  \sfrac{\xi}{2}\Big) \Big|
\label{def:barnustar}
\\
	&&
	\times \Big( \prod_{j=1}^n \delta(\bk_j-\tbk_{\pi(j)}) 
	 M^*(\bk_j) \rd\bk_j\rd\tbk_j\Big)\;\rd\xi \rd\bv\;.
\nonumber
\eey

The rest of the proof goes through until (\ref{trivfin}), just
$\bp_n$ is replaced with $v+\sfrac{\xi}{2}$ and $\tbp_n$ with
$v-\sfrac{\xi}{2}$,  i.e.,
(\ref{newpj}) is used instead of  (\ref{def:pj}),
the $t$-powers are adjusted and $\rd\mu$  is redefined as
$$
	\rd\bar\mu(\xi, v, \ubk): =  \;|\wh J_\e(\xi, v_0)|
	\; \delta\Big(v_0-v-\sum_{j=1}^n\bk_j\Big)\;
	\;\Big|\wh\g_e\Big(v+ \sfrac{\xi}{2},v-  \sfrac{\xi}{2}\Big)\Big|
	\Big( \prod_{j=1}^n M^*(\bk_j) \rd\bk_j\Big) \; \rd\xi\rd v\;.
$$
Let $J_\e^*(\xi):= \sup_\bw |\wh J_\e(\xi, \bw)|$.
After  the $\rd\bk_n$ integration in 
the analogue of (\ref{trivfin}), we arrive at
\bey
	|C_{\um,\utm,\pi}^*(t)|&\leq&
	(C\lambda)^{N+\tN}\Bigg( {t^{M+\tM}\over M!\; \tM!} \Bigg)^{a/2}
	 t^{(|\utm|+|\um|)(1-a/2) + (1-a)n} 
	 \int \rd\xi\rd v  \; J_\e^*(\xi)
	\;\Big|\wh\g_e\Big(v+ \sfrac{\xi}{2},v- \sfrac{\xi}{2}\Big)\Big|
\nonumber\\
	&\leq& {(C_a\lambda^2 t)^{\bar N} \over ( M! \; \tM!)^{a/2}}
	 \int \rd\xi\rd v  \; J_\e^*(\xi)
	\;\Big[\wh\g_e\Big(v- \sfrac{\xi}{2},v-  \sfrac{\xi}{2}\Big)
	+\wh\g_e\Big(v+ \sfrac{\xi}{2},v+  \sfrac{\xi}{2}\Big) \Big]\;,
\label{tveg}
\eey
using  $|\wh\g_e(\bp, \bp')|\leq \sfrac{1}{2}
\Big[ \wh\g_e(\bp, \bp) + \wh\g_e(\bp', \bp')
\Big]$ by $\g_e\ge0$. Then (\ref{trivK}) follows from
(\ref{jeps}) and (\ref{gammadecay}). 

\bigskip

The proof of (\ref{1crossK}) requires very similar modifications
along the proof of the analogous Lemma
\ref{lemma:onecross}. The estimate (\ref{eq:cpinew}) is modified as
$$
	| C^*_{\um, \utm, \pi}(t)| 
	\leq (C\lambda)^{N+\tN} t^{|\um| + |\utm|}\sup_{\usi}
	\int \rd \bar\nu_\pi^* (\xi, v, \ubk, \utbk)
	\int_{-\infty}^\infty \rd \a \prod_{j=0}^n |R_j|
	\int_{-\infty}^\infty \rd \ta \prod_{j=0}^n |\tR_j|\;,
$$
keeping in mind (\ref{newpj}).

It is easy to check that all estimates in the proof of Lemma
\ref{lemma:onecross}
go through, since they were always valid uniformly in the
incoming electron momenta denoted by $\bp$ in (\ref{prot:Ct}),
 (\ref{prot:Clog}), (\ref{prot:Clogalpha}),
(\ref{prot:plus}) and (\ref{prot:Ctdenom}), hence a shift $\pm\sfrac{\xi}{2}$
does not make a difference.
Finally, instead of (\ref{int:alpha3}) we have
\bey
	\lefteqn{\int_{-\infty}^\infty
	 {\rd \a \rd \ta \over \langle \a \rangle \langle \ta \rangle}
	\int\rd\xi\rd v  \; J_\e^*(\xi)
	 {\Big|\wh\g_e\Big(v+ \sfrac{\xi}{2},v-  \sfrac{\xi}{2}\Big)\Big|
	\Big\langle v+\sfrac{\xi}{2}\Big \rangle^2
	\Big\langle v-\sfrac{\xi}{2} \Big\rangle^2 \over
	\Big| \a - e\Big( v +\sfrac{\xi}{2}\Big)+ i\eta\Big|\; 
	\Big| \ta - e\Big( v -\sfrac{\xi}{2}\Big)+ i\eta\Big|}}
\nonumber\\
	&\leq& (C\log^*t)^2\int\rd\xi\rd v  \; J_\e^*(\xi)
	\Bigg[ \wh\g_e\Big(v- \sfrac{\xi}{2},v-  \sfrac{\xi}{2}\Big)
	\langle v - \sfrac{\xi}{2} \rangle^4 
	+  \wh\g_e\Big(v+ \sfrac{\xi}{2}, v+  \sfrac{\xi}{2}\Big)
	\langle v + \sfrac{\xi}{2} \rangle^4 \Bigg]
	 \langle\xi  \rangle^4 
\nonumber\\
	&\leq& (C\log^*t)^2 \;. 
\nonumber\eey
 We again used (\ref{gammadecay}) and 
$$
	\sup_{\e\le1} \int  J_\e^*(\xi)\langle\xi  \rangle^4 \rd \xi=
	\sup_{\e\le1}\int \sup_v |\wh J_\e(\xi, v)| \langle\xi  \rangle^4
	 \rd \xi < \infty\;.
$$
This is  an extension of (\ref{jeps}),
and it is clearly valid for $J\in \cS(\bR^{d}\times \bR^d)$.  $\;\;\;\Box$

\bigskip

{F}rom (\ref{1crossK}) we know that only the direct pairing
counts. To compute $C^*_{\um, \utm, id}$, we use the second formula for
$Y_{\um,\utm,id}$ (\ref{2Y}) and that $\Om_j = \wt\Om_j$
\bey
	 C^*_{\um, \utm, id}(t) : &=&
	\lambda^{2\bar N} \sum_{\s_j\in \{ \pm \}\atop j=1,\ldots, n}
	\int  \rd\bar\nu_{id}(\xi, v, \ubk, \utbk, \usi) 
	e^{2t\eta}
	\int_{-\infty}^\infty \rd\a\; e^{-it\a}
\label{Cstarid}
\\
	&&\times\prod_{j=0}^n R_j^{m_j+1}
	\Big[ \Upsilon_\eta(\alpha-\Om_j, \bp_j) \Big]^{m_j}
	\int \rd\ta \; e^{it\ta} \prod_{j=0}^n
	 \tR_j^{m_j+1}
	\Big[ \overline{\Upsilon}_\eta(\ta- \Om_j, \tbp_j) 
	\Big]^{\tm_j}\;.
\nonumber
\eey

\medskip

Introduce
\be
	\bv_j : = \bv +  \sum_{m =j+1}^n \bk_m\; , \qquad m=1, 2,\ldots, n-1
\label{def:v}\ee
and $\bv_n:=\bv$. Note that $\bp_j = \bv_j + \sfrac{\xi}{2}$.
We show that $\Upsilon_\eta(\alpha-\Om_j, \bp_j)$
and $\Upsilon_\eta(\ta-\Om_j, \tbp_j)$ can be
replaced with
$$
	 \Psi_j:= \Upsilon_{0+}( e(\bv_j), \bv_j)\;,  
	\quad \mbox{and}\quad
	 \ov{\Psi}_j= \ov{\Upsilon_{0+}( e(\bv_j), \bv_j)}
$$
modulo a negligible error (recall (\ref{Upslim})).
 Notice that $\Psi_j$ depend
on $\bv, \xi, \bk_{j+1},\ldots, \bk_n$.

\begin{lemma}\label{lemma:replace}
\be
	 C^*_{\um, \utm, id}(t)  = C^{**}_{\um, \utm, id}(t) 
	+ O\Big((C\lambda^2t)^{\bar N} t^{-1/2}(\log^*t)^4\Big)
\label{replace}
\ee
with
\bey
	C^{**}_{\um, \utm, id}(t) :&=&
	\lambda^{2\bar N}e^{2t\eta} \sum_{\s_j\in \{ \pm \}\atop j=1,\ldots, n}
	\int  \rd\bar\nu_{id}(\xi, v, \ubk, \utbk, \usi)
\label{Cstarstar}
\\
	&&\times \int_{-\infty}^\infty \rd\a\; e^{-it\a} \prod_{j=0}^n
	 R_j^{m_j+1} \Psi_j^{m_j}
	\int \rd\ta \; e^{it\ta} \prod_{j=0}^n
	 \tR_j^{m_j+1} \ov{\Psi}_j^{\tm_j} \;.
\nonumber
\eey
We also have
\be
	\limsup_{L\to\infty} |C^{**}_{\um, \utm, id}(t) |
	\leq {(C_a\lambda^2 t)^{\bar N} \over ( M! \; \tM!)^{a/2}}\;.
\label{trivstarK}
\ee
\end{lemma}

{\it Proof of Lemma \ref{lemma:replace}.}
 The estimate (\ref{trivstarK}) is proven exactly as
(\ref{trivK}).
For (\ref{replace}), we compute the difference  $ C^* - C^{**}$.
We obtain a similar formula as (\ref{Cstarid}), just  
$\prod_j\Upsilon(\ldots)^{m_j}$
is replaced with $\prod_j \Upsilon(\ldots)^{m_j} - \prod_j \Psi_j^{m_j}$
and similarly for the factors with conjugate.
We then estimate this expression exactly as in the proof of
(\ref{trivK}) with $a=0$.

Let $\theta:\bR\to\bR$ be defined as $\theta(s):= |s|$ for 
$|s|\leq 1$ and $\theta(s)\equiv 1$ for $|s|\ge 1$.
Using  (\ref{eq:econd}), (\ref{Upsilonest}) and
 (\ref{derup}), we see by a telescopic estimate that
\be
	\Bigg|\prod_{j=0}^n \Big[\Upsilon_\eta(\alpha-\Om_j, \bp_j)
	\Big]^{m_j} - \prod_{j=0}^n \Psi_j^{m_j}\Bigg|
	\leq C^n\eta^{-1/2} \sum_{j=0}^n \Bigg[\theta\Big( \a - \Om_j-
	e(\bp_j)\Big) + O\Big(|\xi|(\langle \bp_j\rangle +\langle\xi
	\rangle) \Big)\Bigg]\; .
\label{telest}
\ee

\medskip

For the first term in (\ref{telest}) we notice that
$$
	 \theta\Big( \a - \Om_j-e(\bp_j)\Big)\; |R_j|
	\leq {C\over \langle \a -\Om_j- e(\bp_j) \rangle}\;,
$$
in other words, the net effect of a factor $\theta\Big( \a - \Om_j-
e(\bp_j)\Big)$ is that it neutralizes the singularity
of an $|R_j|$ factor and still keeping its decay property.
Hence, effectively, $m_j$ is decreased by one (see (\ref{ttriv})).
We gain a factor $t$ from this decrease
(notice the dependence of the $t$ power on $|\um|$ in (\ref{tveg})),
but we lose $\eta^{-1/2}=t^{1/2}$ in (\ref{telest}).
This gives the extra factor $t^{-1/2}$ in 
the error term of (\ref{replace}) relative to the
robust estimate (\ref{trivK}). 
The special case $m_j=0$  require small modifications which
 we leave  to the reader.

For the second term in (\ref{telest}) we use that $\int |\xi|
\langle\xi\rangle \wh J_\e^*(\xi)
\rd\xi = O(\e)$ to gain $\eta^{-1/2}\e = t^{-1/2}$. The factor
$\langle \bp_j\rangle$ can be absorbed  into the decay
of $M^*$ and $\gamma_e$ as before.
$\;\;\;\Box$.

\bigskip

Combining Proposition \ref{prop:forK},  Lemma \ref{lemma:crossstar}
and Lemma \ref{lemma:replace} we obtain

\begin{proposition}\label{onlydir}
For $K\ge 5$, %J\in \cS(\bR^d\times \bR^d)$
$$
	\limsup_{\e\to0}\limsup_{L\to\infty} \Bigg| \Big\langle 
	J, W^\e_{\g^{main}_K(t)}\Big\rangle
	- \sum_{N, \tN=0}^{K-1}\sum_{n=0}^{\min\{ N, \tN\}}
	\sum_{\um\in \cM(n, N)}\sum_{\utm\in \cM(n, \tN)}
 	C^{**}_{\um, \utm, id}(t) \Bigg| =0\;. \qquad\Box
$$
\end{proposition}

\bigskip

Using the factorials in the estimate (\ref{trivstarK}),
 we can extend the summations over all $\um, \utm$; the error
term goes to zero as $K\to\infty$:
$$
	\limsup_{K\to\infty}
	\limsup_{\e\to0}\limsup_{L\to\infty} \Bigg| \Big\langle 
	J, W^\e_{\g^{main}_K(t)}\Big\rangle
	- \sum_{n=0}^{\infty} C(n,t)
	 \Bigg| =0\;,
$$
where
$$
	C(n,t): = \sum_{m_0,\ldots, m_n=0}^\infty
	\sum_{\tm_0,\ldots, \tm_n=0}^\infty
 	C^{**}_{\um, \utm, id}(t)\;.
$$

The summations over $m_j, \tm_j$ give a geometric series in
(\ref{Cstarstar}). We obtain a shift in
the denominators of $R_j$ ({\it "one loop renormalization"}):
\be
	\sum_{m_j=0}^\infty R_j^{m_j+1}(\lambda^2\Psi_j)^{m_j}
	= {1\over \a - e(\bp_j) -\Om_j -\lambda^2\Psi_j +i\eta}\;.
\label{eq:resum}\ee
We can change back the $\rd\a$ and $\rd\ta$ integrations
into subsequent time integrations
since $\Psi$'s are independent of them (see the similar
identity in (\ref{eq:timeint})).
 We obtain
\bey
	C(n,t)&=& \lambda^{2n} \sum_{\s_j\in \{ \pm \}\atop j=1,\ldots, n}
	\int  \rd\bar\nu_{id}(\xi, v, \ubk, \utbk, \usi)    
	\int_0^{t*} [\rd s_j]_0^n  \int_0^{t*}[\rd \ts_j]_0^n
\label{Cnt}
\\
	&&\times
	\prod_{j=0}^n\exp\Bigg\{
	 -i \Big[s_j \Big( e(\bp_j) +\Om_j +\lambda^2\Psi_j\Big)
	- \ts_j \Big ( e(\tbp_j)
	  +\Om_j +\lambda^2\ov{\Psi}_j\Big)\Big]\Bigg\}\;.
\nonumber
\eey

\bigskip

The rest is similar to the end of Section 4 in \cite{EY2} and we
will skip a few technical details.

\medskip

We introduce the new time variables
 $a_j:= (s_j+\ts_j)/2$, $b_j:=(s_j-\ts_j)/2$ for $j=0,1, \ldots, n$ and
we have
\bey                                       
	\lefteqn{
	C(n,t)= 2^n\lambda^{2n} \sum_{\s_j\in \{ \pm \}\atop j=1,\ldots, n}
	\!\!\int \Big( \prod_{j=0}^n \rd\bv_j\Big)\rd\xi
	 \; \ov{\wh J_\e(\xi, v_0)}
	\int \Big( \prod_{j=0}^{n-1}  M(\bv_{j}-\bv_{j+1}, \s_{j+1}) 
	\Big)
	\wh\g_e\Big(v_n+ \sfrac{\xi}{2},v_n-  \sfrac{\xi}{2}\Big)}
\nonumber\\
	&&\times  \int_0^t  \Big(\prod_{j=0}^n \rd a_j\Big)
	 \delta\Big( t-\sum_{j=0}^n 
	a_j\Big) \exp{\Bigg\{ -i\xi\cdot \sum_{j=0}^n a_j\nabla e(\bv_j)
	+ 2\sum_{j=0}^n a_j \Big(\lambda^2\mbox{Im} \Psi_j + O(\xi^2)\Big) 
	\Bigg\}}
\nonumber\\
	&&\times \prod_{j=0}^n \Big(\int_{-a_j}^{a_j}  \rd b_j\Big)
	 \delta\Big( \sum_{j=0}^n b_j\Big)
	\prod_{j=0}^n \exp\Bigg\{
	 -2i  b_j \Big[ e(\bv_j)+\Om_j +\lambda^2 \mbox{Re} 
	\Psi_j  + O(\xi^2)\Big]\Bigg\}\; .
\nonumber
\eey

Here we used $\bp_j = \bv_j +\sfrac{\xi}{2}$, $\tbp_j = \bv_j -\sfrac{\xi}{2}$
and the second order Taylor expansion of $e(\bv)$
$$
	e(\bv_j \pm \sfrac{\xi}{2}) = e(\bv_j) \pm  \sfrac{\xi}{2}\cdot
	\nabla e(\bv_j) + O(\xi^2)
$$
with a uniform error bound (see (\ref{eq:econd})).

We rescale all microscopic variables into macroscopic ones, i.e.,
$\a_j:= \e a_j$, $\zeta = \e^{-1}\xi$, $t=\e^{-1}T$ and recall
that $\e=\lambda^2$. We define
$$
	\chi_{\ua} (\ub) := \chi\Bigg( -a_0\leq \sum_{j=0}^{n-1}b_j
	\leq a_0\Bigg) \prod_{j=0}^{n-1}\chi(-a_j\leq b_j\leq a_j)
$$
for any $\ua:=(a_0, a_1, \ldots, a_n)$ and $\ub:=(b_0, b_1, \ldots, b_{n-1})$.
Then
\bey
	C(n,t)&=&  \sum_{\s_j\in \{ \pm \}\atop j=1,\ldots, n}
	\int \rd\zeta \int \Big( \prod_{j=0}^n \rd\bv_j\Big)
	 \; \ov{\wh J(\zeta, v_0)}
	\int \Big( \prod_{j=0}^{n-1}  M(\bv_{j}-\bv_{j+1}, \s_{j+1}) 
	\Big)
\nonumber\\
	&&\times  \int_0^T \Big( \prod_{j=0}^n \rd \a_j\Big)
	 \delta\Big( T-\sum_{j=0}^n 
	\a_j\Big) \int_{-\infty}^\infty \rd\ub \; \chi_{\e^{-1}\ualpha} (\ub)
\nonumber\\
	&&\times\prod_{j=0}^{n-1} 2 \exp\Bigg\{
	 -2i b_j \Big[ e(\bv_j)-e(\bv_n)+\Om_j +\e (\mbox{Re} 
	\Psi_j - \mbox{Re} \Psi_n)
	+ O(\e^{2}\zeta^2)\Big]\Bigg\}
\nonumber\\
	&&
	\times \; \exp{\Bigg\{ -i\zeta\cdot \sum_{j=0}^n \a_j\nabla e(\bv_j)
	+2\sum_{j=0}^n \a_j \Big(\mbox{Im} \Psi_j + O(\e\zeta^2)\Big) 
	\Bigg\}}\; \wh W_{\gamma_e}(\e \zeta, \bv_n)\;.
\nonumber
\eey
It is easy to see that the error terms are negligible as $\e\to0$
since $T$ is
fixed and $\zeta$ is essentially bounded by the decay of $\wh J$.
The $\rd b_j$ integrations give $2\pi\delta\Big( e(\bv_j)-e(\bv_n)+\Om_j\Big)$
delta functions on the energy shell as $\e\to0$.

Following the analogous  calculations on pp. 709-710 of \cite{EY2},
using the macroscopic profile of the initial state (\ref{initlimit})
and recalling the definition of the Boltzmann collision kernel (\ref{phonker})
$$ 
	\s(\bV,\bU):=
	 2\pi \sum_{\s =\pm} M(\bV-\bU, \s) \delta\Big(e(\bV)-e(\bU)
	 +\s\omega(\bV-\bU)\Big)\;,
$$
we obtain ($\bV_j\equiv \bv_j$)
\bey
	\lim_{\e\to0} \lim_{L\to\infty} C(n, \e^{-1}T)
	&=& \int \rd\bX \rd\bV_0 \ldots \rd\bV_n \ov{J(\bX, \bV_0)}
	\int_0^T \Bigg( \prod_{j=0}^n 
	e^{2\a_j \mbox{Im}\Psi_j} \rd\a_j\Bigg)
	\delta\Bigg( T-\sum_{j=0}^n \a_j\Bigg)
\nonumber\\
	&&\times \Bigg(\prod_{j=0}^{n-1} \sigma(\bV_j, \bV_{j+1})\Bigg)
	F_0\Bigg( X- \sum_{j=0}^n \a_j \nabla e(\bV_j),  \; \bV_n\Bigg)\;.
\label{integB}
\eey
We also rearranged the energy conservation delta functions as
$$
	\prod_{j=0}^{n-1} \delta\Big( e(\bv_j)-e(\bv_n)+\Om_j\Big)
	=  \prod_{j=0}^{n-1} \delta\Big( e(\bv_j)-e(\bv_{j+1})+\s_{j+1}
	\om(\bv_j-\bv_{j+1})\Big) \;.
$$
Noticing that
$$
	-2 \mbox{Im} \Psi_j = \int \sigma(\bU, \bV_j)\rd \bU
$$
by (\ref{Upszero}), we see that (\ref{integB}) is exactly
the $n$-th order term in the
Dyson series solution of (\ref{eq:Beq}). This completes
the proof of the Main Theorem. $\;\;\;\Box$

\appendix

\section{Case of general $\omega$}\label{sec:staircase}
\setcounter{equation}{0}

Here we show how to modify the proof of Lemma \ref{lemma:main}
if  $\om$ is not constant.
The assumption that $\om= (const.)$ was used only in Sections
\ref{sec:norecbign} and  \ref{sec:onereabs}.

The proof of (\ref{norecest}) for $\om \neq (const.)$
 will be given first. The necessary combinatorial structure
is introduced in Section \ref{sec:gencomb} and the key
analytic estimate (Proposition \ref{prop:omnonconst}) is
proven in Section \ref{sec:indnon}. These sections show
how to modify the argument in Section \ref{sec:norecbign}.
The  modifications for the proof of (\ref{norecampest})
were already explained in Section \ref{sec:ampnorec}
both for constant and nonconstant $\omega$.

Most of the proof of (\ref{1recampestor}) in Section \ref{sec:onereabs}
is valid for arbitrary $\om$. The assumption $\om =(const.)$
was used only when we estimated the nested pairing
in Section \ref{sec:nestconst}  (proof of Lemma
\ref{lemma:A}). The technical modifications for
nonconstant $\om$ are given in Appendix \ref{sec:nest}.

\subsection{Combinatorics for general $\om$}\label{sec:gencomb}

Since the momentum-dependence structure is more complicated, 
in order to perform the succesive integration of Section  \ref{sec:norecbign}
 we need more control on
the structure of the pairing.

\begin{definition} Fix a pairing $\pi\in\Pi_n$.
 A sequence of consecutive numbers $a, a+1, a+2, \ldots, a+h\subset\{ 1, 2,
 \ldots, n\}$
is called a {\bf down-stair} of length $h$ if $\pi(a) >\pi(a+1)
> \ldots > \pi(a+h)$;  and it is called an {\bf up-stair}
 of length $h$ if $\pi(a) <\pi(a+1)
< \ldots< \pi(a+h)$. We always assume that $h\ge 1$.
\end{definition}
 Notice that the elements in a
stair must be consecutive, it is not sufficient
if they  just form a monotonic subsequence.

\begin{definition}\label{def:incrstair}
A set of $\kappa$ up-stairs;
$$
	a_1, a_1 + 1, \ldots, a_1 + h_1\;;
$$
$$
	a_2, a_2 + 1, \ldots, a_2 + h_2\;;
$$
$$
	\vdots
$$
$$
	a_\kappa, a_\kappa + 1, \ldots, a_\kappa + h_\kappa
$$
is called an {\bf increasing $\kappa$-staircase}
if it satisfies the following properties:

(i) $a_j + h_j < a_{j+1}$ for $j=1, 2, \ldots, \kappa-1$;

(ii) Each $a_j+h_j$ is a peak;

(iii) The $\pi$-images of the peaks are increasing: 
$$
	\pi(a_1+h_1)< \pi(a_2+h_2)<\ldots < \pi(a_\kappa+h_\kappa)\;;
$$

(iv) There is no peak $p$ with $a_j < p < a_{j+1}$, $\pi(p)> \pi(a_j+h_j)$,
$j=1, 2, \ldots, \kappa-1$;

(v) The images of the stairs minimally overlap, i.e.,
$$
	\pi(a_{j+1} +1) > \pi(a_{j}+h_j) >
	 \pi(a_{j+1}) \qquad j= 1, 2,
	\ldots, \kappa-1\;.
$$

The numbers $a_1+h_1, a_2+h_2, \ldots, a_\kappa+h_\kappa$
 are  called the {\bf tips} of
the stairs, the numbers $a_1, a_2, \ldots, a_\kappa$ 
are the {\bf bottoms} of the stairs.
\end{definition}

\bigskip\bigskip
\centerline{\epsffile{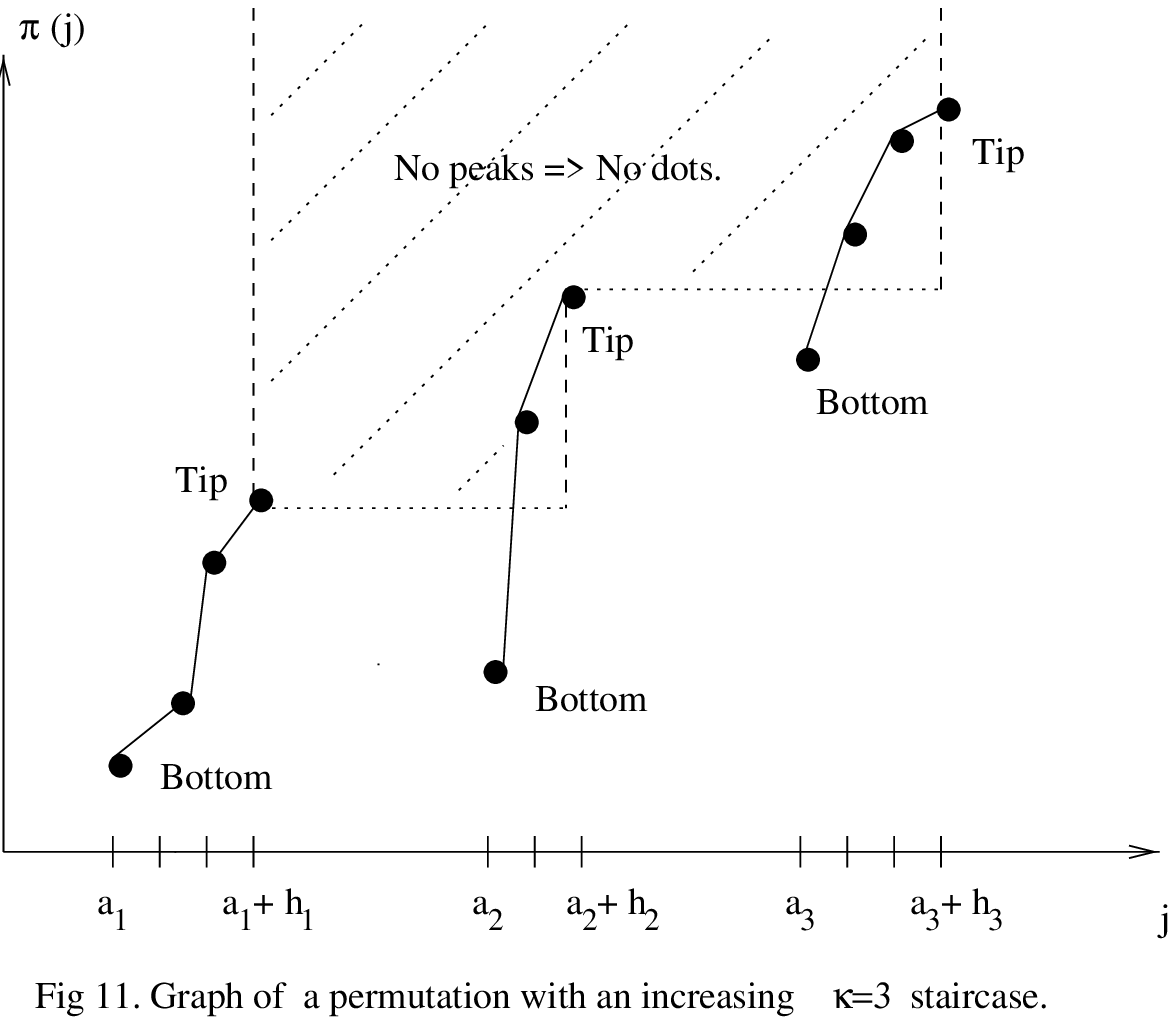}}
\bigskip

In other words, an increasing $\kappa$-staircase consists
of $\kappa$ up-stairs with lengths $h_1, h_2, \ldots, h_\kappa$
in such a way that every up-stair ends just above where
the next one starts, i.e., if we removed the bottom
from each stair then the stairs would not overlap.
(See Fig.~11). Moreover, there is no peak between two
stairs which would be higher than the tip of the lower stair.
This implies, in particular, that there is no dot at
all in the shaded regime, i.e., there is no $p$ with
$a_j < p < a_{j+1}$, $\pi(p)>\pi(a_j+h_j)$, $j=1, 2, \ldots , \kappa-1$.
Note that the bottoms are not necessarily valleys.

Similarly we define:

\begin{definition}\label{def:decrstair}
A set of $\kappa$ down-stairs;
$$
	a_1, a_1 + 1, \ldots, a_1 + h_1\;;
$$
$$
	a_2, a_2 + 1, \ldots, a_2 + h_2\;;
$$
$$
	\vdots
$$
$$
	a_\kappa, a_\kappa + 1, \ldots, a_\kappa + h_\kappa
$$
is called a {\bf decreasing $\kappa$-staircase}
if it satisfies the following properties:

(i)  $a_j + h_j < a_{j+1}$ for $j=1, 2, \ldots, \kappa-1$;

(ii) Each $a_j$ is a peak;

(iii) The $\pi$-images of the peaks are decreasing: 
$$
	\pi(a_1)> \pi(a_2)>\ldots > \pi(a_\kappa)\; ;
$$

(iv) There is no peak $p$ with $a_j < p < a_{j+1}$, $\pi(p)> \pi(a_{j+1})$,
$j=1, 2, \ldots, \kappa-1$;

(v) The images of the stairs minimally overlap, i.e.,
$$
	\pi(a_j + h_j -1) > \pi(a_{j+1}) > \pi(a_j+h_j)\;, \qquad j= 1, 2,
	\ldots, \kappa-1\;.
$$
%The numbers $a_1, a_2, \ldots, a_\kappa$ are called the {\bf tips} of
%the stairs, the numbers $a_1+h_1, a_2+h_2, \ldots, a_\kappa+h_\kappa$ 
%are the {\bf bottoms} of the stairs.
\end{definition}

%In other words, a decreasing $\kappa$-staircase consists
%of $\kappa$ down-stairs with lengths $h_1, h_2, \ldots, h_\kappa$
%in such a way that every down-stair ends just below where
%the next one starts, i.e., if we removed the bottom
%from each stair then the remaining stairs would not overlap
% (see Fig.~12). Moreover, there is no peak between two
%stairs which would be higher than the tip of the lower stair.
%This implies, in particular, that there is no dot at
%all in the shaded regime in Fig.~12, i.e., there is no $p$ with
%$a_j < p < a_{j+1}$, $\pi(p)>\pi(a_{j+1})$, $j=1, 2, \ldots , \kappa-1$.
%Note that the bottoms are not necessarily valleys.

\bigskip\bigskip
\centerline{\epsffile{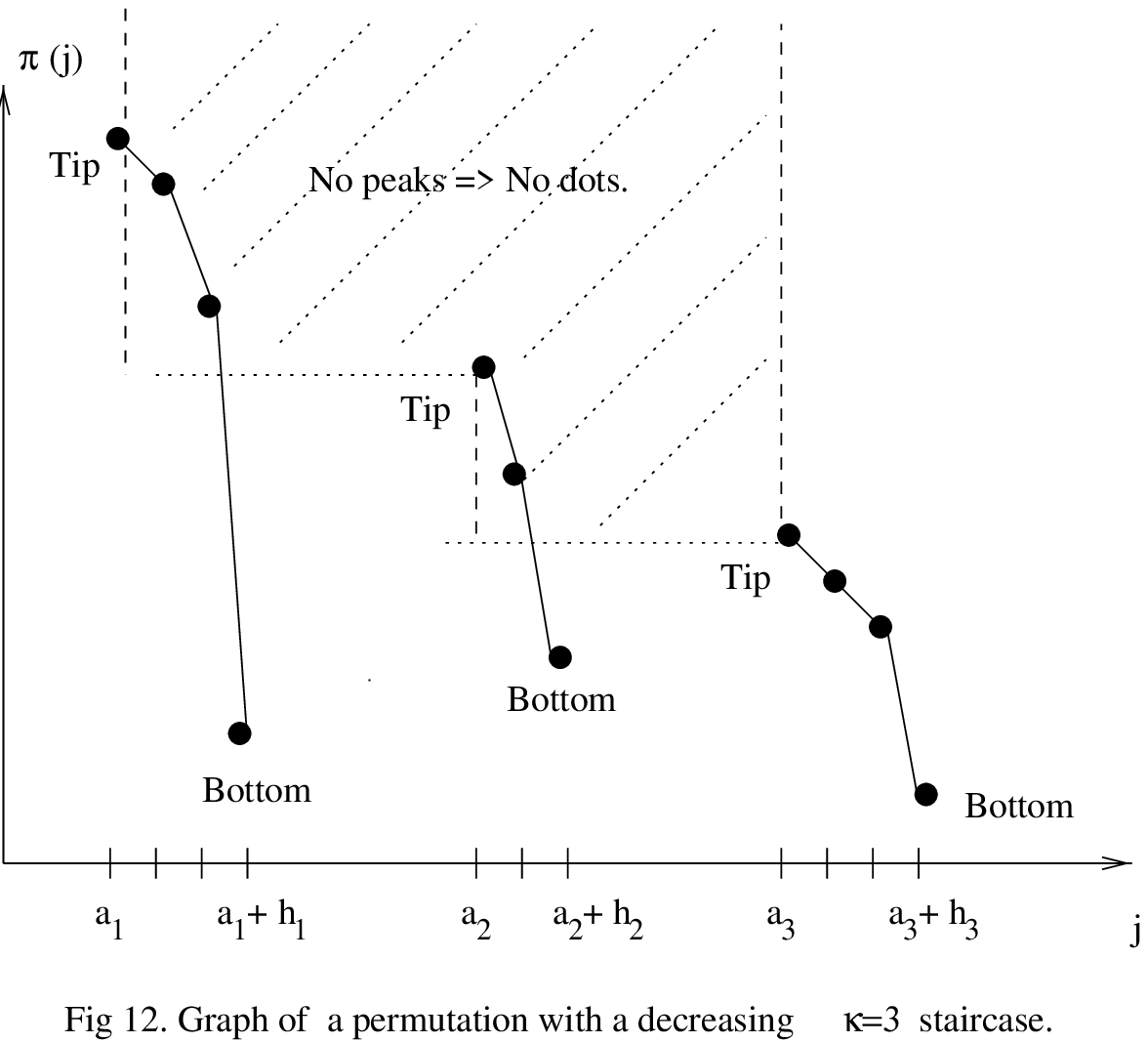}}
\bigskip

The following Proposition shows that essentially every permutation
has either an increasing or a decreasing $\kappa$-staircase.
This is again a Ramsey type theorem.
 However, the estimate is very bad in $\kappa$,
so practically it can  be used for finite $\kappa$ only.

\begin{proposition}\label{prop:count}
For any fixed $\kappa$, with the exception of
$$
	n^{4\cdot 4^{\kappa-1}+3}(2\cdot 4^{\kappa-1}+2)^n
$$
pairings, a pairing $\pi\in \Pi_n$ either contains
an increasing or a decreasing   $\kappa$-staircase (or both).
\end{proposition}

{\it Proof.} We need the following lemma whose proof is 
given in the Appendix \ref{app:comb}.

\begin{lemma}\label{lemma:ramsey}
Suppose that a permutation has at least
${\a+\beta \choose \a} $
peaks. Then it has either an increasing $(\a+1)$-staircase
or it has a decreasing $(\beta +1)$-staircase (or both).
\end{lemma}

{F}rom this Lemma and
 Lemma \ref{lemma:number} the proof of Proposition
\ref{prop:count} follows easily. Choose $\a=\beta = \kappa-1$, and
use that
${2\kappa-2\choose \kappa-1}   <  4^{\kappa-1}$. $\;\;\; \Box$

\bigskip

In the next section we will prove

\begin{proposition}\label{prop:omnonconst} Let $n\leq N$,
$\um,\utm\in \cM(n, N)$.
Let $\om\neq(const.)$, satisfying the assumptions in Section
 \ref{sec:assump},
 and assume  that $\pi\in \Pi_n$
has either an increasing $\kappa$-staircase or
a decreasing  $\kappa$-staircase. Then
$$
	|C_{\um, \utm,\pi}(t)| \leq t^{-\kappa+2}
	 (C\lambda^2 t)^N (\log^* t)^{n+\kappa +2}\;.
$$
\end{proposition}

{F}rom these statements (\ref{norecest}) follows for
the $\om\neq(const.)$ case. Choose $\kappa=8$, then
 all contributions which have a monotonic 8-staircase
 can be included into the second term in  (\ref{norecest}).
The exceptional $n^{4\cdot 4^7 + 3}(2\cdot 4^7 +2)^n \leq C^n$
pairings can be estimated by the apriori
bound from Lemma \ref{lemma:trivbound} 
and be included in the first term
in (\ref{norecest}).

\bigskip

{\it Remark:}
In general, we gain $t^{-\kappa+2}$ with the exception of 
$n^{4\cdot 4^{\kappa-1} + 3}(2\cdot 4^{\kappa-1}+2)^n$
pairings.

\bigskip

\subsection{Estimate of the indirect term
for the nonconstant  $\om$ case.}\label{sec:indnon}

{\it Proof of
 Proposition \ref{prop:omnonconst}.}
The difficulty is that there are momentum dependences
within the argument of $\om$.
This makes it impossible to eliminate all the designated $\tR$ factors
 by integrating out the peak variables as in (\ref{peaks}).
In Section \ref{sec:constom} when we
integrated out $\bp_{a_m-1}$ and $\bp_{a_m}$, we used that
only three denominators (two $R$ and one $\tR$ factors)
depended on these variables 
(see (\ref{int:pam})). In case of nonconstant $\om$,
still only these three denominators depend on
$\bp_{a_m-1}$ and $\bp_{a_m}$ as far as the electron
kinetic energy is concerned, but  due to
the cumulation of $\om$'s 
in the form of $\pm\om(\bp_{\ell+1} -\bp_{\ell+2})
 \pm\om(\bp_{\ell+2} -\bp_{\ell+3})
\ldots$ in $R_\ell$, all $R_j$ factors 
depend on $\bp_{a_m-1}$ and $\bp_{a_m}$ for $j<a_m$.

Hence we have to integrate out all $\bp_j$ in consecutive
order: $\bp_1, \bp_2, \ldots$.
This means that the designated
$\tR_{\pi(a_m)-1}$ factors are eliminated
in an alternating order with the regular $R_j$ factors.
One has to ensure that when we integrate out $\bp_j$ with
some $a_{m-1} < j < a_{m}-1$, the designated
$\tR_{\pi(a_i)-1}$, $i\ge m$, denominators are not affected
since we cannot integrate out a variable which appear in
many $\tR$ factors.

The staircase construction of Section \ref{sec:gencomb} is
designed to ensure these restricted dependences so
that the $\bp_1, \bp_2, \ldots$ integrations could be  done
in this order.

There is one additional difficulty: the staircase controls the ordering
of the momenta between the first and the last stairs, but
it does not control the ordering before and after the
staircase. The natural idea to consider the "first"
staircase does not work (the first staircase could be too short).

So we have to separate the staircase from the rest,
unless $a_1=1$ or $\pi(a_1)=1$.
We use the semigroup property of the measure
$\int^{t*} [\rd s_j]_0^n$:
\be
	\int_0^{t*} [\rd s_j]_0^{n+m}
	= \int_0^t\rd s \Big( \int_0^{s*} [\rd s_j]_0^{n}\Big)
	\Big( \int_0^{(t-s)*} [\rd s_j]_{n+1}^{m}\Big) \;,
\label{semigroup}
\ee
and we will lose a $t$-factor by estimating the outer integral
trivially.
This is why the estimate in Proposition \ref{prop:omnonconst}
is weaker by a $t^2$ factor than that in 
Proposition \ref{prop:omconst}.

\subsubsection{Estimate for increasing staircase}

Assume that $\pi$ has an increasing $\kappa$-staircase
with $a_j$ bottoms and $a_j+h_j$ tips for $1\leq j \le \kappa$
(see Definition \ref{def:incrstair}). 
Our starting point is (\ref{eq:cpi3}). 
We use (\ref{semigroup}) for the time integrals in (\ref{eq:cpi3})
as follows:
\bey
	\lefteqn{\int_0^{t*} [\rd s_j]_0^N
	 \Bigg[ \prod_{j=0}^n \prod_{b\in I_j\cup I_j^c}
	e^{-is_b [e(\bp_{\mu(j)}+\chi_b\bk_b)+
	\Om_j+\chi_b\s_b\om(\bk_b)]} \Bigg]}
\nonumber\\
%$$
%	 \times\int_0^{t*} [\rd \ts_j]_0^N
%	\Bigg[
%	 \prod_{j=0}^n  \prod_{\tb\in \tI_j\cup\tI_j^c}
%	e^{i\ts_\tb [e(\tbp_{\tmu(j)}+\tchi_\tb\tbk_\tb)
%	+\wt\Om_j+\tchi_\tb\s_\tb
%	\om(\tbk_\tb)]} \Bigg]
%$$
	&=& \int_0^t \rd s \Bigg( \int_0^{s*} [\rd s_j]_0^{\mu(a_1)-1}
	 \Bigg [ \prod_{j=0}^{a_1-1} \prod_{b\in I_j\cup I_j^c}
	e^{-is_b [e(\bp_{\mu(j)}+\chi_b\bk_b)+
	\Om_j - \Om_{a_1-1} +\chi_b\s_b\om(\bk_b)]} \Bigg]\Bigg)
	e^{-is \Om_{a_1-1}}
\nonumber\\
	&&\times\Bigg(\int_0^{(t-s)*} [\rd s_j]_{\mu(a_1)}^N
	\Bigg[ \prod_{j=a_1}^n \prod_{b\in I_j\cup I_j^c}
	e^{-is_b [e(\bp_{\mu(j)}+\chi_b\bk_b)+
	\Om_j+\chi_b\s_b\om(\bk_b)]} \Bigg] \Bigg)\; ,
\label{chop}
\eey
%$$
%	\times \int_0^t \rd \ts \Bigg( \int_0^{\ts *} 
%	[\rd \ts_j]_0^{\tmu(\pi(a_1))-1}
%	 \Bigg [ \prod_{j=0}^{\pi(a_1)-1} \prod_{\tb\in \tI_j\cup \tI_j^c}
%	e^{i\ts_\tb [e(\tbp_{\tmu(j)}+\tchi_\tb\tbk_\tb)+
%	\wt\Om_j - \wt\Om_{\pi(a_1)-1} +\tchi_\tb\ts_\tb\om(\tbk_\tb)]}
%	 \Bigg]\Bigg)
%	e^{i\ts \wt\Om_{\pi(a_1)-1}}
%$$
%$$
%	\times\Bigg(\int_0^{(t-s)*} [\rd \ts_j]_{\tmu(\pi(a_1))}^N
%	\Bigg[ \prod_{j=\pi(a_1)}^n \prod_{\tb\in \tI_j\cup \tI_j^c}
%	e^{i\ts_\tb [e(\tbp_{\tmu(j)}+\tchi_\tb\tbk_\tb)+
%	\wt\Om_j+\tchi_\tb\ts_\tb\om(\tbk_\tb)]} \Bigg] \Bigg)\;.
%$$
i.e., we separated the $\rd s_j$, $j< \mu(a_1)$,  integrations from the
rest by prescribing their total sum to be $s$.
We also pulled out a factor $e^{-is\Om_{a_1}-1}$
which shifted all $\Om_j$'s in the first group of propagators
by $\Om_{a_1}-1$.
Notice that 
\be
	\Om_{a_1-1} - \Om_j
	= \sum_{m=j+1}^{a_1-1}\s_m\om(\bk_m)\;  ,  \quad j\leq a_1 -1\;,
\label{Omdep}\ee
depends only on $\bk_{j+1}, \ldots, \bk_{a_1-1}$, and
$\Om_{j}$ for $j\ge a_1$ depends only on $\bk_{a_1+1}, \ldots, \bk_n$.

The decomposition of the $\rd\ts_j$ integrals is similar; we separate
the $\rd\ts_j$, $j< \tmu(\pi(a_1))$, integrations 
 (their sum is $\ts$) and pull out a factor
$e^{i\ts\wt\Om_{\pi(a_1)-1}}$. We again see that
 the $\bk$-dependences of 
$$
	\wt\Om_j - \wt\Om_{\pi(a_1)-1} =	
	\sum_{m\atop j< \pi(m) < \pi(a_1)}
	\s_m\om(\bk_m) \; ,
$$
for $j\leq \pi(a_1)-1$, and those of
$$
	\wt\Om_j = \sum_{m\; : \; \pi(m)>j} \s_m\om(\bk_m) \; ,
$$
for $j\ge \pi(a_1)$, are separated.

Now we use (\ref{eq:timeint}) for each time integral in (\ref{chop})
separately, we integrate out $\bk_b$, $b\in J$, and
similarly for $\tbk_\tb, \tb\in \tJ$ as in (\ref{eq:Y3}).
We then estimate everything by absolute value,
we use (\ref{Upsilonest}) and (\ref{ttriv})
and we estimate the two outside time integrals ($\rd s, \rd \ts$)
trivially. In this last step we lose an extra $t^2$.
We also integrate out $\tbp_n$ and all $\tbk_j$ and express 
everything as a function of $\bk_j$ and $\bp_n$.
The result, similarly to (\ref{eq:cpinew}), is
\bey
	|C_{\um, \utm,\pi}(t)| &\leq& (C\lambda)^{2N} t^{2+2|\um|}
	\sup_{\usi}
	\int \rd\nu_\pi^*(\bp_n, \tbp_n, \ubk, \utbk)
\label{eq:cpinew3}\\
	&&
	\times \int_{-\infty}^\infty \rd \a \rd \beta\prod_{j=0}^{a_1-1} |S_j|
	 \prod_{j=a_1}^n  |R_j|
	\int_{-\infty}^\infty
	 \rd \ta \rd \tbeta \prod_{j=0}^{\pi(a_1)-1} |\tS_j|
	\prod_{j=\pi(a_1)}^n |\tR_j| \;,
\nonumber
\eey
where in addition to recalling the definition of $R_j, \tR_j$ (\ref{def:R}),
we define
$$
	S_j: = S_j(\beta,\bp_j,\Om_j-\Om_{a_1-1},\eta)
	:= {1\over \beta - e(\bp_j)  -  (\Om_j- \Om_{a_1-1}) +i\eta}\;,
$$
$$
	\tS_j: =  \tS_j(\tbeta,\tbp_j,\wt\Om_j-\wt\Om_{\pi(a_1)-1},\eta)
	:= {1\over \tbeta - e(\tbp_j)  - 
	 (\wt\Om_j- \wt\Om_{\pi(a_1)-1}) -i\eta}\;,
$$
and, as usual, $\bp_j, \tbp_j$ are viewed as functions
of $\bp_n$ and $\ubk$ (\ref{def:pj}).

If $a_1=1$ or $\pi(a_1)=1$ then there is no need  for separation, i.e.,
there are no additional $\rd\beta$ or $\rd\tbeta$ integrations.
In this case (\ref{eq:cpinew}) can be used directly. We will not
discuss this simpler case in detail.

We  introduce
$$
	b_j : = a_j + h_j\;, \qquad 1\leq j \leq \kappa
$$
for the location of the tips.
All factors in (\ref{eq:cpinew3}) with tilde  except $ \tR_n$ and
$\tR_{\pi(b_j)-1}$, $j=1, 2, \ldots, \kappa$,
are estimated trivially by $(Ct)$, this gives $(Ct)^{n-\kappa}$.
Moreover, we make sure that we gain a $\langle \ta\rangle^{-1}\langle
\tbeta\rangle^{-2}$
factor from these $L^\infty$-estimates, exactly as
in the constant $\om$ case,  using (\ref{tagain}). Notice that
we need to gain a second $\langle\tbeta\rangle^{-1}$-factor
 to make the $\rd\tbeta$
integration finite.  This is possible only if $\pi(a_1)>1$, but there
is no need for $\rd\tbeta$ integration at all if $\pi(a_1)=1$.
We also insert an explicit extra $\prod_j \langle \bk_j\rangle^{-4}$
decay. Since all $\tbeta$ denominators
are estimated in $L^\infty$ norm, at the end we can perform
the $\rd\tbeta$ integration.
We get
$$
	|C_{\um, \utm,\pi}(t)| \leq  t^{-\kappa+2}(C\lambda^2t)^{N} 
	\sup_{\usi}
	\int\Big(\prod_{j=1}^n {L^*(\bk_j) \delta(\bk_j-\tbk_{\pi(j)})
	\rd \bk_j \rd\tbk_j \over
	\langle \bk_j\rangle^{4}}
	\Big) 
	\int \rd \bp_n \langle \bp_n\rangle^{d+12}\wh\g_e(\bp_n, \bp_n)
$$
$$
	\times \int_{-\infty}^\infty \rd \a \rd \beta
	\Bigg(
	{1\over  \langle \bp_{a_1-2}\rangle^2\langle \bp_{a_1-1}\rangle^{d+1}}
	 \prod_{j=0}^{a_1-1} |S_j|\Bigg)
	\Bigg(
	  {1\over \langle \bp_{b_1-1}\rangle^2}\prod_{j=a_1}^n 
	|R_j|\Bigg)
	\Bigg(\int_{-\infty}^\infty {\rd \ta \over \langle \ta \rangle}
	\prod_{m=1}^\kappa |\tR_{\pi(b_m)-1}|\Bigg)
$$
with
\be
	L^*(\bk): = M^*(\bk) \langle \bk\rangle^{2d+22}
\label{def:Lstar}
\ee
so $L^*(\bk)\leq C \langle \bk \rangle^{-2d-2}$.
Notice that we also inserted $ \langle \bp_{a_1-2}\rangle^{-2}
\langle \bp_{a_1-1}\rangle^{-d-1}\langle \bp_{b_1-1}\rangle^{-2}$
using (\ref{insert}) to prepare for the decays in $\beta$ and $\a$ at the
expense of increasing the $\langle \bp_n\rangle$-power and
using up a few $\prod_j\langle\bk_j\rangle$ power from $M^*$.
If $a_1=1$, then there is no need for  $ \langle \bp_{a_1-2}\rangle^{-2}
$ insertion.

We express the electron momentum in $\tR_{\pi(b_m)-1}$
as $\tbp_{\pi(b_m)-1} = \tbp_{\pi(b_m)}
+\bp_{b_m-1} -\bp_{b_m}$.

\bigskip\bigskip
\centerline{\epsffile{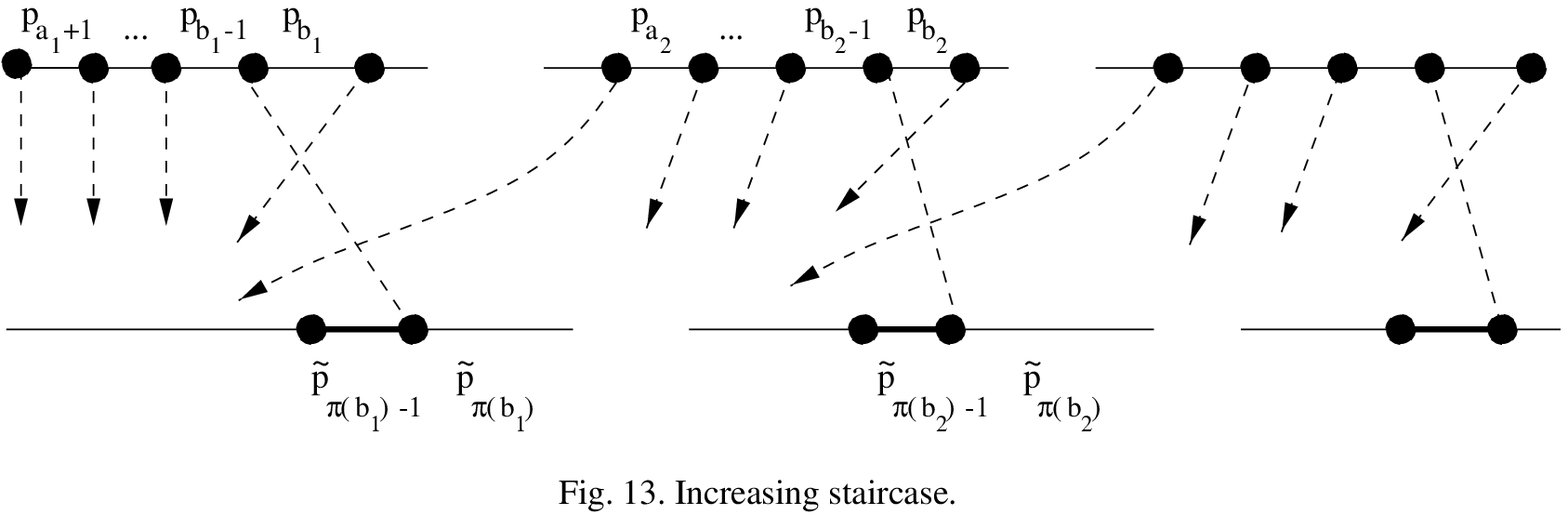}}
\bigskip

Fig. 13 shows an increasing staircase;
 the bold lines indicate the electron momenta in 
those $\tR_j$'s which we kept ($j=\pi(b_m)-1$, $m=1,\ldots , \kappa$).

Now we change variables; we use $\bp_0, \bp_1, \ldots, \bp_{n-1}$
instead of $\bk_1, \bk_2, \ldots, \bk_n$. Notice that for each $m\leq \kappa$
the factor $\tR_{\pi(b_m)-1}$, i.e., the momentum
 $\tbp_{\pi(b_m)-1}$ and
$\wt\Om_{\pi(b_m)-1}$ do not depend on $\bp_j$, $a_1\leq j< b_m-1$.
This follows from the staircase construction.

First we estimate the factor $L^*(\bp_{a_1-1} - \bp_{a_1})$
by a constant. In this way the $\bp$-dependence of the
 remaining $L^*$ factors
are separated. The inserted 
$\langle\bp_{a_1-1}\rangle^{-d-1}$ factor ensures that the
integration of the momenta before the staircase is still finite.

We then integrate out $\bp_{a_1}, \bp_{a_1+1},\ldots,
\bp_{b_1-2}$, these variables do not appear in
$\tR_{\pi(b_m)-1}$, $m=1, \ldots, \kappa$, or in $S_j$, $j\leq a_1-1$
(see (\ref{Omdep})).
We  get a $(C\log^*t)^{h_1-1}$ factor by (\ref{prot:Clog})
and we have eliminated $R_j$, $j=a_1, \ldots, b_1-2$.
For the $\bp_{b_1-1}$ and $\bp_{b_1}$ integrations we are 
in a situation similar to (\ref{int:pam}). We explicitly write out
the factors $R_{b_1-1}$, $R_{b_1}$ and $\tR_{\pi(b_1)-1}$
that are involved and we obtain
\bey
	\int { L^*(\bp_{b_1-1} - \bp_{b_1})
	L^*(\bp_{b_1} -\bp_{b_1+1}) \rd \bp_{b_1-1}\rd\bp_{b_1}\over
	\langle \bp_{b_1-1}\rangle^2
	\Big| \a - e(\bp_{b_1-1}) + i\eta \pm\om(\bp_{b_1-1}-\bp_{b_1})
	\pm \om(\bp_{b_1}-\bp_{b_1+1}) -\Om_{b_1+1}\Big|}&&
\label{int:pam1}
\\
 	\times {1\over \Big|  \a - e(\bp_{b_1}) + i\eta 
	\pm \om(\bp_{b_1}-\bp_{b_1+1}) -\Om_{b_1+1}\Big|}
&&
\nonumber\\
	\times {1\over  \Big| \ta - e\Big(\tbp_{\pi(b_1)} + \bp_{b_1-1} -
	\bp_{b_1}\Big) + i\eta \pm\om(\bp_{b_1-1}-\bp_{b_1})
	 - \wt\Om_{\pi(b_1)-1}\Big|}
	&\leq& {(C\log^* t)^3 \over  \langle\a-\Om_{b_1+1}\rangle}\;.
\nonumber
\eey
Notice that $\tbp_{\pi(b_1)}$ is independent of $\bp_{a_1}, \bp_{a_1+1},
\ldots, \bp_{b_1}$, i.e., of all the variables we have
integrated out so far.
The prototype of this inequality is 
\bey
	\sup_{\bq, \br, \ttheta}\int { \langle \bv-\bu\rangle^{-d-1}
	\langle\bu-\bq\rangle^{-d-1}
	 \rd \bv\rd\bu\over
	\langle \bv\rangle^2
	\Big| \theta - e(\bv) + i\eta \pm\om(\bv-\bu)
	\pm \om(\bu-\bq) \Big|}&&
\label{prot:pamuj}
\\
 	\times {1\over \Big|  \theta - e(\bu) + i\eta 
	\pm \om(\bu-\bq) \Big|} {1\over  \Big| \ttheta - e(\br +\bv-\bu)
	 + i\eta \pm\om(\bv-\bu)\Big|}
	&\leq& {(C\log^* t)^3 \over  \langle\theta\rangle}
\nonumber
\eey
(with $\bv= \bp_{b_1-1}$, $\bu=\bp_{b_1}$,
 $\bq= \bp_{b_1+1}$, $\br=\tbp_{\pi(b_1)}$, $\theta=\a-\Om_{b_1+1}$,
$\ttheta=\ta-\wt\Om_{\pi(b_1)-1}$)
which is proven  similarly to (\ref{prot:pam}).
First we perform the $\rd\bv$ integration using (\ref{prot:twoden}),
then (\ref{prot:plus}) estimates the $\rd\bu$ integration.
We can change the decaying factor $\langle\a-\Om_{b_1+1}\rangle^{-1}$
in (\ref{int:pam1}) into $\langle\a\rangle^{-1}$ at the expense of
the product of $\langle \bk_j\rangle^2 = \langle \bp_{j}-\bp_{j-1}\rangle^2$
factors ($j=b_1-1, \ldots, n$) similarly to (\ref{cOm}).

Next we integrate out $\bp_{b_1+1}, \bp_{b_1+2}, \ldots,
\bp_{a_2}, \ldots, \bp_{a_2+h_2-2} =\bp_{b_2-2}$.
By the staircase property (especially (iv) in Definition
\ref{def:incrstair}), these variables do not appear
in the remaining factors
$\tR_{\pi(b_j)-1}$, $j\ge2$, hence they freely give $C\log^*t$
 factors and they eliminate  $R_j$, $j=b_1+1, \ldots, b_2-2$.
Now we integrate out  $\bp_{b_2-1}$ and $\bp_{b_2}$ exactly
as in (\ref{int:pam1}), etc. 

Once we are done with all $\bp_{b_m-1}$ and $\bp_{b_m}$  
($1\leq m \leq \kappa$)
integrations, then there are
no more $\tR$ factors left. So we integrate out the
remaining variables, $\bp_0, \bp_1, \ldots, \bp_{a_1-1}$,
and $\bp_{b_m+1}, \ldots, \bp_{n-1}$ in this order; it is
easy to see that at each step only one $S_j$ or $R_j$
factors depends on these variables. At the $\bp_{a_1-2}$ and
$\bp_{a_1-1}$ integration we use
\be
	\int { L^*( \bp_{a_1-2}-\bp_{a_1-1})
	\rd \bp_{a_1-2} \over \Big| \beta -e(\bp_{a_1-2})
	+i\eta\pm \om(\bp_{a_1-2}-\bp_{a_1-1})\Big|
	 \; \langle  \bp_{a_1-2}\rangle^2}
	\leq {C\log^*t\over \langle \beta \rangle}
\label{omlog}
\ee
and
\be
	\int {\rd \bp_{a_1-1} \over \Big| \beta -e(\bp_{a_1-1})
	+i\eta\Big| \; \langle  \bp_{a_1-1}\rangle^{d+1}}
	\leq {C\log^*t\over \langle \beta\rangle}\;
\label{trivomlog}
\ee
that follow from (\ref{prot:Clogalpha}) and (\ref{prot:Clog}).
The extra $\langle \beta\rangle^{-2}$ factor
makes the $\rd\beta$ integration convergent.

Finally from the $\a, \ta$ and $\bp_n$ integration we obtain
a factor $(C\log^*t)^2$ similarly to (\ref{int:alpha3}).
This completes the proof of Proposition \ref{prop:omnonconst}
for the case of increasing $\kappa$-staircase.

\subsubsection{Estimate for decreasing staircase}

The proof for the decreasing staircase is very similar.
Here we chop off the end of the expansion.
Let $b_j:= a_j+h_j$ as before, but now these
are the bottoms of the stairs. Then
instead of (\ref{chop}) we split the $s_j$ variables into
two groups: $j=0, 1, \ldots, \mu(b_\kappa)-1$ and
$j=\mu(b_\kappa), \ldots, N$. The result, analogous to
(\ref{eq:cpinew3}), is
%$$
%	\int_0^{t*} [\rd s_j]_0^n 
%	 \Big(\prod_{j=0}^n e^{-is_j[\bp_j^2/2+ \Om_j(\bk,\usi)]} \Big)
%	 \int_0^{t*} [\rd \ts_j]_0^n 
%	 \Big(\prod_{j=0}^n e^{i\ts_j[\tbp_j^2/2 + \Om_j (\tbk, \usi_\pi)]}
%	\Big)
%$$
%$$
%	= \int_0^t \rd s \Bigg( \int_0^{s*} [\rd s_j]_0^{b_\kappa-1}
%%	 \prod_{j=0}^{b_\kappa-1} e^{-is_j[\bp_j^2/2+ \Om_j- 
%	\Om_{b_\kappa-1}]}\Bigg)
%	e^{-is \Om_{b_\kappa-1}}
%	\Bigg(\int_0^{(t-s)*} [\rd s_j]_{b_\kappa}^n
%	 \prod_{j=b_\kappa}^n e^{-is_j[\bp_j^2/2+ \Om_j]} \Bigg)
%%$$
%$$
%	\times \int_0^t \rd \ts \Bigg( \int_0^{\ts *} 
%	[\rd \ts_j]_0^{\pi(b_\kappa)-1} \prod_{j=0}^{\pi(b_\kappa)-1}
%	 e^{i\ts_j[\tbp_j^2/2+ \wt\Om_j - \wt\Om_{\pi(b_\kappa)-1}]}
%	\Bigg)
%	 e^{ i\ts \wt\Om_{\pi(b_\kappa)-1}}
%	\Big(\int_0^{(t-\ts )*} [\rd \ts_j]_{\pi(b_\kappa)}^n
%	 \prod_{j=\pi(b_\kappa)}^n
%	 e^{i\ts_j[\tbp_j^2/2+ \wt\Om_j]} \Big)
%$$
\bey
	|C_{\um, \utm,\pi}(t)| &\leq& (C\lambda)^{2N} t^{2+2|\um|}
	\sup_{\usi}
	\int \rd\nu^*_\pi(\bp_n, \tbp_n, \ubk, \utbk)
\nonumber\\
	&&
	\times \int_{-\infty}^\infty \rd \a \rd \beta\prod_{j=0}^{b_\kappa-1}
	|S_j| \prod_{j=b_\kappa}^n |R_j|
	 \int_{-\infty}^\infty
	 \rd\ta \rd \tbeta \prod_{j=0}^{\pi(b_\kappa)-1}
	|\tS_j|
	\prod_{j=\pi(b_\kappa)}^n|\tR_j|\;,
\nonumber
\eey
where $S_j$ and $\tS_j$ have been redefined accordingly:
$$
	S_j: = {1\over \beta - e(\bp_j) - 
	 (\Om_j- \Om_0)+i\eta},\qquad
	\tS_j :=
	{1\over \tbeta
	 -e(\tbp_j)-  (\wt\Om_j - \wt\Om_{\pi(b_\kappa)-1})-i\eta}\;.
$$
Notice that we pulled out the factor $e^{-is\Om_0}$ and 
$e^{i\ts \wt\Om_{\pi(b_\kappa)-1}}$ in the analogue of (\ref{chop}). In this
way, $S_j$ depends on $\bp_0, \ldots, \bp_j$ and $\tS_j$ depends
on $\tbp_j, \ldots, \tbp_{\pi(b_\kappa)-1}$.

All factors with tilde except $\tR_n$ and $\tS_{\pi(a_m)-1}$,
 $m=1, 2, \ldots, \kappa$, 
are estimated trivially, and  we can ensure the decaying
$\langle \ta \rangle^{-1}
\langle \tbeta \rangle^{-2}$  terms by inserting approriate
factors at the expense of $\langle\bp_n\rangle$ and 
$\langle\bk_j\rangle$ powers as before.
We get
$$
	|C_{\um, \utm,\pi}(t)| \leq t^{-\kappa+2} (C\lambda^2 t)^{N}
	\sup_{\usi}
	\int\Big(\prod_{j=1}^n { L^*(\bk_j)\delta(\bk_j-\tbk_{\pi(j)})\rd \bk_j
	\rd\tbk_j
	\over\langle\bk_j\rangle^4}  \Big)
	\int \rd \bp_n \langle \bp_n\rangle^{2d+12}\wh\g_e(\bp_n, \bp_n)
$$
$$
	\times \int_{-\infty}^\infty
	 {\rd \a \rd \beta \rd\ta \rd\tbeta\over \langle \ta \rangle
	 \langle \tbeta \rangle^2
	\;  \langle \bp_{a_1-2}\rangle^{d+1}
	\langle \bp_{b_{\kappa}-1}\rangle^{d+1}
	\langle \bp_{a_\kappa-1}\rangle^2  \langle \bp_{b_\kappa}\rangle^2}
	\Bigg(\prod_{j=0}^{b_\kappa-1} |S_j|\Bigg)
	\Bigg(  \prod_{j=b_\kappa}^n 
	|R_j|\Bigg) \; |\tR_n|
	\prod_{m=1}^\kappa |\tS_{\pi(a_m)-1}|
%	{1\over \Big|\tbeta - e\Big(\tbp_{\pi(a_m)}
%	+\bp_{a_m-1} -\bp_{a_m}\Big)+i\eta -  
%	\wt\Om_{\pi(a_m)-1}
%	\Big|}
$$
and we will again use the variables $\bp_0, \bp_1, \ldots, \bp_{n-1}$
instead of $\bk_1, \ldots, \bk_n$.

%\bigskip\bigskip
%\centerline{\epsffile{fig14ph.eps}}
%\bigskip

%Fig. 14 shows a decreasing staircase. The pairing lines corresponding
%to the bottoms of the stairs are solid.

We first estimate the factors $L^*(\bp_{a_1-2}-\bp_{a_1-1})$ 
and $L^*(\bp_{b_\kappa-1}-\bp_{b_{\kappa}})$
by a constant. Then the integrations of $\bp_{b_\kappa-1},
\ldots, \bp_{a_\kappa+1}$ can be done successively
in this order; no $\tS_j$ factor
depends on these and we collect a $(C\log^*t)$ factor from
eliminating each $S_j$.
 We then 
perform $\rd\bp_{a_\kappa-1}, \rd\bp_{a_\kappa}$ integrations,
 very similarly to (\ref{int:pam1}):
\bey
	\int {L^*(\bp_{a_\kappa-2}-\bp_{a_\kappa-1})	
	 L^*(\bp_{a_\kappa-1} - \bp_{a_\kappa})
	 \rd \bp_{a_\kappa-1}\rd\bp_{a_\kappa}\over
	\langle \bp_{a_\kappa-1}\rangle^2
	\Big| \beta - e(\bp_{a_\kappa-1}) + 
	i\eta \pm\om(\bp_{a_\kappa+1}-\bp_{a_\kappa})
	\pm\om(\bp_{a_\kappa}-\bp_{a_\kappa-1}) - \Om_{a_\kappa+1} \Big|}&&
\label{int:pamnew}
\\
	\times{1\over \Big|  \beta
	 - e(\bp_{a_\kappa}) + i\eta  \pm\om(\bp_{a_\kappa+1}-\bp_{a_\kappa})
	-\Om_{a_\kappa}\Big|}
&&
\nonumber\\
	\times {1\over  \Big| \ta - e\Big(\tbp_{\pi(a_\kappa)} + 
	\bp_{a_\kappa-1} -
	\bp_{a_\kappa}\Big) + i\eta 
	\pm \om( \bp_{a_\kappa} -\bp_{a_\kappa-1})
	-\wt \Om_{\pi(b_\kappa)-1} \Big|}
	&\leq& {(C\log^* t)^3 \over  \langle\beta-\Om_{a_k+1} \rangle} \;.
\nonumber
\eey
Again, we used that $\tbp_{\pi(a_\kappa)}$ does not depend on
the momenta integrated out so far.
We perform $\bp_{a_\kappa-2}, \bp_{a_\kappa-3}, \ldots$, 
$ \bp_{b_{\kappa-1}},
\ldots, \bp_{a_{\kappa-1}+1}$ successively; no remaining $\tS_j$ factors
depend on them. We then  perform $\bp_{a_{\kappa-1}}, \bp_{a_{\kappa-1}-1}$
exactly as in (\ref{int:pamnew}), etc.
Once we are done with  $\bp_{a_1}, \bp_{a_1-1}$, there are no $\tS_j$
factors left, and the rest
can be integrated out successively: $\bp_0, \bp_1, \ldots, \bp_{n-1}$.
The $\rd\bp_{a_1-2}$ integration uses the $\langle 
\rd\bp_{a_1-2}\rangle^{-d-1}$ decay since there is no $L^*$ factor left.
We also gain a $\langle\beta\rangle^{-1}$ factor from this integration
similarly to (\ref{trivomlog}).
The decay $\langle\beta-\Om_{a_k+1} \rangle^{-1}$ is changed to
$\langle \beta\rangle^{-1}$  using the extra decaying factors similarly to
(\ref{cOm}). This ensures the finiteness of
the $\beta$ integration. From the $\bp_{b_\kappa}$ integration we gain
an extra $\langle \a \rangle^{-1}$ because of the inserted
$\langle \bp_{b_\kappa}\rangle^{-2}$.

Finally the $\a, \ta, \bp_n$ 
integrations are done exactly as in (\ref{int:alpha3}),
which completes the proof of Proposition \ref{prop:omnonconst}.
$\;\;\;\Box$

\subsection{Estimate of the nested term
for the nonconstant  $\om$ case.}\label{sec:nest}

In Section \ref{sec:allpair} we proved
(\ref{smallcomest}) for $a\ge 3$ and for $a=2$ with $\om = (const.)$.
Here we outline the argument for the remaining
case when $\om\neq (const.)$, $a=2$.
 Only the proof of
Lemma \ref{lemma:A} needs to be modified, the rest of the
argument from Section \ref{sec:allpair} is unchanged.

{\it Proof of Lemma \ref{lemma:A} for $\om\neq (const.)$.}
%The technical
%difficulty is that even after a change of variables (see (\ref{chang}))
% we have
%$$
%	\Upsilon_\eta(\a, \bp) = \sum_{\pm}\int{M(\bk-\bp)\rd\bk\over
%	\a - e(\bk)\pm\om (\bk-\bp)+i\eta}\;, 
%$$
%i.e., the denominator of the integrand also depends on $\bp$.
%When computing $\nabla^\ell_\bp \Upsilon_\eta$, one has to raise
%the almost-singular denominator to a high power.
%It is easy to check from stationary phase asymptotics that 
%high derivatives of $\Upsilon_\eta$ do not remain bounded 
%as $\eta\to0+0$ unlike (\ref{Upsder}) for the constant $\om$ case.
%On the other hand, the integration $\rd\bk_2$ in (\ref{def:A}) must use
%a cancellation of the type of a singular integral; estimating the integrand
%by absolute value is not sufficient. Hence some differentiation on $\Upsilon$
%is necessary. The proof below uses only one derivative which
%is still controllable, unlike stationary phase formulas which
%use high derivatives. We note that the argument of Section
%\ref{sec:nestconst} together with an $\eta$-dependent stationary
%phase bound on $|\nabla^\ell_\bp \Upsilon_\eta|$ would be sufficient for
%high dimensions, but not in $d=3$.
The proof is an extension of the argument  in
 Section 3.4 of \cite{EY2} and we only indicate the main steps.
We write
\be
	A = i^{m+1}\int M(\bk)
	\int_0^\infty s^m e^{is(\a - \Phi(\bp,\bk) +i\eta)}
	\Big[\Upsilon_\eta(\a- \om(\bk), \bp+\bk)\Big]^m\rd s \rd \bk\;,
\label{Auj}
\ee
where we omitted $\s_2$ and we
let $m=m_1$, $p=p_2$,  and $\Om_2=0$ for simplicity.
By a partition of unity we divide  the space
into cubes $Q$ of size $\wt\varrho$.
Correspondingly, we can replace  $M(\bk)$ in (\ref{Auj}) by $M_Q(\bk)$
that  is supported on a cube $Q$ and $M=\sum_Q M_Q$.
We can assume that $ M_Q$ is as smooth as $M$.
 Moreover, $\| M_Q\|_\infty \leq C\langle \mbox{dist}(0,Q) \rangle^{-2d}$
and similar estimates are valid for its derivatives, since $M_Q$
inherits the size of $M$ on $Q$.
We denote the corresponding expression by $A_Q$
$$
	A_Q=A_Q(\a, \bp): = i^{m+1}\int M_Q(\bk)
	\int_0^\infty s^m e^{is(\a - \Phi(\bp,\bk) +i\eta)}
	\Big[\Upsilon_\eta(\a- \om(\bk), \bp+\bk)\Big]^m\rd s \rd\bk\;,
$$
and we neglect the $\bp$ variable which we consider fixed.

After a diffeomorphic
change of the $\bk$ variable we arrive at one of the following
normal forms depending on whether $Q$ contains a critical point or not:
\be
	A_Q= \int E_Q(\ta,\bk)
	\int_0^\infty s^m e^{is(\ta - \bk^2+i\eta)}\rd s \rd\bk\;;
\label{AD}
\ee
or
$$
	 A_Q = \int E_Q(\ta,\bk)
	\int_0^\infty s^m e^{is(\ta - \bu\cdot \bk +i\eta)}\rd s \rd\bk\;,
$$
where $\bu=\bu_Q$ is a constant unit vector, $\ta=\ta_Q\in\bR$, and we defined
$$
	E_Q(\ta, \bk): =  i^{m+1}M_Q(\psi(\bk))
	\Big[\Upsilon_\eta(\ta-\om(\psi(\bk)), \psi(\bk)+\bp )\Big]^m J(\bk)\;.
$$
Here the function $\psi$ expresses the change of variable,
$J$ is the corresponding Jacobian.
These are as smooth  functions
as  $\Phi$ and $\om$ and they depend on $Q$ and $\bp$
with uniform bounds.

We will discuss the more complicated first case, the
other case is similar. We rewrite the $\rd\bk$ integration as follows
$$
	\int  e^{-is\bk^2}E_Q(\ta,\bk)\rd\bk
	= (2\pi)^{-d/2}\int_0^\infty e^{-is\vartheta}
	\Bigg( \int_{\bk^2 = \vartheta} E_Q(\ta,\bk)
	d\s(\bk)\Bigg)\vartheta^{d/2-1}\rd \vartheta\;,
$$
where $d\s(\bk)$ is the normalized surface measure on the sphere 
$\{\bk \;: \; \bk^2=\vartheta\}$.
Let
\bey
	H_Q(\a, \bk):&=& -i \vartheta^{-d/2+1} 
	{\partial\over \partial	\vartheta}
	\Bigg( \vartheta^{d/2-1}\int_{\tbk^2 = \vartheta}
	 E_Q(\ta, \tbk) d\s(\tbk)
	\Bigg)\Bigg|_{\vartheta:=\bk^2}
\label{Hdef}
\\
	&=& -i(\sfrac{d}{2}-1)\bk^{-2} \int_{\tbk^2 = \bk^2} E_Q(\ta, \tbk) 
	d\s(\tbk)
	-i   {\partial\over \partial	\vartheta}
	\Bigg( \int_{\tbk^2 = \vartheta} E_Q(\ta, \tbk) d\s(\tbk)\Bigg)
	\Bigg|_{\vartheta:=\bk^2}\;.
\nonumber
\eey
Using integration by parts
$$
	\int  e^{-is\bk^2}E_Q(\ta, \bk)\rd\bk = s^{-1} 
	\int  e^{-is\bk^2}H_Q(\ta, \bk)\rd\bk\;,
$$
hence after undoing the $\rd s$ integration in (\ref{AD}),
$$
	A_Q = i^m\int { H_Q(\ta, \bk) \rd \bk\over
	(\ta - \bk^2+i\eta)^m}\;.
$$
We  show that
\be
	|H_Q(\ta, \bk)|\leq C_Q(\eta^{-1/2}  + |\bk|^{-2})\;. 
\label{Hest}
\ee
 $H_Q$ consists of two terms (\ref{Hdef}), the
 first one is easy since $E_Q(\ta, \bk)$ is
uniformly bounded from (\ref{Upsilonest}). 
For the second term, we use the bound on the derivative of
$\Upsilon_\eta (\a, \bp)$ in $\bp$ and $\a$ (\ref{derup}).
All the other factors in $E_Q$ have derivatives bounded by a constant.

Since $\bk\mapsto H_Q(\ta, \bk)$ is compactly supported, (\ref{Hest})
gives an estimate of order $C_Qt^{m-1/2}$ for $ A_Q$ (with $\eta:=t^{-1}$),
where the constant $C_Q$
depends on the cube $Q$ and it behaves as $C_Q\leq  
C\langle \mbox{dist}(0,Q) \rangle^{-2d}$. In particular, these
estimates are summable over all the cubes $Q$ of the partition of unity.
This will complete the proof of Lemma \ref{lemma:A}.
 $\;\;\;\Box$

\section{Proofs of the combinatorial Lemmas}\label{app:comb}
\setcounter{equation}{0}

{\it Proof of Lemma \ref{lemma:number}.}
We  consider a pairing $\pi \in \Pi_n$ that has exactly $K$ peaks
at the locations $a_1, a_2, \ldots, a_K$. Such pairing has
$K-1, K$ or $K+1$ valleys at $b_0, b_1, \ldots, b_{K}$
such that
$$
	b_0 < a_1 < b_1 < a_2 < \ldots <b_{K-1} < a_K < b_K\;,
$$
where $b_0$ and $b_K$ might not be present.

The number of possible  peak-and-valley configurations, including
 their heights is at most $n^{4K+2}$, since one only has
to prescribe the values $a_j, \pi(a_j)$ and $b_j, \pi(b_j)$.

Once the peak-and-valleys are fixed, we consider the set
$$
	S = \{ 1, 2, \ldots , n\} \setminus \bigcup_j \Big(\pi(a_j) \cup
	 \pi(b_j) \Big)\;.
$$
We define the following sequences:
\bey
	D_j :& =& \Big( \pi(a_j+1), \pi(a_j+2) , \ldots, \pi(b_j-1)\Big)
\nonumber\\
	U_j : &=&  \Big( \pi(b_j+1), \pi(b_j+2) , \ldots, \pi(a_{j+1}-1)\Big)
\nonumber
\eey
for $j=0, 1, \ldots, K$ and $a_0 =0$, $a_{K+1}=n+1$ for definiteness.
Clearly
$S = \bigcup_j \Big( U_j \cup D_j\Big)$
is a partition; and $U_j$ ($D_j$) is a monotonically increasing
(decreasing) subsequence. This partition consists of at most
$2K+2$ non-empty sets, we label them with distinct labels.
For any permutation $\pi$ with fixed peak-and-valleys,
we assign to every element in $S$ the label of its set
in the partition. This can be done by at most
$(2K+2)^{|S|}\leq (2K+2)^n$ ways.
Notice that once the peak-and-valleys and this
assigment are fixed, the permutation is
uniquely determined, since the exact order within $U_j$ and $D_j$
are determined by the monotonicity.
Hence the number of such permutations is not more
than $n^{4K+2} (2K+2)^n$. If we consider 
pairings with at most $K$ peaks, then we have to
add these numbers up for $K=1, 2, \ldots $, which
gives a total number at most $n^{4K+3}(2K+2)^n$
 $\;\;\;\Box$

\bigskip

{\it Proof of Lemma \ref{lemma:ramsey}}.
Let $f(\a,\beta )$ be the smallest  number
such that  any permutation
with at least $f(\a,\beta)$  peaks has either 
an increasing $(\a+1)$-staircase
or  a decreasing $(\beta+1)$-staircase.
 We show the recursion
\be
	f(\a, \beta +1) \leq  f(\a-1, \beta+1)+ f(\a, \beta) 
\label{recursion}
\ee
and the relations
\be
	f(\a, 1) = \a+1, \qquad f(1,\beta)= \beta+1\; .
\label{initial}
\ee
{F}rom these recursive relations we easily obtain the estimate
$f(\a, \beta) \leq {\a+\beta \choose \a}$.
%follows, since it is true for the initial values and
%$$
%	{\a+ \beta+1\choose \a} =
%	{\a+\beta \choose \a-1}+{\a+\beta \choose \a}\;.
%$$
We first  notice the following fact:

\bigskip

{\it Observation:} If we have an increasing $\kappa$-staircase
with tips at $a_1 < a_2 < \ldots <a_\kappa$,
 and if there is a peak $a>a_\kappa$
which is higher than the highest tip, i.e., $\pi(a_\kappa)< \pi(a)$, 
then we also have an increasing $(\kappa+1)$-staircase.
Similarly, if we have a decreasing $\kappa$-staircase
with tips at $a_1 > a_2 > \ldots > a_\kappa$, and if there is a peak $a<a_1$
which is higher than highest tip, i.e., $\pi(a_1)< \pi(a)$, 
then we also have an decreasing $(\kappa+1)$-staircase.
\bigskip

For the proof of this observation in the  increasing case, suppose
 that we have a peak  $a>a_\kappa$ with 
$\pi(a_\kappa)< \pi(a)$. Then,
we simply choose the smallest number $b>a_\kappa$, such that $b$ is
a peak and $\pi(b)> \pi(a_\kappa)$.
 Let $b$ be the tip of
the $(\kappa+1)$-th up-stair and let $a_\kappa<c<b$ be the biggest
number such that $\pi(c)< \pi(a_\kappa)$. Such number exists,
since any valley between $a_\kappa$ and $b$ must lie below
$\pi(a_\kappa)$, otherwise there would also be a peak somewhere strictly
between $a_\kappa$ and $b$ higher than $\pi(a_\kappa)$ which
contradicts to the choice of $b$. Now we simply choose
$c$ to be the bottom of the $(\kappa+1)$-th up-stair.
The proof of the decreasing statement is similar.

% We simply choose the
%biggest number $b< a_1$ such that $b$ be a peak and $\pi(b)> \pi(a_1)$.
%Let $b$ be the tip of the "zeroth" down-stair and let $b< c < a_1$
%be the smallest number such that $\pi(c)< \pi(a_1)$.
%Such number exists since any valley between $b$ and $a_1$
%must lie below
%$\pi(a_1)$, otherwise there would also be a peak somewhere strictly
%between $b$  and $a_1$  higher than $\pi(a_1)$ which
%contradicts to the choice of $b$. Now we simply choose
%$c$ to be the bottom of the zeroth up-stair.

\medskip

Now $f(\a, 1)=\a +1$ is proven as follows. Suppose that
there are two peaks in decreasing position; $a_1< a_2$
and $\pi(a_1) > \pi(a_2)$, then it gives rise to
a decreasing $2$-staircase by the observation above.
Hence if we have $\a + 1$ peaks, and there is no decreasing
$2$-staircase, then the peaks must be monotonically
increasing,  which gives rise to an increasing  $(\a + 1)$-staircase
inductively applying the observation above.
The relation $f(1, \beta) = \beta +1 $ is analogous.

\medskip

The proof of (\ref{recursion}) is done by induction on $\a+\beta$.
Consider a permutation that has at least $f(\a, \beta+1)$
peaks. We have to show that it either  has an increasing $(\a+1)$-staircase
or a decreasing $(\beta+2)$-staircase. 

The  permutation up to the first  $f(\a-1, \beta+1)$ peaks
either has a decreasing $(\beta+2)$-staircase
 or it has an increasing $\a$-staircase.
In the latter case let $a_1 < a_2 < \ldots < a_{\a}$
be the location of the tips.
Now  we look at the next $f(\a, \beta)$ peaks.  If they have
an increasing $(\a+1)$-staircase (within themselves), we are done, so
can assume that it has a decreasing $(\beta+1)$-staircase,
let $b_1 < b_2 <\ldots < b_{\beta +1}$ be the locations of the tips
and certainly $a_\a < b_1$.

Now if $\pi(a_\a) < \pi( b_1)$, 
then we found a new peak after and above $\pi(a_\a)$, hence
by the observation on increasing peaks
 there is an increasing $(\a+1)$-staircase.

If $\pi(a_\a) > \pi( b_1)$, then we  found a 
higher peak before $\pi(b_1)$, hence by the observation on the
decreasing peaks, there is a decreasing
$(\beta+2)$-staircase.
This shows that if the permutation has at least
$f(\a-1, \beta+1)+ f(\a, \beta)$ peaks, then it 
either has a $(\beta+2)$-staircase  or it has an increasing $\a$-chain.
$\;\;\Box$

\end{document}